\documentclass[times,authoryear,3p,final]{elsarticle}
\usepackage{framed,multirow}

\usepackage{amssymb}
\usepackage{latexsym}
\usepackage{mathtools}

\usepackage{makecell}

\usepackage{pdflscape}
\usepackage{caption}
\usepackage{tabularx}
\usepackage{longtable}

\usepackage{amsmath}                           
\usepackage{amssymb}                          
\usepackage{amsthm}
\usepackage{bbm}
\usepackage{graphicx} 
\usepackage[dvipsnames]{xcolor}
\usepackage{url}   
\usepackage{paralist}
\usepackage{subfigure}
\usepackage{todonotes}
\usepackage{setspace}
\usepackage{tablefootnote}

\usepackage{pifont}
\newcommand{\cmark}{\ding{51}}
\newcommand{\xmark}{\ding{55}}
\DeclareMathOperator*{\argmax}{arg\,max} 
\DeclareMathOperator*{\argmin}{arg\,min}
\newcommand{\cmmnt}[1]{}
\newcommand{\revision}[1]{{#1}}
\newcommand{\typo}[1]{{#1}}

\usepackage{adjustbox}
\usepackage{tikz}
\usetikzlibrary{positioning}
\usetikzlibrary{mindmap}
\usetikzlibrary{arrows,shapes,positioning,shadows,trees,calc}

\usepackage[final]{changes}
\usepackage{comment}

\usepackage[hidelinks,colorlinks=true]{hyperref}

\definecolor{newcolor}{rgb}{.8,.349,.1}

\journal{Medical Image Analysis}

\usepackage{fancyhdr}

\begin{document}


\begin{frontmatter}

\title{A Survey on Deep Learning for Skin Lesion Segmentation}%
\author[1]{Zahra {Mirikharaji}\fnref{fn1}}
\author[1]{Kumar {Abhishek}\fnref{fn1}}
\fntext[fn1]{Joint first authors}
\author[2]{Alceu {Bissoto}}
\author[3]{Catarina {Barata}}
\author[2]{Sandra {Avila}}
\author[4]{Eduardo {Valle}}
\author[5]{M. Emre {Celebi}\fnref{fn2}\corref{cor1}}
\author[1]{Ghassan {Hamarneh}\fnref{fn2}\corref{cor1}}
\fntext[fn2]{Joint senior authors}
\cortext[cor1]{Corresponding authors: ecelebi@uca.edu (M. Emre Celebi) and hamarneh@sfu.ca (Ghassan Hamarneh)}

\address[1]{Medical Image Analysis Lab, School of Computing Science, Simon Fraser University, Burnaby V5A 1S6, Canada}
\address[2]{Institute for Systems and Robotics, Instituto Superior Técnico, Avenida Rovisco Pais, Lisbon 1049-001, Portugal}

\address[3]{RECOD.ai Lab, Institute of Computing, University of Campinas, Av. Albert Einstein 1251, Campinas 13083-852, Brazil}
\address[4]{RECOD.ai Lab, School of Electrical and Computing Engineering, University of Campinas, Av. Albert Einstein 400, Campinas 13083-952, Brazil}
\address[5]{Department of Computer Science and Engineering, University of Central Arkansas, 201 Donaghey Ave., Conway, AR 72035, USA}

\begin{abstract}
Skin cancer is a major public health problem that could benefit from computer-aided diagnosis to reduce the burden of this common disease. Skin lesion segmentation from images is an important step toward achieving this goal. However, the presence of natural and artificial artifacts (e.g., hair and air bubbles), intrinsic factors (e.g., lesion shape and contrast), and variations in image acquisition conditions make skin lesion segmentation a challenging task. Recently, various researchers have explored the applicability of deep learning models to skin lesion segmentation. In this survey, we cross-examine \revision{$177$} research papers that deal with deep learning\typo{-}based segmentation of skin lesions. We analyze these works along several dimensions, including input data (datasets, preprocessing, and synthetic data generation), model design (architecture, modules, and losses), and evaluation aspects (data annotation requirements and segmentation performance). We discuss these dimensions both from the viewpoint of select seminal works, and from a systematic viewpoint, examining how those choices have influenced current trends, and how their limitations should be addressed. \revision{To facilitate comparisons, w}e summarize all examined works in a comprehensive table \revision{as well as an interactive table available online\footnote{ \url{https://github.com/sfu-mial/skin-lesion-segmentation-survey}}} 
\end{abstract}

\end{frontmatter}

\section{Introduction}

Segmentation is a challenging and critical operation in the automated skin lesion analysis workflow.  Rule-based skin lesion diagnostic systems, popular in the clinical setting, rely on an accurate lesion segmentation for the estimation of diagnostic criteria such as asymmetry, border irregularity, and lesion size, as needed for implementing the \textsf{ABCD} algorithm (Asymmetry, Border, Color, Diameter of lesions)~\citep{friedman1985early,nachbar1994abcd} and its derivatives: \textsf{ABCDE} (\textsf{ABCD} plus Evolution of lesions)~\citep{abbasi2004early} and \textsf{ABCDEF} (\textsf{ABCDE} plus the ``ugly duckling” sign)~\citep{jensen2015abcdef}. By contrast, in machine learning-based diagnostic systems, restricting the areas within an image, thereby focusing the model on the interior of the lesion, can improve the robustness of the classification. For example, recent studies have shown the utility of segmentation in improving the deep learning (DL)-based classification performance for certain diagnostic categories by {regularizing attention maps~\citep{yan2019melanoma},} allowing the cropping of lesion images~\citep{yu2017,142mahbod2020effects,liu2020automatic,singh2023empirical}, tracking the evolution of lesions~\citep{navarro2018accurate} and the removal of imaging artifacts~\citep{maron2021reducing,bissoto2022artifact}. In a DL-based skin lesion classification framework, presenting the delineated skin lesion to the user can also help with interpreting the DL black box \citep{jaworek2021interpretability}, and thus may either instill trust, or raise suspicion, in computer-aided diagnosis (\textsf{CAD}) systems for skin cancer.

Lesion detection and segmentation are also useful as preprocessing steps when analyzing wide-field images with multiple lesions \citep{birkenfeld2020computer}. Additionally, radiation therapy and image-guided human or robotic surgical lesion excision require localization and delineation of lesions~\citep{cancer2023}. Ensuring fair diagnosis that is unbiased to minority groups, a pressing issue with the deployment of these models and the trust therein, requires the estimation of lesion-free skin tone, which in turn also relies upon the delineation of skin lesions~\citep{kinyanjui2020fairness}. However, despite the importance of lesion segmentation, manual delineation of skin lesions remains a laborious task that suffers from significant inter- and intra-observer variability and consequently, a fast, reliable, and automated segmentation algorithm is needed.

Skin cancer and its associated expenses\typo{, }$\$8.1$ billion annually in U.S.~\citep{guy2015prevalence}\typo{,} have grown into a major public health issue in the past decades. In the \textsf{USA} alone, \revision{$97,610$} new cases of melanoma are expected in \revision{2023~\citep{siegel2023cancer}}. Broadly speaking, there are two types of skin cancer: melanomas and non-melanomas, the former making up just $1\%$ of the cases, but the majority of the deaths due to its aggressiveness.
Early diagnosis is critical for a good prognosis: melanoma can be cured with a simple outpatient surgery if detected early, but its five-year survival rate drops from \revision{over} $99\%$ to \revision{$32\%$} if it is diagnosed at an advanced stage~\citep{cancer2023}.

\cmmnt{Visual inspection by clinicians is the primary step of clinical screening for skin cancers. Two popular strategies for lesion analysis commonly used by experts are intra-patient comparative analysis (\textsf{IPCA}) and lesion-focused analysis (\textsf{LFA}). The ``ugly duckling” sign is the strategy used in \textsf{IPCA} which compares the individual lesions to detect outliers as suspicious spots. On the other hand, \textsf{LFA} utilizes an algorithm to look into the morphological criteria of lesions~\citep{gaudy2017ugly}. \textsf{ABCD} rules (Asymmetry, Border, Color, Diameter of moles)~\citep{nachbar1994abcd}, \textsf{ABCDE} rules (\textsf{ABCD} plus Evolution of moles)~\citep{abbasi2004early} and 7-point checklist~\citep{argenziano1998epiluminescence} are widely used diagnostic algorithms.}

Two imaging modalities are commonly employed in automated skin lesion analysis~\citep{Daneshjou22}: dermoscopic (microscopic) images and clinical (macroscopic) images. While dermoscopic images allow the inspection of lesion properties that are invisible to the naked eye, they are not always accessible even to dermatologists~\citep{engasser2010dermatoscopy}. On the other hand, clinical images acquired using conventional cameras are easily accessible but suffer from lower quality. Dermoscopy is a non-invasive skin imaging technique that aids in the diagnosis of skin lesions by allowing dermatologists to visualize sub-surface structures~\citep{kittler2002diagnostic}. However, even with dermoscopy, diagnostic accuracy\typo{ can vary widely, ranging} from $24\%$ to $77\%$\typo{,} depending on the clinician's level of expertise~\citep{tran2005assessing}. Moreover, dermoscopy may actually lower the diagnostic accuracy in the hands of inexperienced dermatologists~\citep{binder95epiluminescence}. Therefore, to minimize the diagnostic errors that result from the difficulty and \typo{the} subjectivity of visual interpretation and to reduce the burden of skin diseases and limited access to dermatologists, the development of \textsf{CAD} systems is crucial.

\begin{figure}[!ht]
\centering
 \subfigure[Hairs]{\label{fig:prob_a}\includegraphics[width=0.3\columnwidth,draft=false]{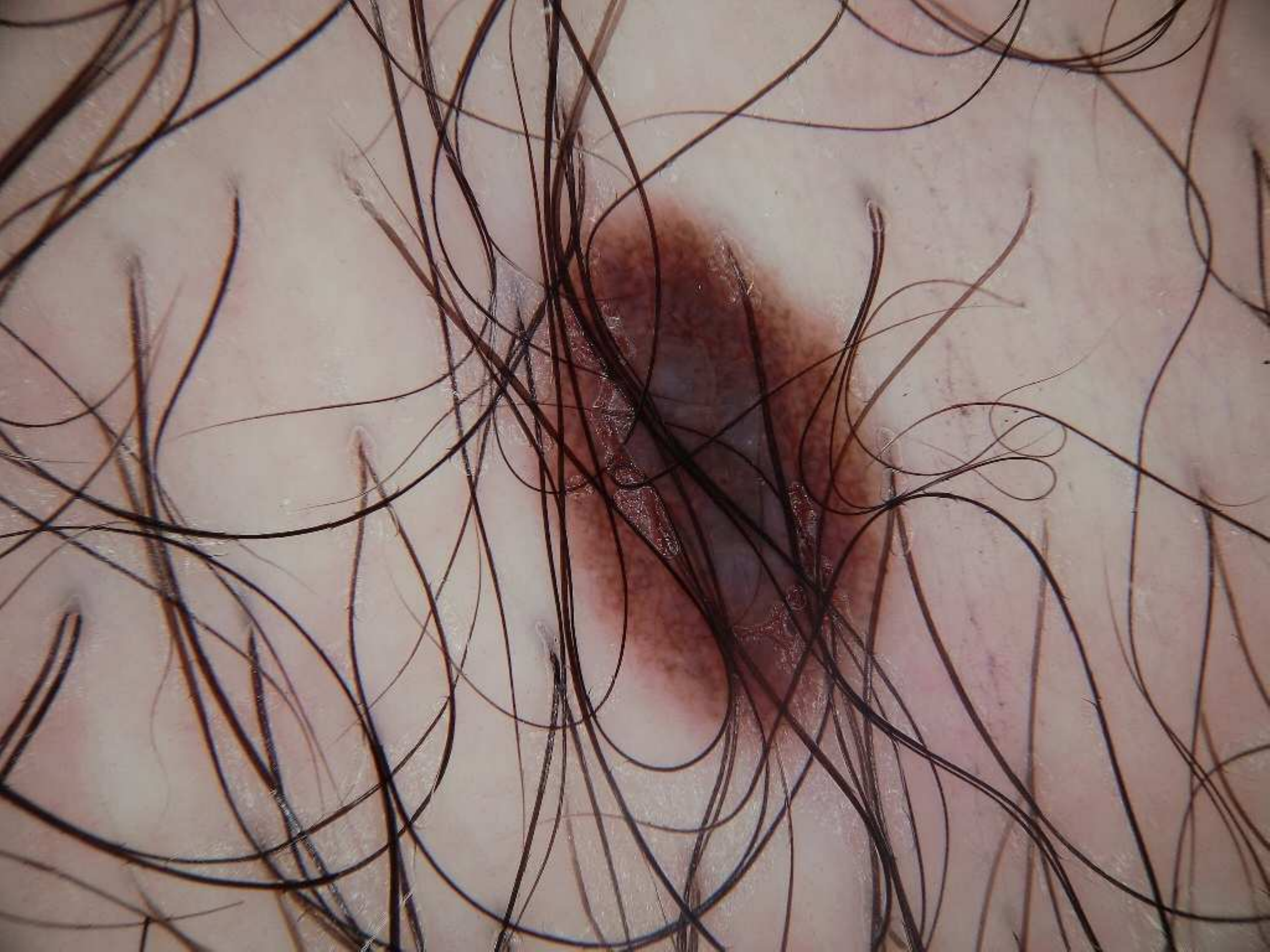}}
 \subfigure[Blood vessels]{\label{fig:prob_b}\includegraphics[width=0.3\columnwidth,draft=false]{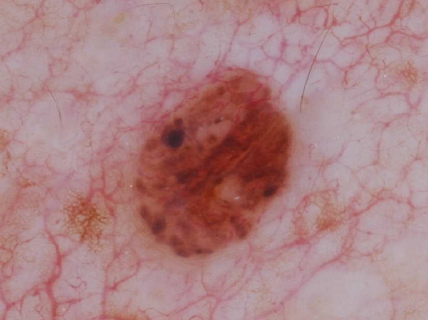}}
 \subfigure[Surgical marking]{\label{fig:prob_c}\includegraphics[width=0.3\columnwidth,draft=false]{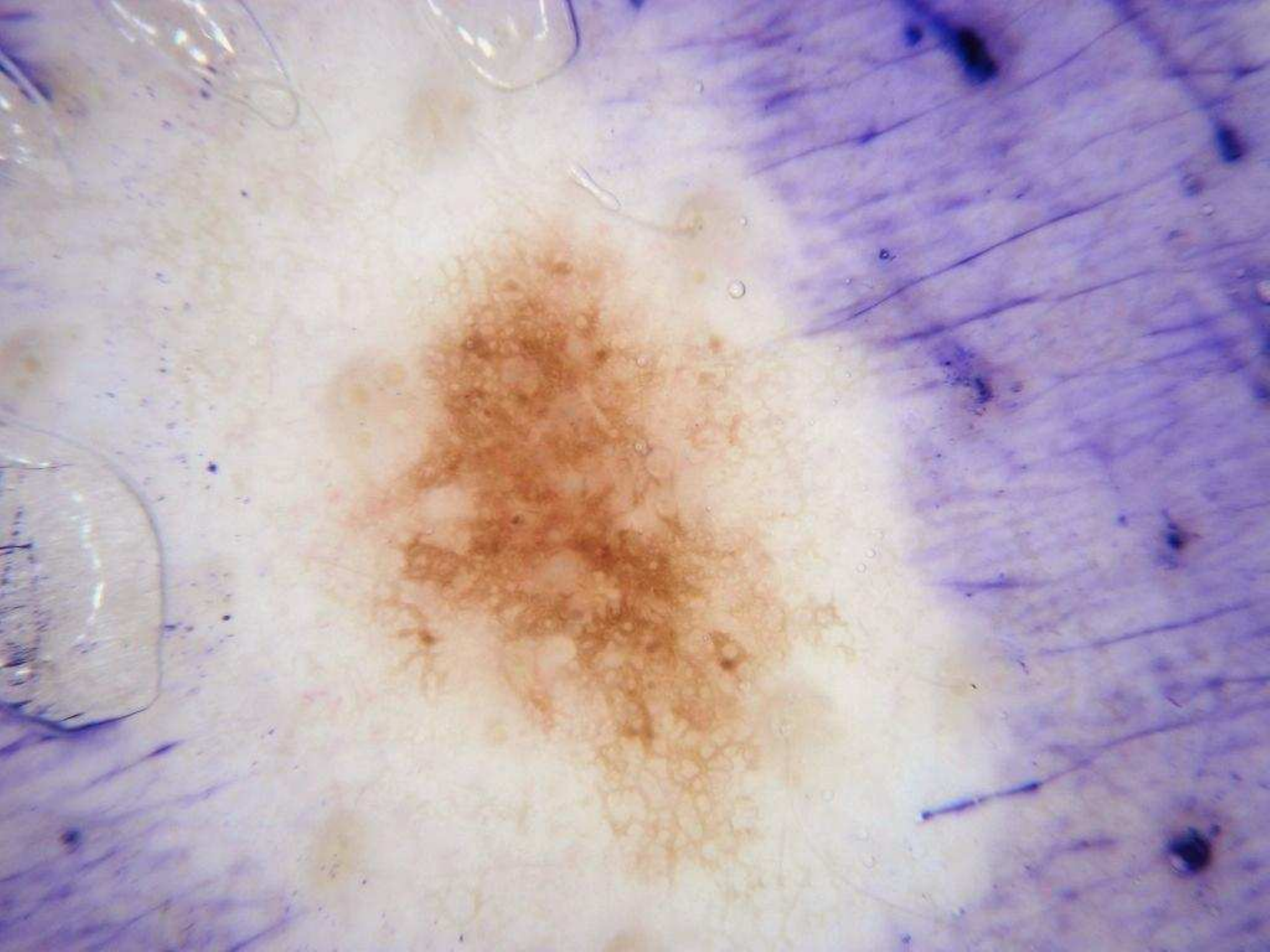}}
 \\
 \subfigure[Irregular border and black frame]{\label{fig:prob_d}\includegraphics[width=0.3\columnwidth,draft=false]{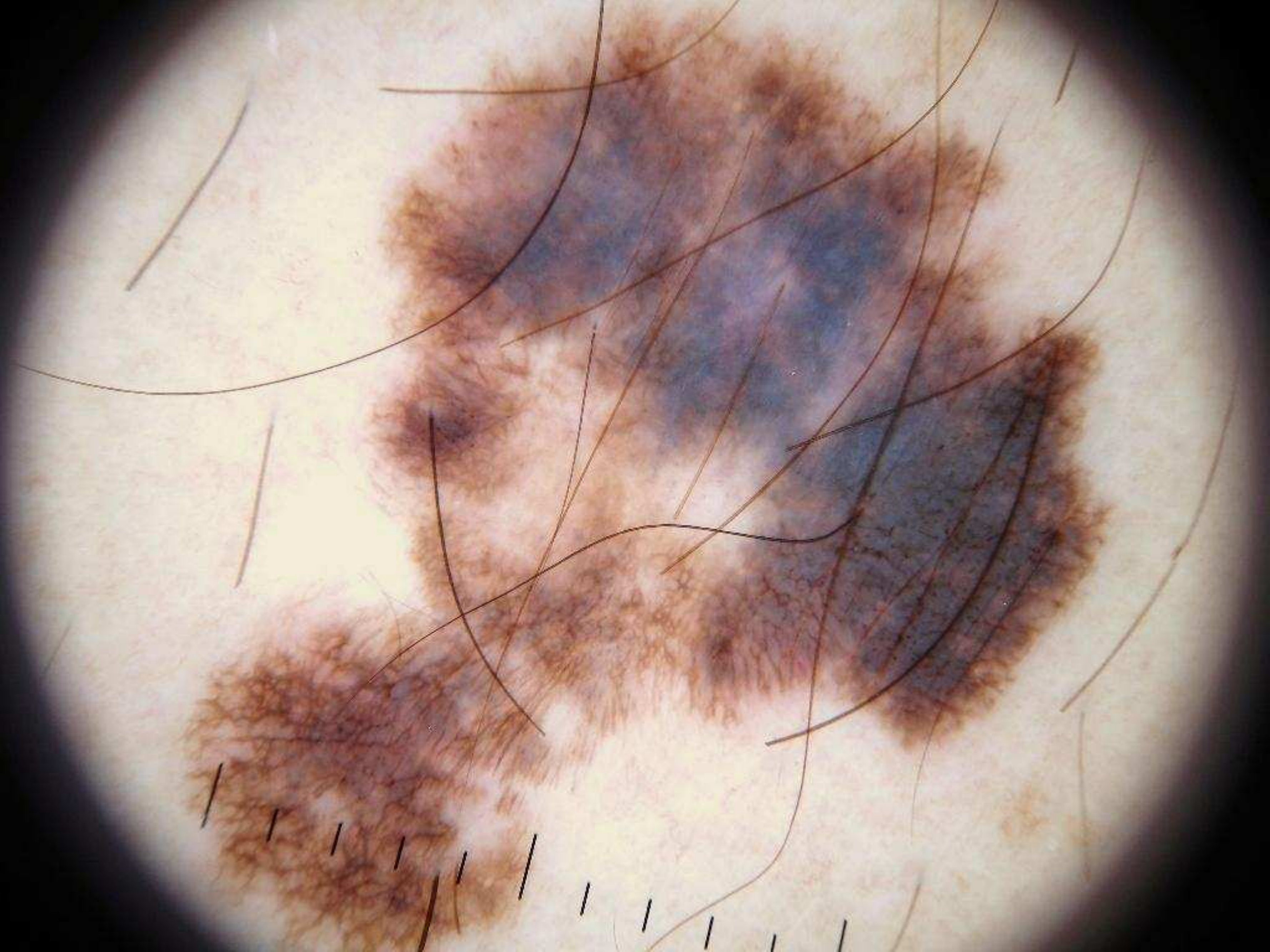}}
 \subfigure[Bubbles]{\label{fig:prob_e}\includegraphics[width=0.3\columnwidth,draft=false]{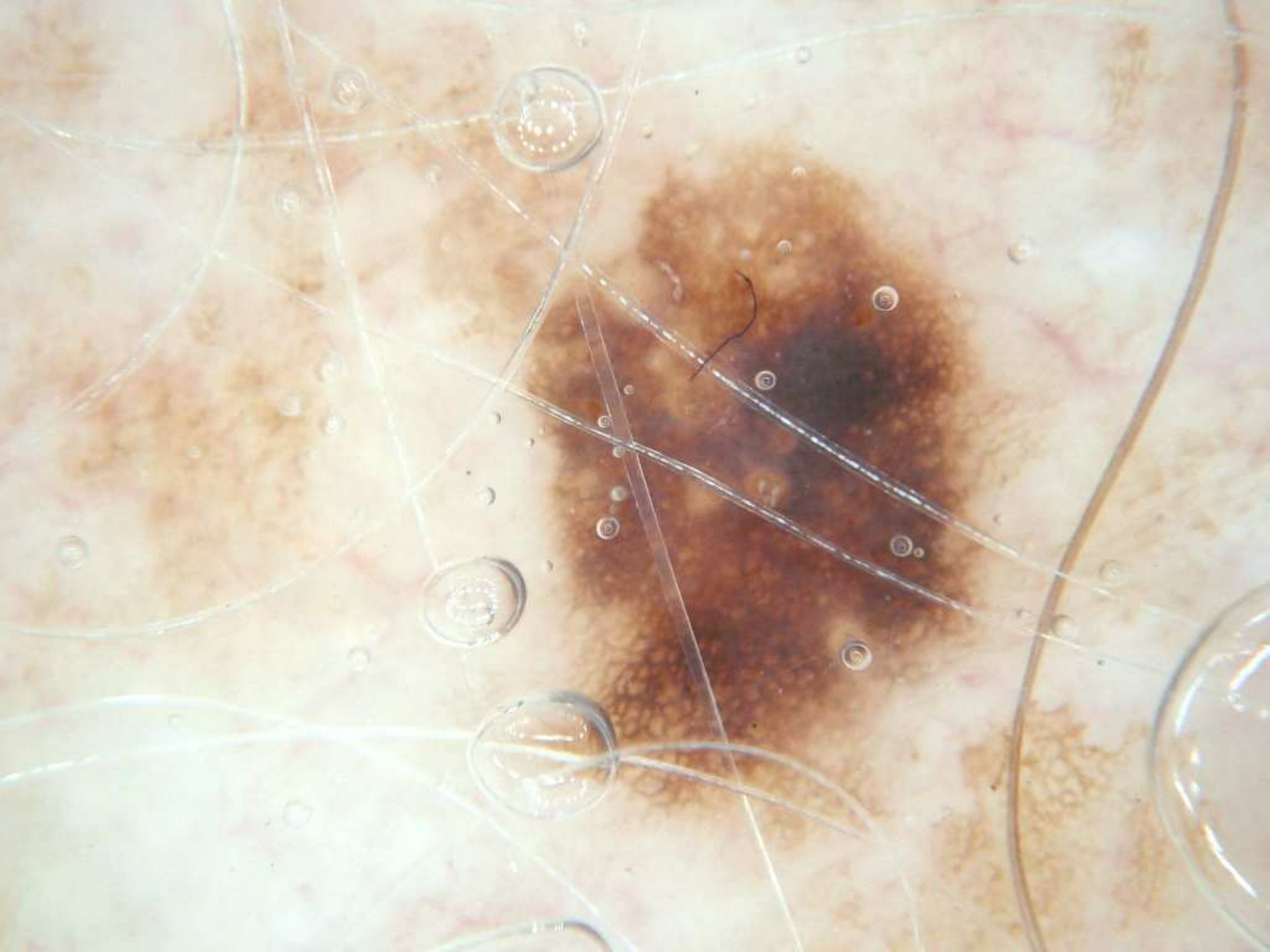}}
 \subfigure[Very small lesion]{\label{fig:prob_f}\includegraphics[width=0.3\columnwidth,draft=false]{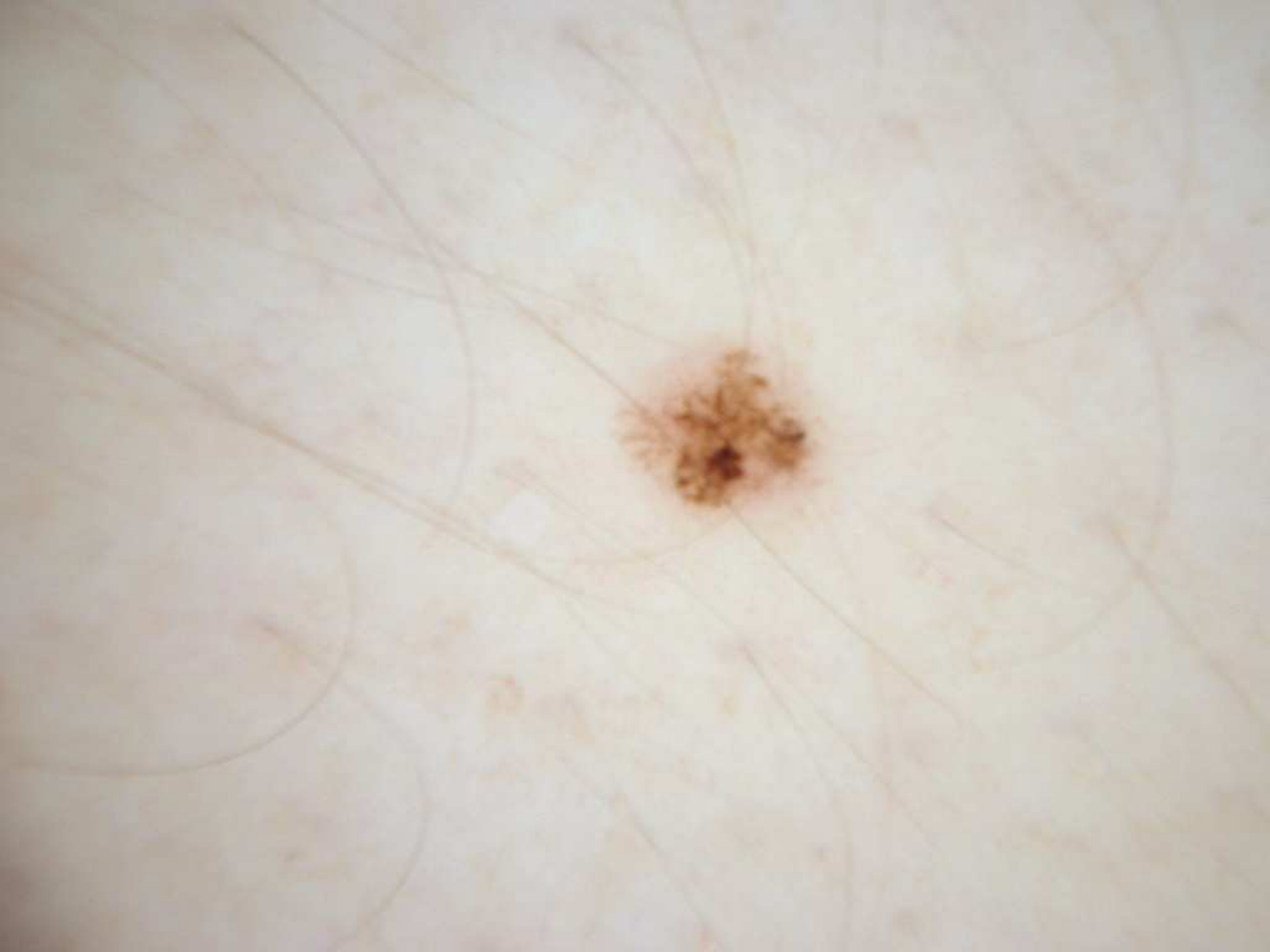}}
 \\
 \subfigure[Very large lesion]{\label{fig:prob_g}\includegraphics[width=0.3\columnwidth,draft=false]{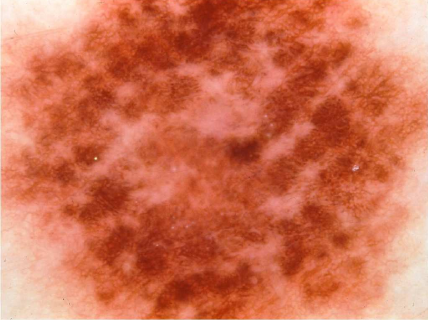}}
 \subfigure[Fuzzy border and variegated
 coloring]{\label{fig:prob_h}\includegraphics[width=0.3\columnwidth,draft=false]{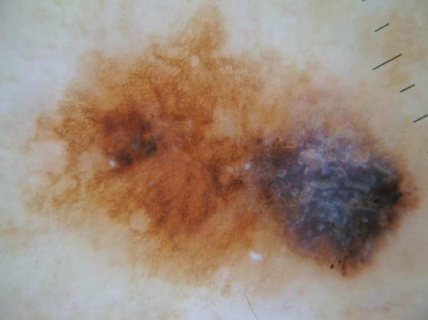}} 
 \subfigure[Low contrast and color calibration chart]{\label{fig:prob_i}\includegraphics[width=0.3\columnwidth,draft=false]{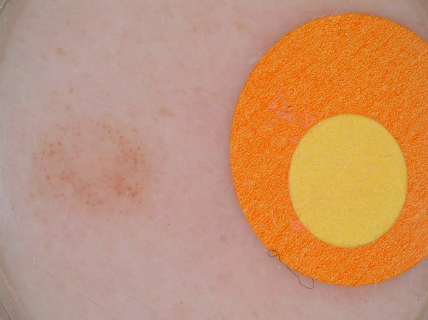}}
 \\
 \caption{Factors that complicate dermoscopy image segmentation (image source: \textsf{ISIC} 2016 dataset~\citep{Gutman16}).}
 \label{fig:probs}
\end{figure}

Segmentation is the partitioning of an image into meaningful regions. Semantic segmentation, in particular, assigns appropriate class labels to each region. For skin lesions, the task is almost always binary, separating the lesion from the surrounding skin. Automated skin lesion segmentation is hindered by illumination and contrast issues, intrinsic inter-class similarities and intra-class variability, occlusions, artifacts, and the diversity of imaging tools used. The lack of large datasets with ground-truth segmentation masks generated by experts compound\typo{s} the problem, impeding both the training of models and their reliable evaluation. Skin lesion images are occluded by natural artifacts such as hair (Fig.~\ref{fig:prob_a}), blood vessels (Fig.~\ref{fig:prob_b}), and artificial ones such as surgical marker annotations (Fig.~\ref{fig:prob_c}), lens artifacts (dark corners) (Fig.~\ref{fig:prob_d}), and air bubbles (Fig.~\ref{fig:prob_e}). Intrinsic factors such as lesion size and shape variation (Fig.~\ref{fig:prob_f} and \ref{fig:prob_g}), different skin colors (Fig.~\ref{fig:prob_h}), low contrast (Fig.~\ref{fig:prob_i}), and ambiguous boundaries (Fig.~\ref{fig:prob_h}) complicate the automated segmentation of skin lesions. 

\begin{figure*}[t]
    \centering
    \begin{adjustbox}{width=13cm, keepaspectratio}
    \begin{tikzpicture}[mindmap, grow cyclic, every node/.style=concept, concept color=orange!40, align=center,
    level 1/.append style={level distance=5cm,sibling angle=90},
    level 2/.append style={level distance=3cm,sibling angle=45}]

\node{DL-based\\Skin Lesion\\Segmentation}
    child [concept color=green!40] { node {Future\\Research\\\S\ref{sec:conc_future}}
    }
    child [concept color=teal!40] { node {Evaluation\\\S\ref{sec:evaluation}}
        child { node {Metrics\\\S\ref{subsec:metrics}}}
        child { node {Inter-\\Annotator\\Agreement\\\S\ref{subsec:agreement}}}
        child { node {Segmentation\\Annotation\\\S\ref{subsec:annotation}}}
    }
    child [concept color=red!30] { node {Model Design\\\& Training\\\S\ref{sec:model}}
    child { node {Loss Functions\\\S\ref{sebsec:loss}}}
    child { node {Model\\Architecture\\\S\ref{subsec:architecture}}}
    }
    child [concept color=blue!30] { node {Input Data\\\S\ref{sec:input}}
        child { node {Image\\Preprocessing\\\S\ref{subsec:preproc}}}
        child { node {Supervision\\\S\ref{subsec:semi}}}
        child { node [scale=1.] {Synthetic Data\\\S\ref{subsec:aug}}}
        child { node {Datasets\\\S\ref{subsec:datasets}}}
    };
\end{tikzpicture}
\end{adjustbox}
\caption{\revision{An overview of the various components of this review. We structure the review based on the different elements of a DL-based segmentation pipeline and conclude it with discussions on future potential research directions.}}
\label{fig:paper_layout}
\end{figure*}
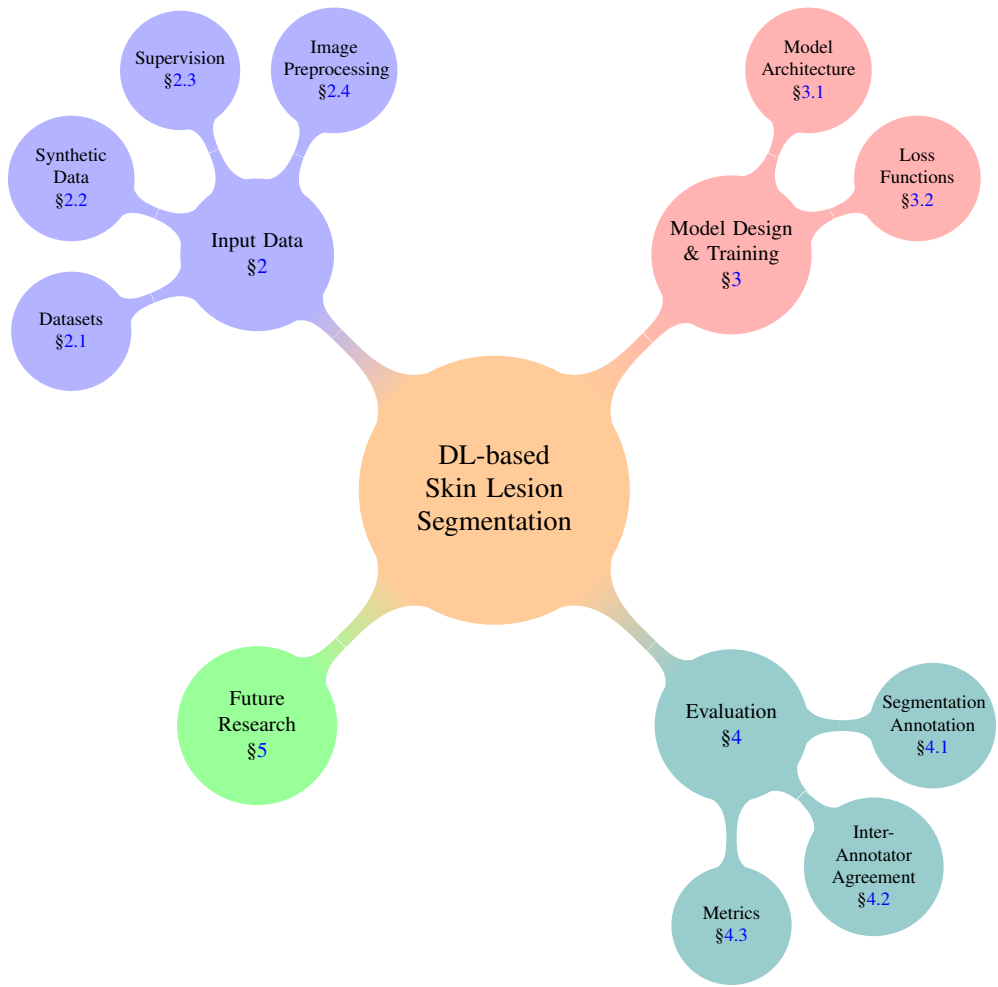

Before the deep learning (\textsf{DL}) revolution, segmentation was based on classical image processing and machine learning techniques such as adaptive thresholding~\citep{green1994computer,Celebi13}, active contours~\citep{Erkol05}, region growing~\citep{Iyatomi06,Celebi07b}, unsupervised clustering~\citep{gomez2007independent}, and support vector machines~\citep{zortea2011automatic}. These approaches depend on hand-crafted features, which are difficult to engineer and often limit invariance and discriminative power from the outset. As a result, such conventional segmentation algorithms do not always perform well on  larger and more complex datasets. In contrast, \textsf{DL} integrates feature extraction and task-specific decision seamlessly, and \typo{does} not just cope with, but actually require\typo{s} larger datasets.

\smallskip
\noindent \emph{Survey of surveys.}
\citet{celebi09b} reviewed 18 skin lesion segmentation algorithms for dermoscopic images, published between 1998 and 2008, with their required preprocessing and postprocessing steps. \citet{Celebi15a} later extended their work with 32 additional algorithms published between 2009 and 2014, discussing performance evaluation and computational requirements of each approach, and suggesting guidelines for future works. Both surveys appeared before \textsf{DL} was widely adopted for skin lesion segmentation, but  cover all the important works based on classical image processing and machine learning. \citet{adegun2020deep} reviewed the literature on \textsf{DL}\typo{-}based skin image analysis, with an emphasis on the best-performing algorithms in the \textsf{ISIC} (International Skin Imaging Collaboration) Skin Image Analysis Challenges 2018~\citep{Codella19} and 2019~\citep{Tschandl18,Codella18,Combalia19}. \revision{However, since their review focused on the ISIC Challenges 2018 and 2019, it is more general as it covers both lesion classification and segmentation. Consequently, the number of papers surveyed for skin lesion segmentation by \citet{adegun2020deep} is almost an order of magnitude smaller than that in this review.}

\smallskip
\noindent \emph{Main contributions.}
No existing survey approaches the present work in breadth or depth, as we cross-examine \revision{$177$} research papers that deal with the automated segmentation of skin lesions in clinical and dermoscopic images. We analyze the works along several dimensions, including input data (datasets, preprocessing, and synthetic data generation), model design (architecture, modules, and losses), and evaluation (data annotation and evaluation metrics). We discuss these dimensions both from the viewpoint of select seminal works, and from a systematic viewpoint, examining how those choices have influenced current trends, and how their limitations should be addressed. We summarize all examined works in a comprehensive table to facilitate comparisons.

\smallskip
\noindent 
\emph{Search strategy.}
We searched DBLP and Arxiv Sanity Preserver for all scholarly publications: peer-reviewed journal papers, papers published in the proceedings of conferences or workshops, and non-peer-reviewed preprints from 2014 to \revision{2022}. The DBLP search query was \texttt{(conv* | \revision{trans* |} deep | neural | learn*) (skin | derm*) (segment* | delineat* | extract* | localiz*)}, thus restricting our search to \textsf{DL}-based works involving skin and segmentation. We use DBLP for our literature search because (a) it allows for customized search queries and lists, and (b) we did not find any relevant publications on other platforms (Google Scholar and PubMed) that were not indexed by DBLP. For unpublished preprints, we also searched on Arxiv Sanity Preserver using a similar query\footnote{\textbf{Arxiv Sanity Preserver}: \url{https://www.arxiv-sanity-lite.com/search?q=segmentation+skin+melanoma+deep+learning+convolution+transformer}}. We filtered our search results to remove false positives \revision{($31$ papers)} and included only papers related to skin lesion segmentation. We excluded papers that focused on general skin segmentation \typo{and} general skin conditions (e.g., psoriasis, acne, or certain sub-types of skin lesions). We also included unpublished preprints from arXiv, which \revision{(a)} passed minimum quality checks levels \revision{and (b) had at least 10 citations,} and excluded those that were clearly of low quality. In particular, papers that had one or more of the following were excluded from this survey: (a) missing quantitative results, (b) missing important sections such as Abstract or Methods, (c) conspicuously poor writing quality, and (d) no methodological contribution. \revision{This led to the filtering out of papers of visibly low quality ((a-c) criteria above; $18$ papers) and those with no methodological contribution ($20$~papers).}

The remaining text is organized as follows\typo{: i}n Section~\ref{sec:input}, we introduce the publicly available datasets and discuss preprocessing and synthetic data generation\typo{; i}n Section~\ref{sec:model}, we review the various network architectures used in deep segmentation models and discuss how deep models benefit from these networks. We also describe various loss functions designed either for general use or specifically for skin lesion segmentation. In Section~\ref{sec:evaluation}, we detail segmentation evaluation techniques and measures. Finally, in Section~\ref{sec:conc_future}, we discuss the open challenges in \textsf{DL}\typo{-}based skin lesion segmentation and conclude our survey. \revision{A visual overview of the structure of this survey is presented in Fig.~\ref{fig:paper_layout}.}

\section{Input Data}
\label{sec:input}

Obtaining data in sufficient quantity and quality is often a significant obstacle to developing effective segmentation models. State-of-the-art segmentation models have a huge number of adjustable parameters that allow them to generalize well, provided they are trained on massive labeled datasets~\citep{sun2017revisiting,Buslaev20}. Unfortunately, skin lesion datasets---like most medical image datasets\revision{~\citep{asgari2021deep}}---tend to be small~\citep{Curiel19} due to issues \typo{such as} copyright, patient privacy, acquisition \typo{and} annotation cost\typo{,} standardization, and scarcity of many pathologies of interest. The two most common modalities used in the training of skin lesion segmentation models are \textit{clinical images}, which are close-ups of the lesions acquired using conventional cameras, and \textit{dermoscopic images}, which are acquired using dermoscopy, a non-invasive skin imaging through optical magnification, and either liquid immersion and low angle-of-incidence lighting, or cross-polarized lighting. Dermoscopy eliminates skin surface reflections~\citep{kittler2002diagnostic}, reveals subsurface skin structures, and allows the identification of dozens of morphological features such as atypical pigment networks, dots/globules, streaks, blue-white areas, and blotches~\citep{Menzies03}.

Annotation is often the greatest barrier for increasing the amount of data. Objective evaluation of segmentation often requires laborious \emph{region-based annotation}, in which an expert manually outlines the region where the lesion (or a clinical feature) appears in the image. By contrast, \emph{textual annotation} may involve diagnosis (e.g., melanoma, carcinoma, benign nevi), presence/absence/score of dermoscopic features (e.g., pigment networks, blue-white areas, streaks, globules), diagnostic strategy (e.g., pattern analysis, \textsf{ABCD} rule, 7-point checklist, 3-point checklist), clinical metadata (e.g., sex, age, anatomic site, familial history), and other details (e.g., timestamp, camera model)~\citep{Caffery18}. We \typo{extensively discuss the image annotation issue} in Section~\ref{subsec:annotation}. 

\subsection{Datasets}
\label{subsec:datasets}

The availability of larger, more diverse, and better-annotated datasets is one of the main driving factors for the advances in skin image analysis in the past decade~\citep{Marchetti18,Celebi19b}. 
Works in skin image analysis date back to the 1980s \citep{Vanker84,Dhawan84}, but until the mid-2000s, \typo{these works} used small, private datasets, containing \typo{a} few hundred images. 

The \emph{Interactive Atlas of Dermoscopy} (sometimes called the \emph{Edra Atlas}, in reference to the publisher) by \citet{Argenziano00} included a \textsf{CD-ROM} with $1,039$ dermoscopy images ($26\%$ melanomas, $4\%$ carcinomas, $70\%$ nevi) of $1,024 \times 683$ pixels, acquired by three European university hospitals (University of Graz, Austria, University of Naples, Italy, and University of Florence, Italy). The works of \citet{Celebi07a,Celebi08} popularized the dataset in the dermoscopy image analysis community, where it became a \textit{de facto} evaluation standard for almost a decade, until the much larger \textsf{ISIC} Archive datasets (see below) became available. Recently, \citet{Kawahara19} placed this valuable dataset, along with additional textual annotations based on the 7-point checklist, in public domain under the name \emph{derm7pt}. Shortly after the publication of the Interactive Atlas of Dermoscopy, \citet{Menzies03} published \emph{An Atlas of Surface Microscopy of Pigmented Skin Lesions: Dermoscopy}, with a CD-ROM containing $217$ dermoscopic images ($39\%$ melanomas, $7\%$ carcinomas, $54\%$ nevi) of $712 \times 454$ pixels, acquired at the Sydney Melanoma Unit, Australia.

The $\mathsf{PH}^2$ dataset, released by \citet{Mendonca13} and detailed by \citet{Mendonca15}, was the first public dataset to provide region-based annotations with segmentation masks, and masks for the clinically significant colors (white, red, light brown, dark brown, blue-gray, and black) present in the images. The dataset contains $200$ dermoscopic images ($20\%$ melanomas, $40\%$ atypical nevi, and $40\%$ common nevi) of $768 \times 560$ pixels, acquired at the Hospital Pedro Hispano, Portugal. The Edinburgh DermoFit Image Library~\citep{ballerini2013color} also provides region-based annotations for $1,300$ clinical images (10 diagnostic categories including melanomas, seborrhoeic keratosis, and basal cell carcinoma) of sizes ranging from $177 \times 189$  to $2,176 \times 2,549$ pixels. The images were acquired with a Canon EOS 350D SLR camera, in controlled lighting and at a consistent distance from the lesions, resulting in a level of quality atypical for clinical images.

The \textsf{ISIC} Archive contains the world's largest curated repository of dermoscopic images. \textsf{ISIC}, an international academia-industry partnership sponsored by \textsf{ISDIS} (International Society for Digital Imaging of the Skin), aims to “facilitate the application of digital skin imaging to help reduce melanoma mortality”~\citep{ISICArchive}. At the time of writing, the archive contains more than \revision{$240,000$} images, of which \revision{more than $71,000$} are publicly available. These images were acquired in leading worldwide clinical centers, using a variety of devices. 

In addition to curating the datasets that collectively form the “\textsf{ISIC} Archive”, \textsf{ISIC} has released standard archive subsets as part of its \emph{Skin Lesion Analysis Towards Melanoma Detection} Challenge, organized annually since 2016. The 2016, 2017, and 2018 challenges comprised segmentation, feature extraction, and classification tasks, while the 2019 and 2020 challenges featured only classification. Each subset is associated with a challenge (year), one or more tasks, and has two (training/test) or three (training/validation/test) splits. The \textsf{ISIC} Challenge 2016~\citep{Gutman16} (\textsf{ISIC} 2016, for \typo{brevity}) contains $1,279$ images split into $900$ for training ($19\%$ melanomas, $81\%$ nevi), and $379$ for testing ($20\%$ melanomas, $80\%$ nevi). There is a large variation in image size, ranging from $0.5$ to $12$ megapixels. All tasks used the same images. 
The \textsf{ISIC} 2017~\citep{Codella18} dataset more than doubled, with $2,750$ images split into $2,000$ for training ($18.7\%$ melanomas, $12.7\%$ seborrheic keratoses, $68.6\%$ nevi), $150$ for validation ($20\%$ melanomas, $28\%$ seborrheic keratoses, $52\%$ nevi), and $600$ for testing ($19.5\%$ melanomas, $15\%$ seborrheic keratoses, $65.5\%$ nevi). Again, image size varied markedly, ranging from $0.5$ to $29$ megapixels, and all tasks used the same images.

\textsf{ISIC} 2018 provided\typo{,} for the first time\typo{,} separate datasets for the tasks, with $2,594$ training ($20$\% melanomas, $72$\% nevi, and $8$\% seborrheic keratoses) and $100$/$1,000$ for validation/test images ranging from $0.5$ to $29$ megapixels, for the tasks of segmentation and feature extraction~\citep{Codella19}, and $10,015$/$1,512$ training/test images for the classification task, all with $600 \times 450$ pixels. The training dataset for classification was the \textsf{HAM10000} dataset~\citep{Tschandl18}, acquired over a period of $20$ years at the Medical University of Vienna, Austria and the private practice of Dr.~Cliff Rosendahl, Australia. It allowed a five-fold increase in training images in comparison to 2017 and comprised seven diagnostic categories: melanoma ($11.1\%$), nevus ($66.9\%$), basal cell carcinoma ($5.1\%$), actinic keratosis or Bowen's disease ($3.3\%$), benign keratosis (solar lentigo, seborrheic keratosis, or lichen planus-like keratosis, $11\%$), dermatofibroma ($1.1\%$), and vascular lesion ($1.4\%$). \revision{As a part of a 2020 study of human-computer collaboration for skin lesion diagnosis involving dermatologists and general practitioners~\citep{tschandl2020human}, the lesions in the \textsf{HAM10000} dataset were segmented by a single dermatologist and consequently released publicly \citep{ham10ksegmentations}, making this the single largest publicly available skin lesion segmentation dataset (Table~\ref{datasets}).}

\begin{table}
\centering
\caption{Public skin lesion datasets with \revision{lesion} segmentation annotations. \revision{All the datasets contain RGB images of skin lesions.}}
\resizebox{\textwidth}{!}{
\setlength{\tabcolsep}{0.5em}
\def\arraystretch{1.95}
\begin{tabular}{|c|c|c|c|c|c|c|}
\hline

dataset& year &modality & size & training/validation/test & class distribution &additional info\\
\hline
DermQuest\tablefootnote{DermQuest was deactivated on December 31, 2019. However, 137 of its images are publicly available~\citep{Skin_Cancer_Detection_data}.}~\citep{dermquest} &2012 & clinical & $137$ & -- & \makecell{61 non-melanomas\\76 melanomas} & \makecell{acquired with different cameras\\ under various lighting conditions}\\
\hline
DermoFit~\citep{ballerini2013color} & 2013 & clinical & $1,300$ & -- & \makecell{$1,224$ non-melanomas\\76 melanomas} & \makecell{sizes ranging from $177 \times 189$ \\ to $2,176 \times 2,549$ pixels}\\
\hline
 \makecell{Pedro Hispano Hospital (PH$^2$)\\~\citep{Mendonca13}} &2013 &dermoscopy & $200$ & -- & \makecell{160 benign nevi\\40 melanomas} & \makecell{sizes ranging from $553 \times763$ \\ to $577 \times769$ pixels\\ acquired at $20\times$ magnification}\\
\hline
 ISIC2016~\citep{Gutman16} & 2016 &dermoscopy & $1,279$ & $900$/--/$379$ & \makecell{Training:\\ 727 non-melanomas\\ 173 melanomas\\
Test:\\ 304 non-melanomas\\ 75 melanomas} & \makecell{sizes ranging from $566 \times679$\\ to $2,848 \times4,288$ pixels}\\
\hline
 ISIC2017~\citep{Codella18} & 2017 &dermoscopy & $2,750$ & $2,000$/$150$/$600$ & \makecell{Training:\\ $1,626$ non-melanomas\\374 melanomas
\\Test:\\ 483 non-melanomas\\ 117 melanomas
} & \makecell{sizes ranging from $540 \times 722$ \\to $4,499 \times 6,748$ pixels}\\
\hline
 ISIC2018~\citep{Codella19} &2018 &dermoscopy & $3,694$ & $2,594$/$100$/$1,000$ & -- & \makecell{sizes ranging from $540\times576$\\ to $4,499 \times 6,748$ pixels}\\
 \hline
 \makecell{HAM10000\\\citep{Tschandl18}\\\citep{tschandl2020human}\\\citep{ham10ksegmentations}} &\revision{2020} &\revision{dermoscopy} & \revision{$10,015$} & \revision{--} & \revision{\makecell{$1,113$ non-melanomas\\$8,902$ melanomas}} & \revision{\makecell{all images of $600 \times 450$ pixels}}\\
\hline
\end{tabular}}
\label{datasets}
\end{table}

\textsf{ISIC} 2019~\citep{Codella18,Tschandl18,Combalia19} contains $25,331$ training images ($18\%$ melanomas, $51\%$ nevi, $13\%$ basal cell carcinomas, $3.5\%$ actinic keratoses, $10\%$ benign keratoses, $1\%$ dermatofibromas, $1\%$ vascular lesions, and $2.5\%$ squamous cell carcinomas) and $8,238$ test images (diagnostic distribution unknown). The images range from $600 \times 450$ to $1,024 \times 1,024$ pixels.

\textsf{ISIC} 2020~\citep{Rotemberg21} contains $33,126$ training images ($1.8\%$ melanomas, $97.6\%$ nevi, $0.4\%$ seborrheic keratoses, $0.1\%$ lentigines simplex, $0.1\%$ lichenoid keratoses, $0.02\%$ solar lentigines, $0.003\%$ cafe-au-lait macules, $0.003\%$ atypical melanocytic proliferations) and $10,982$ test images (diagnostic distribution unknown), ranging from 0.5 to 24 megapixels. Multiple centers, distributed worldwide, contributed to the dataset, including the Memorial Sloan Kettering Cancer Center \typo{(}\textsf{USA}\typo{)}, the Melanoma Institute, the Sydney Melanoma Diagnostic Centre, and the University of Queensland \typo{(}Australia\typo{)}, the Medical University of Vienna \typo{(}Austria\typo{)}, the University of Athens \typo{(}Greece\typo{)}, and the Hospital Clinic Barcelona \typo{(}Spain\typo{)}. An important novelty in this dataset is the presence of multiple lesions per patient, with the express motivation of exploiting intra- and inter-patient lesion patterns, e.g., the so-called “ugly-ducklings”, lesions whose appearance\typo{s are} atypical for a given patient, and which present an increased risk of malignancy~\citep{gachon2005first}.

\revision{There is, however, an overlap among these ISIC Challenge datasets. \cite{abhishek2020input} analyzed all the lesion segmentation datasets from the ISIC Challenges (2016-2018) and found considerable overlap between these 3 datasets, with as many as $1,940$ images shared between at least 2 datasets and $706$ images shared between all 3 datasets. In a more recent analysis of the ISIC Challenge datasets for the lesion diagnosis task from 2016 through 2020, \cite{cassidy2022analysis} found overlap between the datasets as well as the presence of duplicates within the datasets. Using a duplicate removal strategy, they curated a new set of $45,590$ training images ($8.61\%$ melanomas, $91.39\%$ others) and $11,397$ validation images ($8.61\%$ melanomas, $91.39\%$ others), leading to a total of $56,987$ images. Additionally, since the resulting dataset is highly imbalanced (melanomas versus others in a ratio of $1:10.62$), the authors also curated a balanced dataset with $7,848$ training images (50\% melanoma, 50\% others) and $1,962$ validation images (50\% melanoma, 50\% others).}

\begin{figure*}
\centering
\includegraphics[width=6in]{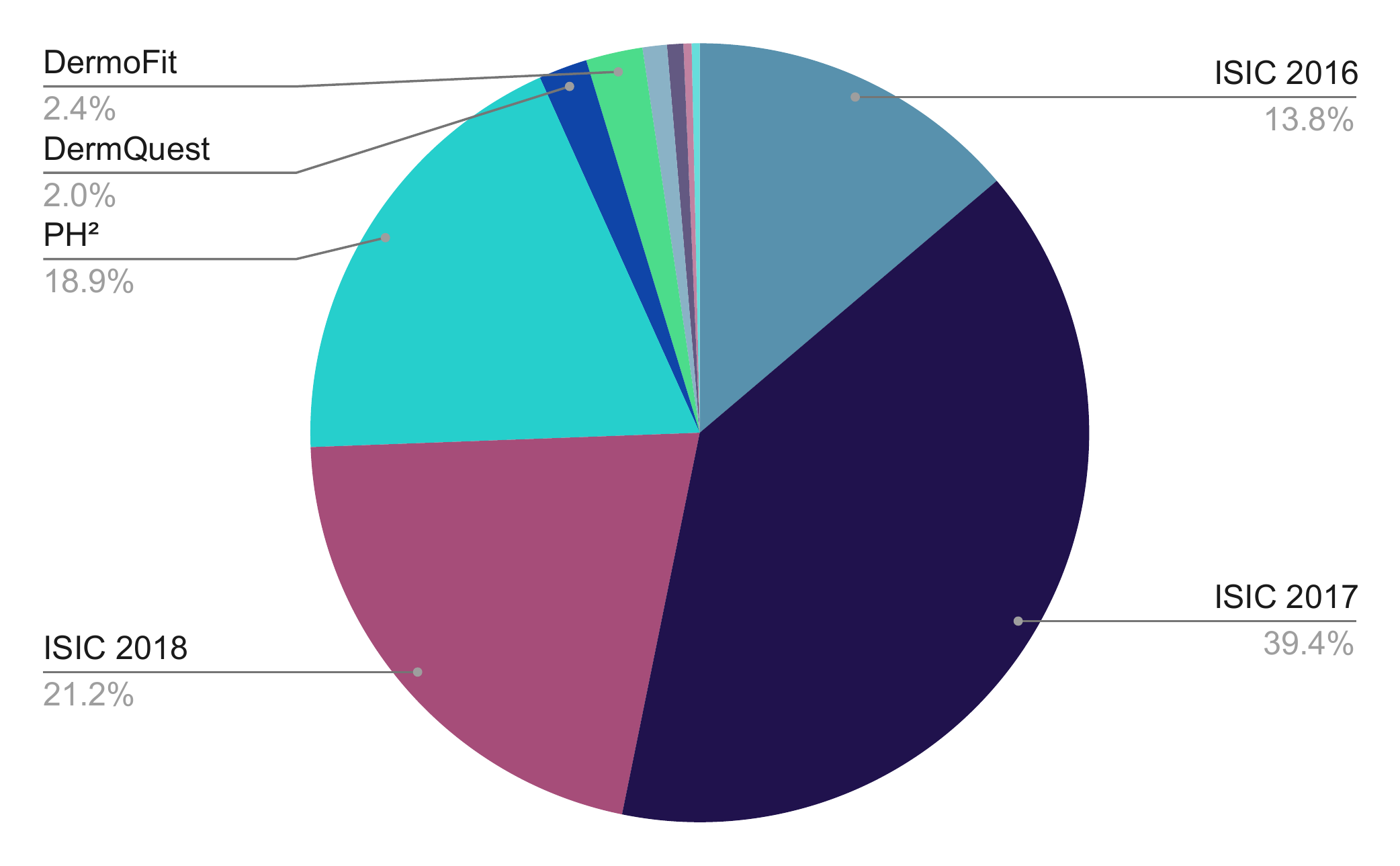}
\caption{The frequency of utilization of different skin lesion segmentation datasets in the surveyed studies. \revision{We found that $82$ papers evaluated on more than $1$ dataset, with $36$ papers opting for cross-dataset evaluation (\textsf{CDE} in Table~\ref{tab:main}). \textsf{ISIC} datasets (\textsf{ISIC} 2016, \textsf{ISIC} 2017, \textsf{ISIC} 2018, and \textsf{ISIC} Archive) are used in the majority of papers, with $168$ of $177$ papers using at least one \textsf{ISIC} dataset and the \textsf{ISIC} 2017 dataset being the most popular ($117$ papers). The \textsf{PH}$^2$ dataset is the second most widely used ($56$ papers) following \textsf{ISIC} datasets.}}
\label{data_eval_freq}
\end{figure*}

Table~\ref{datasets} shows a list of publicly available skin lesion datasets with pixel-wise annotations, image modality, sample size, original split sizes, and diagnostic distribution. Fig.~\ref{data_eval_freq} shows how frequently these datasets appear in the literature. It is also worth noting that several other skin lesion image datasets have not been described in our survey as they do not provide the corresponding skin lesion segmentation annotations. However, these datasets, including \textsf{SD-198}~\citep{sun2016benchmark}, \textsf{MED-NODE}~\citep{giotis2015med}, derm7pt~\citep{Kawahara19}, Interactive Dermatology Atlas~\citep{usatine2013interactive}, Dermatology Information System~\citep{DermIS}, DermWeb~\citep{DermWeb}, DermNet New Zealand~\citep{DermNetNZ}, may still be relevant for skin lesion segmentation research (see Section~\ref{sec:conc_future}).

Biases in \typo{c}omputer \typo{v}ision datasets are a constant source of issues~\citep{Torralba2011_CVPR}, which is compounded \typo{in} medical imag\typo{ing} due to the small\typo{er} number of samples, insufficient \typo{image} resolution, lack of geographical or ethnic diversity, or statistics  unrepresentative of clinical practice. All existing skin lesion datasets suffer to a certain extent from one or more of the aforementioned issues, to which we add the specific issue of the availability and reliability of annotations. For lesion classification, many samples lack \typo{the gold standard} histopathological confirmation, and  ground-truth segmentation, even when available, is inherently noisy (Section~\ref{subsec:agreement}). The presence of artifacts (Fig.~\ref{fig:probs}) may lead to spurious correlations, an issue that \citet{Bissoto_2019_CVPR} attempted to quantify for classification models.

\subsection{Synthetic Data Generation}
\label{subsec:aug}

\begin{figure}[!ht]
\centering
 \subfigure[Original]{\label{fig:aug_a}\includegraphics[width=0.3\columnwidth,draft=false]{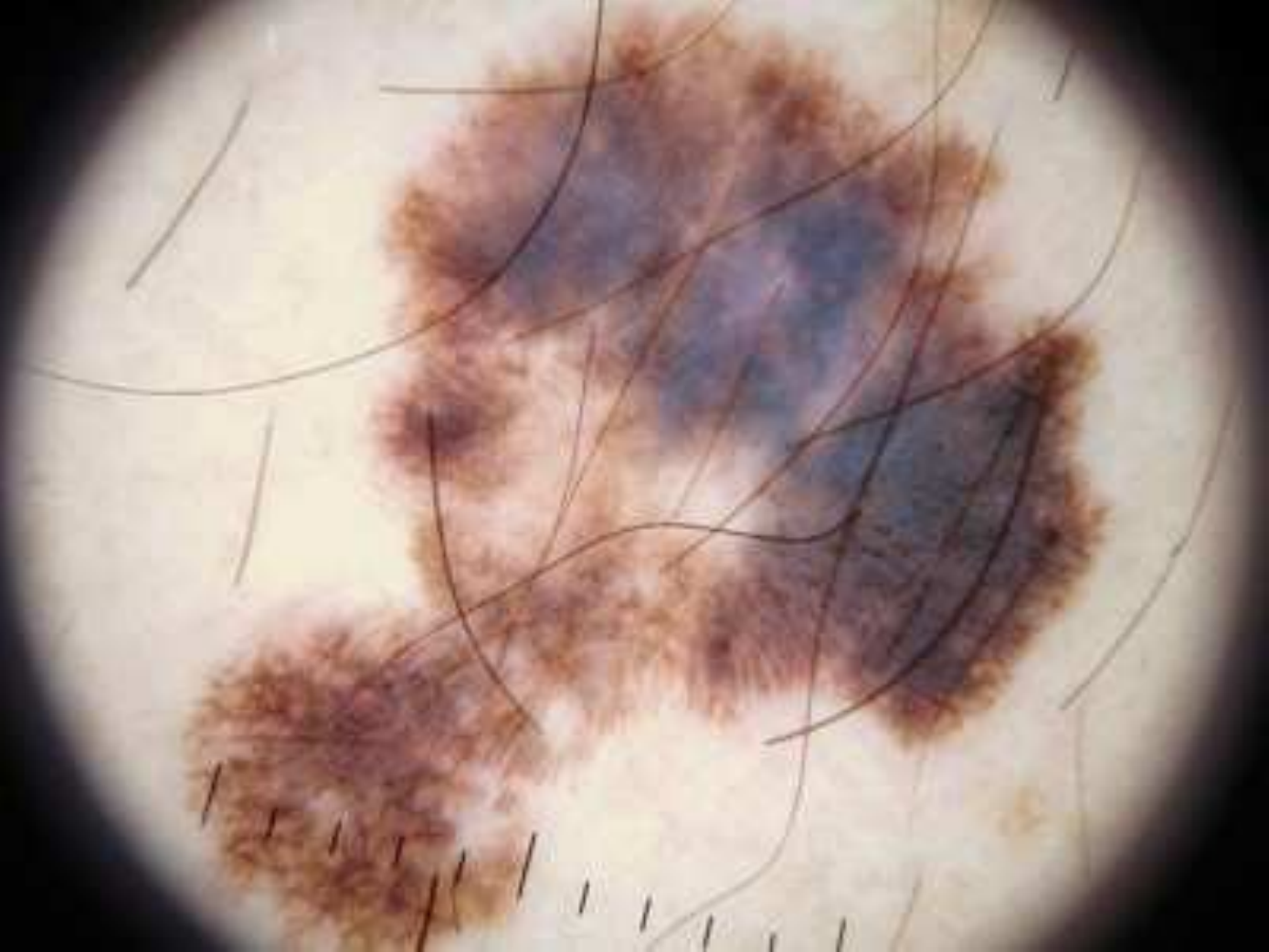}}
 \subfigure[Affine deformation]{\label{fig:aug_b}\includegraphics[width=0.3\columnwidth,draft=false]{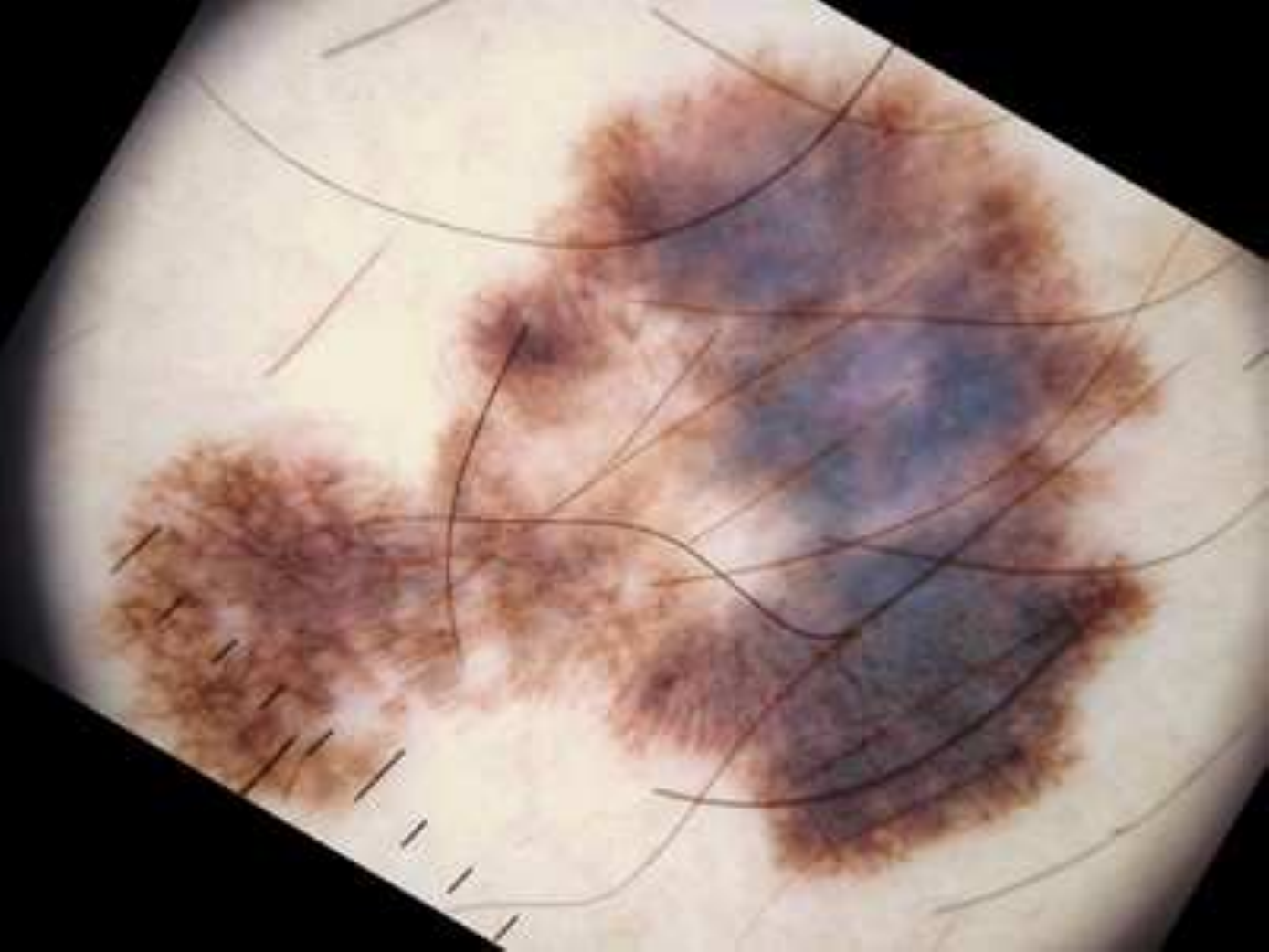}}
 \subfigure[Elastic deformation ]{\label{fig:aug_c}\includegraphics[width=0.3\columnwidth,draft=false]{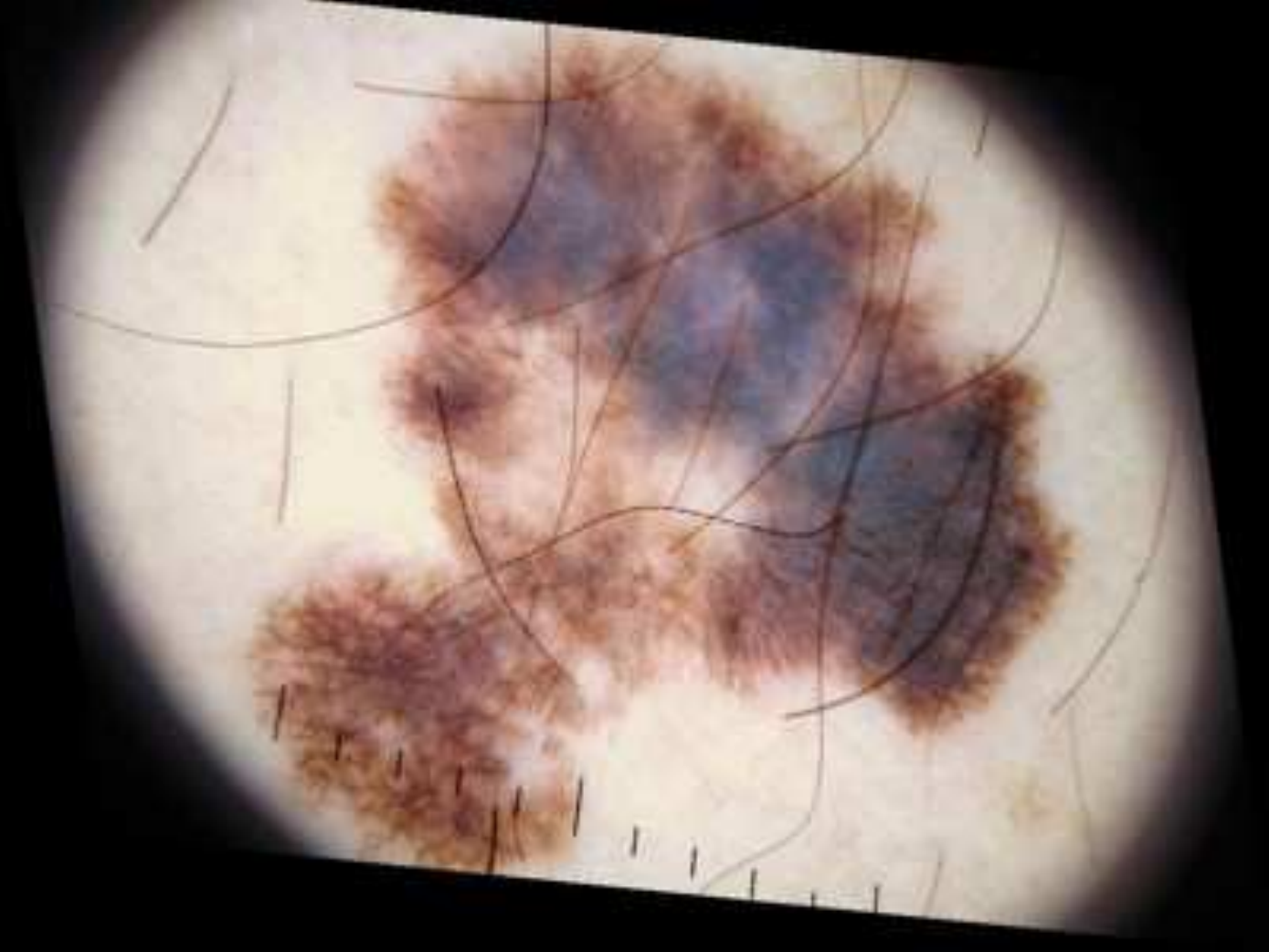}}
 \\
 \subfigure[Histogram equalization]{\label{fig:aug_d}\includegraphics[width=0.3\columnwidth,draft=false]{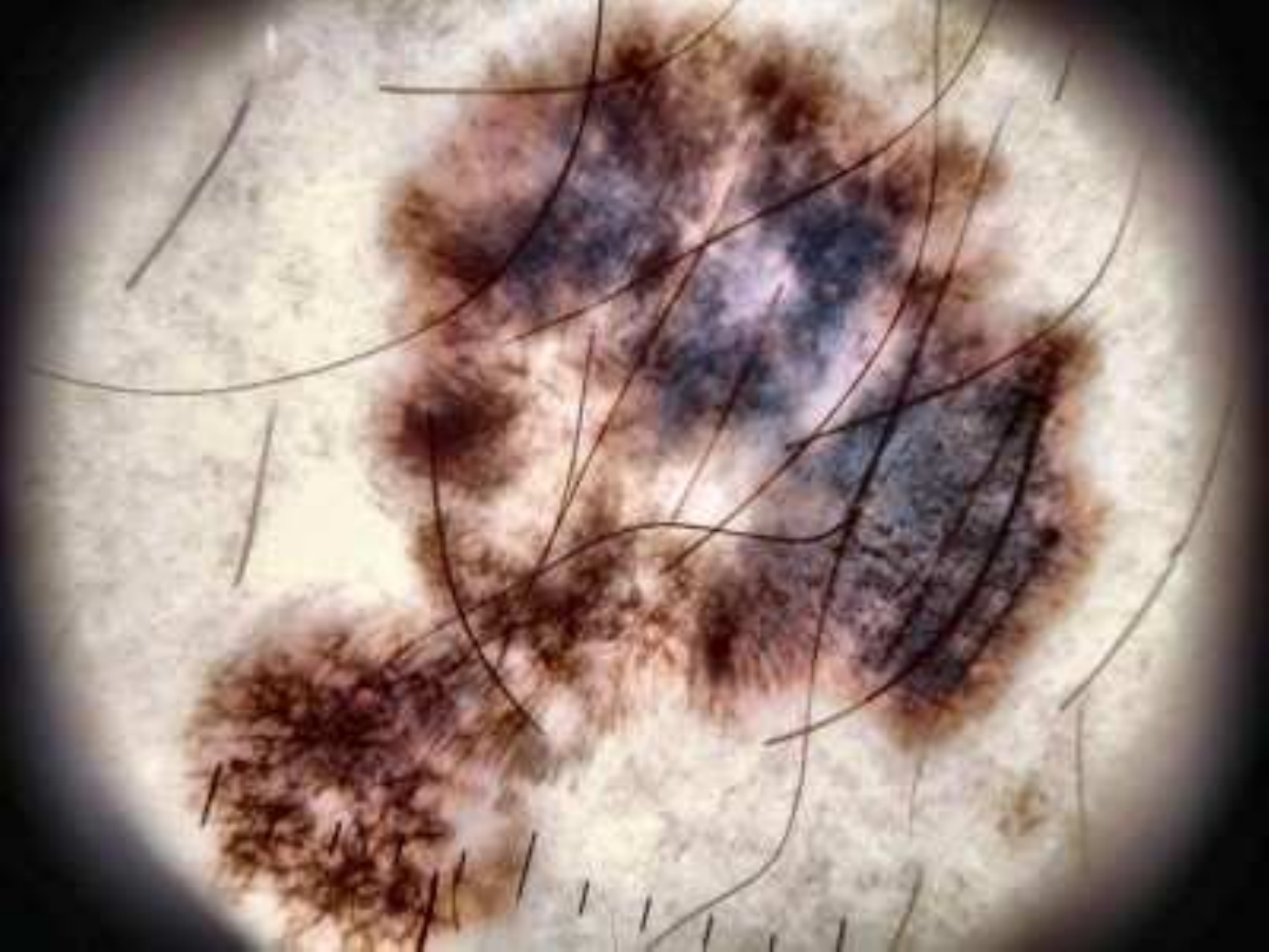}}
	\subfigure[\textsf{HSV} shift]{\label{fig:aug_e}\includegraphics[width=0.3\columnwidth,draft=false]{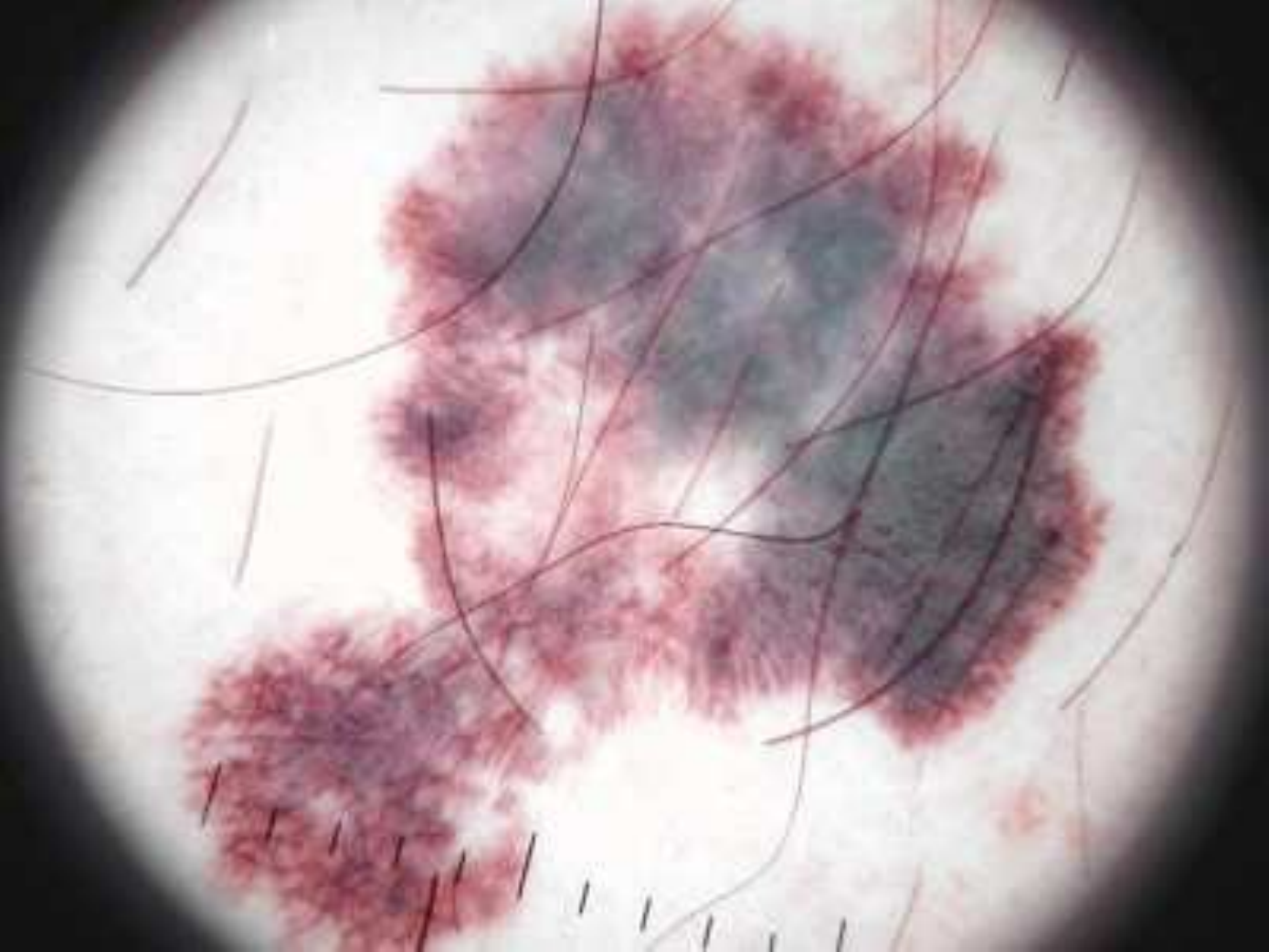}}
	\subfigure[\textsf{RGB} shift]{\label{fig:aug_f}\includegraphics[width=0.3\columnwidth,draft=false]{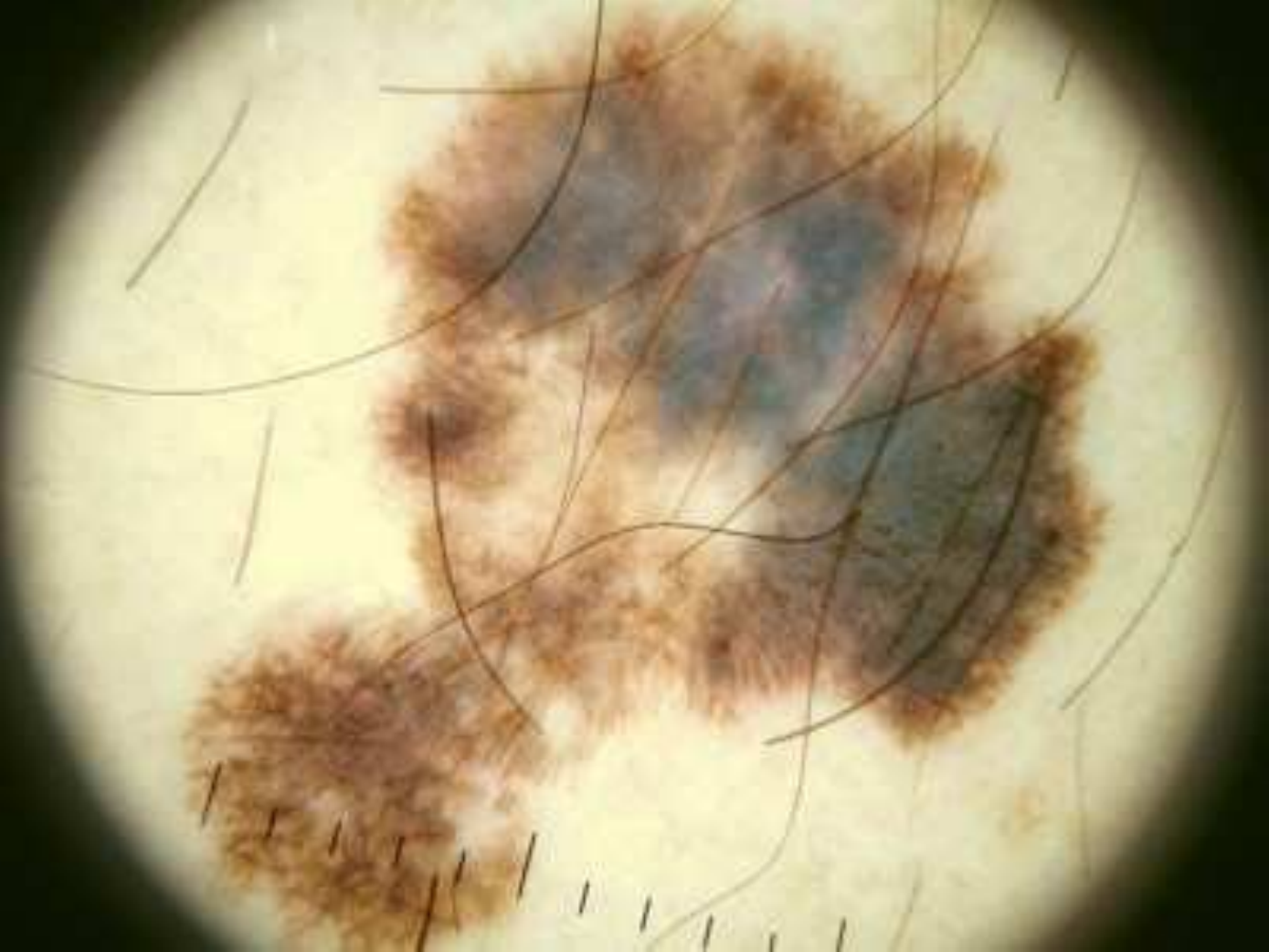}}
 \caption{Various data augmentation transformations \revision{applied to a dermoscopic image} (image source: \textsf{ISIC} 2016 dataset~\citep{Gutman16}) \revision{using the Albumentations library~\citep{Buslaev20}}.}
 \label{fig:augs}
\end{figure}

Data augmentation---synthesizing new samples from existing ones---is commonly employed in the training of \textsf{DL} models. Augmented data serve as a regularizer, increase the amount and diversity of data~\citep{Shorten19}, induce desirable invariances on the model, and alleviate class imbalance. Traditional data augmentation applies simple geometric, photometric, and colorimetric transformations on the samples, including mirroring, translation, scaling, rotation, cropping, random region erasing, affine or elastic deformation, modifications of hue, saturation, brightness, and contrast. Usually, several transformations are chosen at random and combined. Fig.~\ref{fig:augs} exemplifies the procedure, as applied to a dermoscopic image with Albumentations~\citep{Buslaev20}, a state-of-the-art open-source library for image augmentation.

As mentioned earlier, augmented training data induce invariance on the models: random translations \typo{and} croppings, for example, help induce a translation-invariant model. This has implications for skin lesion analysis, e.g., data augmentation for generic datasets (such as ImageNet~\citep{deng2009imagenet}) forgo vertical mirroring and large-angle rotations, because natural scenes have a strong vertical anisotropy, while skin lesion images are isotropic. \typo{In addition, a}ugmented \textit{test} data (test-time augmentation) \typo{may} also improve generalization by combining the predictions of several augmented samples through, \typo{for example}, average pooling or majority voting~\citep{Shorten19}. \citet{Perez18} have systematically evaluated the effect of several data augmentation schemes for skin lesion classification, finding that the use of both training and test augmentation is critical for performance, surpassing, in some cases, increases of real data without augmentation. \citet{Valle2020} found, in a very large-scale experiment, that test-time augmentation was the second most influential factor for classification performance, after training set size. No systematic study of this kind exists for skin lesion segmentation. 

Although traditional data augmentation is crucial for training \textsf{DL} models, it falls short of providing samples that are both diverse and plausibl\typo{e} from the same distribution as real data. Thus, modern data augmentation~\citep{tajbakhsh2020embracing} employs generative modeling, learning the probability distribution of the real data, and sampling from that distribution. Generative \typo{a}dversarial \typo{n}etworks (\textsf{GAN}s)~\citep{Goodfellow20} are the most promising approach in this direction~\citep{Shorten19}, especially for medical image analysis~\citep{Yi19,Kazeminia20,Shamsolmoali21}. \textsf{GAN}s employ an adversarial training between a generator, which attempts to generate realistic fake samples, and a discriminator, which attempts to differentiate real samples from fake ones. When the procedure converges, the generator output is surprisingly convincing, but \textsf{GAN}s are computationally expensive and difficult to train~\citep{Creswell18}. 

Synthetic generation of skin lesions has received some recent interest, especially in the context of improving classification. Works can be roughly divided into those that use \textsf{GAN}s to create new images from a Gaussian latent variable~\citep{Baur18,74pollastri2019augmenting,Abdelhalim21}, and those that implement \textsf{GAN}s based on image-to-image translation~\citep{80abhishek2019mask2lesion, bissoto2018, Ding21}.

Noise-based \textsf{GAN}s, such as \textsf{DCGAN}~\citep{yuun2017}, \textsf{LAPGAN}~\citep{denton2015}, and \textsf{PGAN}~\citep{karras2018}, learn to decode a Gaussian latent variable into an image that belongs to the training set distribution. The main advantage of these techniques is the ability to create more, and more diverse images, as, in principle, any sample from a multivariate Gaussian distribution may become a different image. The disadvantage is that the images tend to be of  lower quality, and, in the case of segmentation, one needs to generate plausible pairs of images and segmentation masks.

Image-to-image translation \textsf{GAN}s, such as pix2pix~\citep{isola2017} and pix2pixHD~\citep{wang2018high}, learn to create new samples from a semantic segmentation map. They have complementary advantages and disadvantages. Because the procedure is deterministic (one map creates one image), they have much less freedom in the number of samples available, but the images tend to be of higher quality (or more “plausible”). There is no need to generate separate segmentation maps because the generated image is intrinsically compatible with the input segmentation map. 

The two seminal papers on \textsf{GAN}s for skin lesions \citep{Baur18, bissoto2018} evaluate several models. 
\citet{Baur18} compare the noise-based \textsf{DCGAN}, \textsf{LAPGAN}, and \textsf{PGAN} for the generation of $256 \times 256$-pixel images using both qualitative and quantitative criteria, finding that the \textsf{PGAN} gives considerably better results. They further examine the \textsf{PGAN} against a panel of human judges, composed by dermatologists and \textsf{DL} experts, in a “visual Turing test”, showing that both had difficulties in distinguishing the fake images from the true ones.
\citet{bissoto2018} adapt the \textsf{PGAN} to be class-conditioned on diagnostic category, and the image-to-image pix2pixHD to employ the semantic annotation provided by the feature extraction task of the \textsf{ISIC} 2018 dataset (Table~\ref{datasets}), comparing those to an unmodified \textsf{DCGAN} on $256 \times 256$-pixel images, and finding the modified pix2pixHD to be qualitatively better. They use the performance improvement on a separate classification network as a quantitative metric, finding that the use of samples from both \textsf{PGAN} and pix2pixHD leads to the best improvements. They also showcase images of size up to $1,024 \times 1,024$ pixels generated by the pix2pixHD-derived model. 

\citet{74pollastri2019augmenting} extended \textsf{DCGAN} and \textsf{LAPGAN} architectures to generate the segmentation masks (in the pairwise scheme explained above), making their work the only noise-based \textsf{GAN}s usable for segmentation to date.  
\citet{31bi2019improving} introduced stacked adversarial learning to \textsf{GAN}s to learn class-specific skin lesion image generators given the ground-truth segmentations.
\citet{80abhishek2019mask2lesion} employ pix2pix to translate a binary segmentation mask into a dermoscopic image \revision{and use the generated image-mask pairs to augment skin lesion segmentation training datasets, improving segmentation performance}.

\citet{Ding21} feed 
a segmentation mask and an instance mask 
to a conditional \textsf{GAN} generator, where the instance mask states the diagnostic category to be synthesized. 
In both cases, the discriminator receives different resolutions of the generated image and is required to make a decision for each of them. 
\citet{Abdelhalim21} is a recent work that also conditions \textsf{PGAN} on the class label \revision{and uses the generated outputs to augment a melanoma diagnosis dataset}.

Recently, \citet{Bissoto_2021_CVPR} cast doubt on the power of \textsf{GAN}-synthesized data augmentation to reliably improve skin lesion classification. Their evaluation, which included four \textsf{GAN} models, four datasets, and several augmentation scenarios, showed improvement only in a severe cross-modality scenario (training on dermoscopic and testing on clinical images). To the best of our knowledge, no corresponding systematic evaluation exists for skin lesion segmentation. 

\subsection{Supervised, Semi-supervised, Weakly supervised, Self-supervised learning}
\label{subsec:semi}

Although supervised \textsf{DL} has achieved outstanding performance in various medical image analysis applications, its dependency on high-quality annotations limits its applicability, as well as its generaliza\typo{bility} to unseen, out-of-distribution data. Semi-supervised techniques attempt to learn from both labeled and unlabeled samples. Weakly supervised techniques attempt to exploit partial annotations like image-level labels or bounding boxes, often in conjunction with a subset of pixel-level fully-annotated samples. 

\begin{figure*}
\centering
\includegraphics[width=6in]{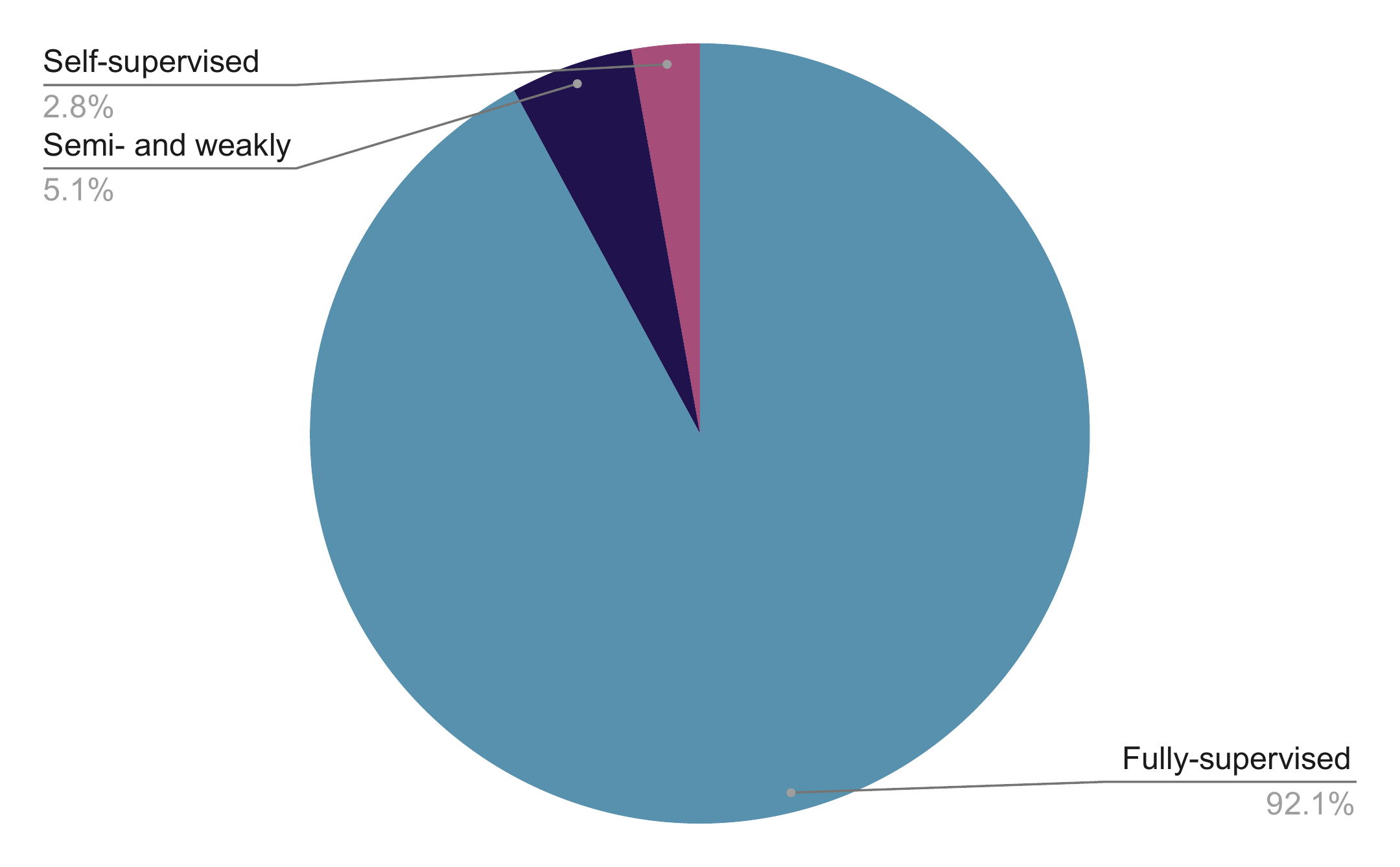}
\caption{\revision{A breakdown of different levels of supervision used in the $177$ surveyed works. Fully supervised models continue to make up the majority of the literature ($163$ papers), with semi-supervised and weakly supervised methods appearing in only $9$ papers. Self-supervision in skin lesion segmentation is fairly new, with all the $5$ papers appearing from 2020 onwards.}}
\label{semi_chart}
\end{figure*}

Since pixel-level annotation of skin lesion images is costly, there is a trade-off between annotation precision and efficiency. In practice, the annotations are intrinsically noisy, which can be modeled explicitly to avoid over-fitting. (We discuss the issue of annotation variability in Section~\ref{subsec:agreement}.)
To deal with label noise, \citet{mirikharaji2019learning} learn a model robust to annotation noise, making use of a large set of unreliable annotations and a small set of perfect clean annotations. They propose to learn a spatially adaptive weight map corresponding to each training data, assigning different weights to noisy and clean pixel-level annotations while training the deep model. 
To remove the dependency on having a set of perfectly clean annotations, \citet{redekop2021uncertainty} propose to alter noisy ground-truth masks during training by considering the quantification of aleatoric uncertainty \citep{der2009aleatory,gal2016uncertainty,depeweg2018decomposition,kwon2020uncertainty} to obtain a map of regions of high and low uncertainty. Pixels of ground-truth masks in highly uncertain regions are flipped, progressively increasing the model's robustness to label noise.
\citet{153ribeiro2020less} deal with noise by discarding inconsistent samples and annotation detail during training time, showing that the model generalizes better even when detailed annotations are required in test time.

When there is a labeled dataset, even if the number of labeled samples is far less than that of unlabeled samples, semi- and self-supervision techniques can be applied. \citet{24li2019transformation} propose a semi-supervised approach, using a transformation-consistent self-ensemble to leverage unlabeled data and to regularize the model. They minimize the difference between the network predictions of different transformations (random perturbations, flipping, and rotation) applied to the input image and the transformation of the model prediction for the input image.
Self-supervision attempts to exploit intrinsic labels by solving proxy tasks, enabling the use of a large\typo{,} unlabeled corpus of data to pretrain a model before fine-tuning it on the target task. 
An example is to artificially apply random rotations in the input images, and train the model to predict the exact degree of rotation \citep{komodakis2018unsupervised}. Note that the degree of rotation of each image is known\typo{, }since it was artificially applied\typo{,} and thus, can be used as a label during training. Similarly, for skin lesion segmentation, \citet{100li2020multi} propose to exploit the color distribution information, the proxy task being to predict values from blue and red color channels while having the green one as input. They also include a task to estimate the red and blue color distributions to improve the model's ability to extract global features. After the pretraining, they use a smaller set of labeled data to fine-tune the model.

\subsection{Image Preprocessing}

\label{subsec:preproc}

Preprocessing may facilitate the segmentation of skin lesion images. Typical preprocessing operations include:
\begin{itemize}
   \item \textbf{Downsampling}: Dermoscopy is typically a high-resolution technique, resulting in large image sizes, while many \typo{c}onvolutional \typo{n}eural \typo{n}etwork (\textsf{CNN}) architectures\typo{,} e.g., LeNet, AlexNet, \textsf{VGG}, GoogLeNet, ResNet\typo{, etc.,} require fixed-size input images, usually $224 \times 224$ or $299 \times 299$ pixels, and even those \textsf{CNN}s that can handle arbitrary-sized images (e.g., fully-convolutional networks \typo{(}\textsf{FCN}\typo{s)}) may benefit from downsampling for computational reasons. Downsampling is common in the skin lesion segmentation literature~\citep{Codella17,yu2017,93yuan2017automatic,17al2018skin,26zhang2019automatic,74pollastri2019augmenting}. 
   
   \item \textbf{Color space transformations}: \textsf{RGB} images are expected by most models, but some works~\citep{Codella17,17al2018skin,9yuan2019improving,74pollastri2019augmenting,Pour20} employ alternative color spaces~\citep{Busin08}, such as \textsf{CIELAB}, \textsf{CIELUV}, and \textsf{HSV}. Often, one or more channels of the transformed space are combined with the \textsf{RGB} channels for reasons including\typo{, but not limited to,} increasing the class separability, decoupling luminance and chromaticity, ensuring (approximate) perceptual uniformity, achieving invariance to illumination or viewpoint, and eliminating highlights.
   
  \item
  \textbf{Additional inputs}: In addition to color space transformations, recent works incorporate more focused and domain-specific inputs to the segmentation models, such as Fourier domain representation using the discrete Fourier transform~\citep{tang2021introducing} and inputs based on the physics of skin illumination and imaging~\citep{137abhishek2020illumination}.
   
   \item \textbf{Contrast enhancement}: \typo{I}nsufficient contrast (Fig.~\ref{fig:prob_i}) is a prime reason for segmentation failures~\citep{Bogo15}, leading some works~\citep{Saba19, Schaefer11} to enhance the image contrast prior to segmentation. 
   
   \item \textbf{Color normalization}: \typo{V}arying illumination~\citep{Barata15a,Barata15b} may lead to inconsistencies \typo{in skin lesion segmentation}. This problem can be addressed by color normalization~\citep{2goyal2019automatic}.
   
   \item \textbf{Artifact removal}: \typo{D}ermoscopic images often present artifacts, among which hair (Fig.~\ref{fig:prob_g}) is the most distracting~\citep{Abbas11}, leading some studies~\citep{66unver2019skin,124zafar2020skin,li2021digital} to remove it prior to segmentation.
\end{itemize}

Classical machine learning models (e.g., nearest neighbors, decision trees, support vector
machines~\citep{Celebi07a,Celebi08,Iyatomi08,Barata14,Shimizu15}), which rely on hand-crafted features~\citep{Barata19}, tend to benefit more from preprocessing than \textsf{DL} models, which, when properly trained, can learn from the data how to bypass input issues~\citep{Celebi15b, Valle2020}. However, preprocessing may still be helpful when dealing with small or noisy datasets.

\section{Model Design and Training}
\label{sec:model}

Multi-\typo{l}ayer \typo{p}erceptrons (\textsf{MLP}s) for pixel-level classification~\citep{Gish89,Katz89} appeared soon after the publication of the seminal backpropagation paper~\citep{Rumelhart86}, but these shallow feed-forward networks had many drawbacks~\citep{LeCun98}, including an excessive number of parameters, lack of invariance, and disregard for the inherent structure present in images. 

\typo{\textsf{CNN}s} are deep feedforward neural networks designed to extract progressively more abstract features from multidimensional signals ($1$-D signals, $2$-D images, $3$-D video, etc.)~\citep{LeCun15}. Therefore, in addition to addressing the aforementioned problems of \textsf{MLP}s, \textsf{CNN}s automate \emph{feature engineering}~\citep{Bengio13}, that is, the design of algorithms that can transform raw signal values to discriminative features. Another advantage of \textsf{CNNs} over traditional machine learning classifiers is that they require minimal preprocessing of the input data. Due to their significant advantages, \textsf{CNN}s have become the method of choice in many medical image analysis applications over the past decade~\citep{Litjens17}. The key enablers in this deep learning revolution were:
\begin{inparaenum}[(i)]
   \item the availability of massive data sets;
   \item the availability of powerful and inexpensive graphics processing units;
   \item the development of better network architectures, learning algorithms, and regularization techniques; and
   \item the development of open-source deep learning frameworks.
\end{inparaenum}

Semantic segmentation may be understood as the attempt to answer the parallel and complementary questions “what” and “where” in a given image. The former is better answered by translation-invariant global features, while the latter requires well-localized features, posing a challenge to deep models. \textsf{CNN}s for pixel-level classification first appeared in the mid-2000s~\citep{Ning05}, but their use accelerated after the seminal paper on \textsf{FCN}s by \citet{long2015fully}, which\revision{, along with U-Net~\citep{ronneberger2015}, have} become the basis for many state-of-the-art segmentation models. In contrast to classification \textsf{CNN}s (e.g., LeNet, AlexNet, \textsf{VGG}, GoogLeNet, ResNet), \textsf{FCN}s easily cope with arbitrary-sized input images.

\subsection{Architecture}
\label{subsec:architecture}

An ideal skin lesion segmentation algorithm is accurate, computationally inexpensive, invariant to noise and input transformations, \typo{requires little training data} and is easy to implement and train. Unfortunately, no algorithm has, so far, been able to achieve these conflicting goals. \textsf{DL}\typo{-}based segmentation tends towards accuracy and invariance at the cost of computation and training data. Ease of implementation is debatable: on the one hand\typo{,} the algorithms often forgo cumbersome preprocessing, postprocessing, and feature engineering steps. On the other hand, tuning and optimizing them is often a painstaking task.

\begin{figure*}[t]
    \centering
    \begin{adjustbox}{width=13cm, keepaspectratio}
        \begin{tikzpicture}[mindmap, grow cyclic, every node/.style=concept, concept color=red!30,
            level 1/.append style={level distance=5cm,sibling angle=90},
            level 2/.append style={level distance=3cm,sibling angle=45}]
        
        \node{Segmentation\\Model\\Architectures\\\S\ref{subsec:architecture}}
            child [concept color=orange!40] { node {Transformer Models\\\S\ref{transfomer_models}}
            }
            child [concept color=purple!60] { node {Hybrid Feature Models\\\S\ref{hybrid_feature_models}}
            }
            child [concept color=gray!40] { node {Multiple Network Models\\\S\ref{multiple_network_models}}
                child { node {GANs\\\S\ref{GAN_models}}}
                child { node {Multi-task Models\\\S\ref{multi_task_models}}}
                child { node {Ensembles\\\S\ref{standard_ensembles}}}
            }
            child [concept color=Goldenrod!60] { node {Single Network Models\\\S\ref{sec:single_net}}
                child { node {Recurrent CNNs\\S\ref{rcnn_modules}}}
                child { node {Attention Modules\\\S\ref{attention_modules}}}
                child { node {Multi-scale Modules\\\S\ref{multi_scale}}}
                child { node [scale=1.] {Conv. Modules\\\S\ref{conv_mod}}}
                child { node {Shortcut Connections\\\S\ref{shortcut}}}
            };
        \end{tikzpicture}
\end{adjustbox}
\caption{Taxonomy of \textsf{DL}-based skin lesion segmentation model architectures. \revision{}}
\label{fig:arch_taxonomy}
\end{figure*}
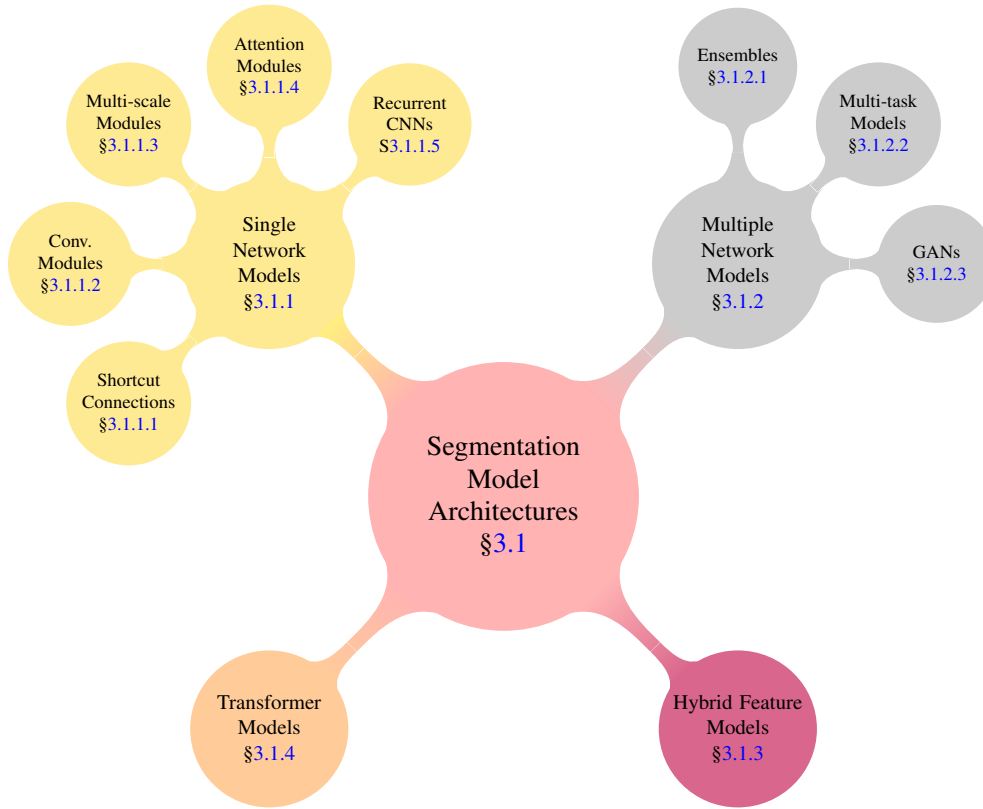

\begin{figure*}
\centering
\includegraphics[width=0.7\linewidth]{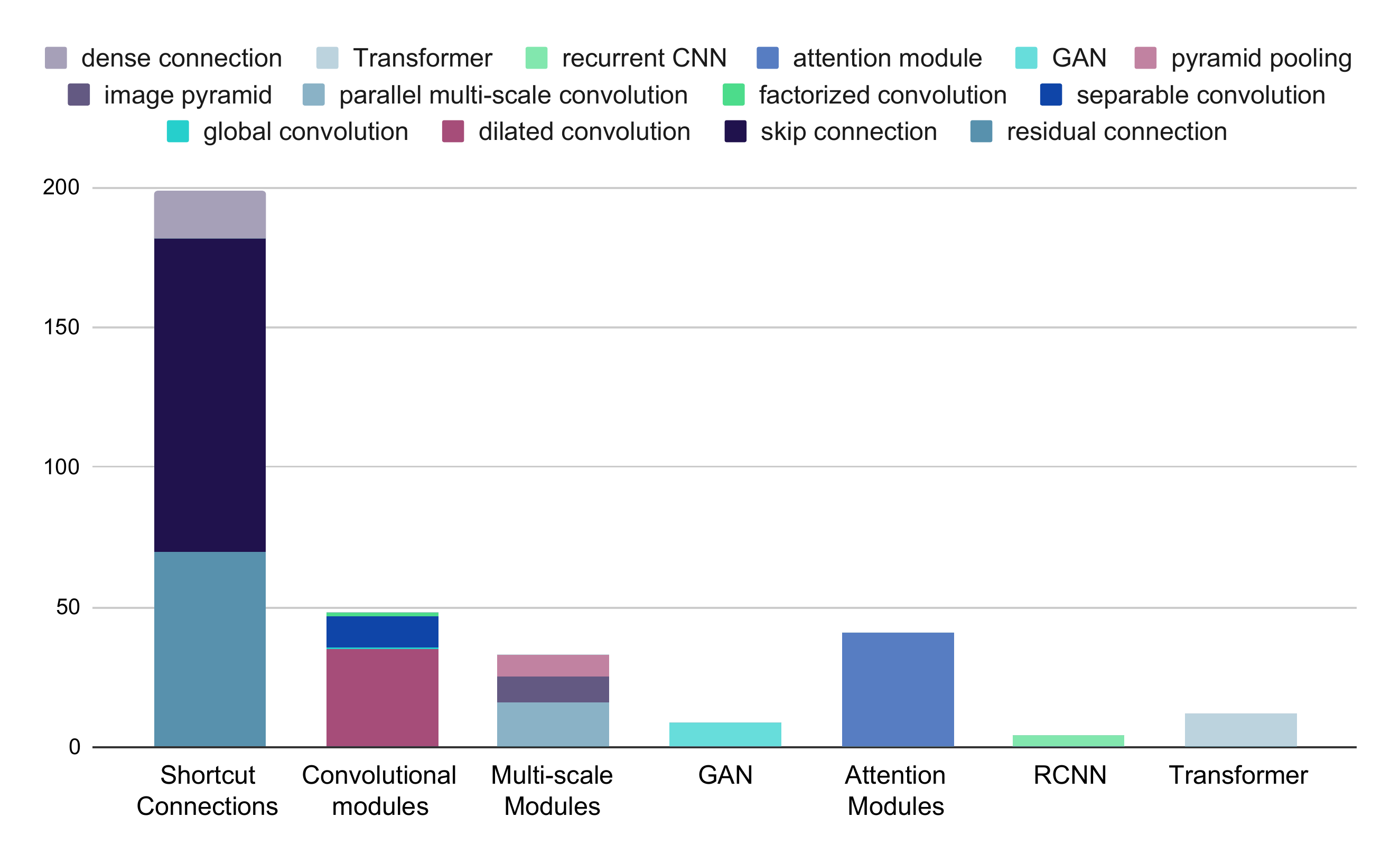}
\caption{The frequency of utilization of different architectural modules in the surveyed studies. \revision{Shortcut connections, particularly, skip connections ($112$ papers) and residual connections ($70$ papers) are the two most frequent components in \textsf{DL}-based skin lesion segmentation models. Attention mechanisms learn dependencies between elements in sequences, either spatially or channel-wise, and are therefore used by several encoder-decoder-style segmentation models ($41$ papers). Dilated convolutions help expand the receptive field of \textsf{CNN}-models without any additional parameters, which is why they are the most popular variant of convolution in the surveyed studies ($35$ papers). Finally, papers using Transformers ($12$ papers) started appearing from 2021 onwards and are on the rise.}}
\label{arch_modules}
\end{figure*}

\revision{As shown in Fig.~\ref{fig:arch_taxonomy},} we have classified \typo{the} existing literature into single-network models, multiple-network models, hybrid-feature models, \revision{and Transformer models}. The first and second groups are somewhat self-descriptive, but notice that the latter is further divided into ensembles of models, multi-task methods (often \typo{performing} simultaneous classification and segmentation), and \textsf{GAN}s. Hybrid-feature models combine \textsf{DL} with hand-crafted features. \revision{Transformer models, as the name suggests, employ Transformers either with or without CNNs for segmentation, and have started being used for skin lesion segmentation only recently.} We classified works according to their most relevant feature, but the architectural improvements discussed in Section~\ref{sec:single_net} also appear in the models listed in the other sections. \revision{In Fig.~\ref{arch_modules}, we show how frequently different architectural modules appear in the $177$ surveyed works, grouped by our taxonomy of model architectures (Fig.~\ref{fig:arch_taxonomy}).}

Table \ref{tab:main} summarizes all the \revision{$177$} surveyed works \revision{in this review, with the following attributes for each work: type of publication, datasets, architectural modules, loss functions, and augmentations used, reported Jaccard index, whether the paper performed cross-dataset evaluation (\textsf{CDE}) and postprocessing (\textsf{PP}), and whether the corresponding code was released publicly. For papers that reported segmentation results on more than 1 dataset, we list all of them and list the performance on only one dataset, formatting that particular dataset in bold. Since \textsf{ISIC} 2017 is the most popular dataset (Fig.~\ref{data_eval_freq}), wherever reported, we note the performance (Jaccard index) on \textsf{ISIC} 2017. For papers that do not report the Jaccard index and instead report the Dice score, we compute the former based on the latter and report this computed score denoted by an asterisk.
Cross-dataset evaluation (\textsf{CDE}) refers to when a paper trained model(s) on one dataset but evaluated on another}.

\subsubsection{Single Network Models}
\label{sec:single_net}

The approaches in this section employ a single \textsf{DL} model, usually an \textsf{FCN}, following an \textit{encoder-decoder} structure, where the encoder extracts increasingly abstract features, and the decoder outputs the segmentation mask. In this section, we discuss these architectural choices for designing deep models for skin lesion segmentation.

Earlier \typo{\textsf{DL}-based skin lesion segmentation} works adopted either \textsf{FCN}~\citep{long2015fully} or U-Net \citep{ronneberger2015}. \textsf{FCN} originally comprised a backbone of \textsf{VGG}16~\citep{simonyan2014} \textsf{CNN} layers in the encoder and a single deconvolution layer in the encoder. The original paper proposes three versions, two with skip connections (\textsf{FCN}-8 and \textsf{FCN}-16), and one without them (FCN-32). \text{U-Net}~\citep{ronneberger2015}, originally proposed for segmenting electron microscopy images, was rapidly adopted in the medical image segmentation literature. As its name suggests, it is a U-shaped model, with an encoder stacking convolutional layers that double in size filterwise, intercalated by pooling layers, and a symmetric decoder with pooling layers replaced by up-convolutions. Skip connections between corresponding encoder-decoder blocks improve the flow of information between layers, preserving low-level features lost during pooling and producing detailed segmentation boundaries.

U-Net \typo{frequently appears} in the skin lesion segmentation \typo{literature} both in its original form~\citep{Codella17,74pollastri2019augmenting,ramani2019} and modified forms ~\citep{tang2019,8alom2019recurrent,28hasan2020dsnet}, discussed below. Some works introduce their own models~\citep{93yuan2017automatic,17al2018skin}.  

\paragraph{Shortcut Connections}
\label{shortcut}

Connections between early and late layers in \textsf{FCN}s have been widely explored to improve both the forward and backward (gradient) information flow in the models, facilitating the training. The three most popular types of connections are described below. 
\smallskip
 
\noindent \textit{Residual connections}: \typo{C}reating non-linear blocks that add their unmodified inputs to their outputs~\citep{he2016deep} alleviates gradient degradation in very deep networks. It provides a direct path for the gradient to flow through to the early layers of the network, while still allowing for very deep models. The technique appears often in skin lesion segmentation, in the implementation of the encoder~\citep{16sarker2018slsdeep,27baghersalimi2019dermonet,yu2017} or both encoder and decoder~\citep{3he2017skin,5venkatesh2018deep,32li2018dense,61tu2019dense,48zhang2019dsm,82he2018dense,87xue2018adversarial}. Residual connections have also appeared in recurrent units~\citep{8alom2019recurrent,73alom2019skin}, dense blocks~\citep{41song2019dense}, chained pooling~\citep{3he2017skin,32li2018dense,82he2018dense}, and 1-D factorized convolutions~\citep{42singh2019fca}.
\smallskip
   
\noindent \textit{Skip connections} appear in encoder-decoder architectures, connecting high-resolution features from the encoder's contracting path to the semantic features on the decoder's expanding path~\citep{ronneberger2015}. These connections help preserve localization, especially near region boundaries, and combine multi-scale features, resulting in sharper boundaries in the predicted segmentation. Skip connections are very popular in skin lesion segmentation because they are effective and easy to implement~\citep{48zhang2019dsm,27baghersalimi2019dermonet,41song2019dense,65wei2019attention,5venkatesh2018deep,7azad2019bi,3he2017skin,8alom2019recurrent,16sarker2018slsdeep,22zeng2018multi,32li2018dense,61tu2019dense,yu2017,42singh2019fca, 82he2018dense,87xue2018adversarial,73alom2019skin,49vesal2018skinnet,75liu2019skin}.
\smallskip

\noindent \textit{Dense connections} expand the convolutional layers by connecting each layer to all its subsequent layers, concatenating their features~\citep{huang2017densely}. Iterative reuse of features in dense connections maximizes information flow forward and backward. 
Similar to deep supervision (Section~\ref{Deep_supervision}), the gradient \typo{is propagated backwards} directly through all previous layers. Several works~\citep{22zeng2018multi,41song2019dense,24li2019transformation,61tu2019dense,49vesal2018skinnet} integrated dense blocks in both the encoder and the decoder. \citet{27baghersalimi2019dermonet}, \citet{28hasan2020dsnet} and \citet{65wei2019attention} used multiple dense blocks iteratively in only the encoder, while \citet{32li2018dense} proposed dense deconvolutional blocks to reuse features from the previous layers. \citet{7azad2019bi} encoded densely connected convolutions into the bottleneck of their encoder-decoder to obtain better features.

\paragraph{Convolutional Modules} 
\label{conv_mod}

As mentioned earlier, convolution not only provides a structural advantage, respecting the local connectivity structure of images in the output futures, but also dramatically improves parameter sharing since the parameters of a relatively small convolutional kernel are shared by all patches of a large image. Convolution is a critical element of deep segmentation models. In this section, we discuss some new convolution variants, which have enhanced and diversified this operation, appearing in the skin lesion segmentation literature.
\smallskip

\noindent \textit{Dilated convolution}: In contrast to requiring full-resolution outputs in dense prediction networks, pooling and striding operations have been adopted in deep convolutional neural networks (\textsf{DCNN}\typo{s}) to increase the receptive field and diminish the spatial resolution of feature maps. Dilated or atrous convolutions are designed specifically for the semantic segmentation task to exponentially expand the receptive fields while keeping the number of parameters constant~\citep{yu2016multi}. Dilated convolutions are convolutional modules with upsampled filters containing zeros between consecutive filter values. \citet{16sarker2018slsdeep} and \citet{29jiang2019decision} utilized dilated residual blocks in the encoder to control the image field-of-view explicitly and incorporated multi-scale contextual information into the segmentation network. SkinNet~\citep{49vesal2018skinnet} used dilated convolutions at the lower level of the network to enlarge the field-of-view and capture non-local information. \citet{75liu2019skin} introduced dilated convolutions to the U-Net architecture, \typo{significantly improving} the segmentation performance. \typo{Furthermore}, different versions of the DeepLab architecture~\citep{Chen17,chen2017rethinking,chen2018encoder}, which replace standard convolutions with dilated ones, have been used in skin lesion segmentation~\citep{14goyal2019skin,2goyal2019automatic,38cui2019ensemble,53chen2018multi,71canalini2019skin}.
\smallskip

\noindent \textit{Separable convolution}: Separable convolution or depth-wise separable convolution~\citep{chollet2017xception} is a spatial convolution operation that convolves each input channel with its corresponding kernel. This is followed by a $1\times1$ standard convolution to capture the channel-wise dependencies in the output of depth-wise convolution. Depth-wise convolutions are designed to reduce the number of parameters and the computation of standard convolutions while maintaining \typo{the} accuracy. DSNet~\citep{28hasan2020dsnet} and separable-Unet~\citep{tang2019} utilized depth-wise separable convolutions in the model to have a lightweight network with a reduced number of parameters. Adopted from the DeepLab architecture, \citet{2goyal2019automatic}, \citet{38cui2019ensemble} and, \citet{71canalini2019skin} incorporated depth-wise separable convolutions in conjunction with dilated convolution to improve the speed and accuracy of dense predictions.
\smallskip

\noindent \textit{Global convolution}: State-of-the-art segmentation models remove densely connected and global pooling layers to preserve spatial information required for full-resolution output recovery. As a result, by keeping high-resolution feature maps, segmentation models become more suitable for localization and, in contrast, less suitable for per-pixel classification, which needs transformation invariant features. To increase the connectivity between feature maps and classifiers, large convolutional kernels should be adopted. However, such kernels have a large number of parameters, which renders them computationally expensive. To tackle this, global convolutional network (\textsf{GCN}) modules adopt a combination of symmetric parallel convolutions in the form of $1 \times k + k \times 1$ and $k \times 1 + 1 \times k$ to cover a $k \times k$ area of feature maps~\citep{peng2017large}. \textsf{SeGAN}~\citep{87xue2018adversarial} employed \textsf{GCN} modules with large kernels in the generator's decoder to reconstruct segmentation masks and in the discriminator architecture to optimally capture a larger receptive field.
\smallskip

\noindent \textit{Factorized convolution}: Factorized convolutions~\citep{wang2017factorized} are designed to reduce the number of convolution filter parameters as well as \typo{the} computation time through kernel decomposition when a high-dimensional kernel is substituted with a sequence of lower-dimensional convolutions. Additionally, by adding non-linearity between the composited kernels, the network's capacity may improve. \textsf{FCA}-Net~\citep{42singh2019fca} and MobileGAN~\citep{60sarker2019mobilegan} utilized residual 1-D factorized convolutions (a sequence of $k \times 1$ and $1 \times k$ convolutions with \textsf{ReLU} non-linearity) in their segmentation architecture.
   
\paragraph{Multi-scale Modules}
\label{multi_scale}

In \textsf{FCN}s, taking semantic context into account when assigning per-pixel labels leads to a more accurate prediction~\citep{long2015fully}. Exploiting multi-scale contextual information, effectively combining them as well as encoding them in deep semantic segmentation have been widely explored. 
\smallskip

\noindent \textit{Image Pyramid}: 
RefineNet~\citep{3he2017skin} and its extension~\citep{82he2018dense}, \textsf{MSFCDN}~\citep{22zeng2018multi}, \textsf{FCA-Net}~\citep{42singh2019fca}, and \citet{33abraham2019novel} fed a pyramid of multi-resolution skin lesion images as input to their deep segmentation network to extract multi-scale discriminative features. RefineNet~\citep{3he2017skin,82he2018dense}, \typo{Factorized channel attention network (}FCA-Net~\citep{42singh2019fca}\typo{)} and \citet{33abraham2019novel} applied convolutional blocks to different image resolutions in parallel to generate features which are then up-sampled in order to fuse multi-scale feature maps. \typo{Multi-scale fully convolutional DenseNets} \typo{(}\textsf{MSFCDN}~\citep{22zeng2018multi}\typo{)} gradually integrated multi-scale features extracted from the image pyramid into the encoder's down-sampling path. Also, ~\citet{85jafari2016skin,86jafari2017extraction} extracted multi-scale patches from clinical images to predict semantic labels and refine lesion boundaries by deploying local and global information. While aggregating the feature maps computed at various image scales improves the segmentation performance, it also increases the computational cost of the network.
\smallskip

\noindent \textit{Parallel multi-scale convolutions}: Alternatively, given a single image resolution, multiple convolutional filters with different kernel sizes~\citep{48zhang2019dsm,59wang2019automated,56jahanifar2018segmentation} or multiple dilated convolutions with different dilation rates~\citep{14goyal2019skin,2goyal2019automatic,38cui2019ensemble,53chen2018multi,71canalini2019skin} can be adopted in parallel paths to extract multi-scale contextual features from images. DSM~\citep{48zhang2019dsm} integrated multi-scale convolutional blocks into the skip connections of an encoder-decoder structure to handle different lesion sizes. \cite{59wang2019automated} utilized multi-scale convolutional branches in the bottleneck of an encoder-decoder architecture\typo{,} followed by attention modules to selectively aggregate the extracted multi-scale features.
\smallskip

\noindent \textit{Pyramid pooling}: Another way of incorporating multi-scale information into deep segmentation models is to integrate a pyramid pooling (\textsf{PP}) module in the network architecture~\citep{Zhao17}. \textsf{PP} fuses a hierarchy of features extracted from different sub-regions by adopting parallel pooling kernels of various sizes\typo{,} followed by up-sampling and concatenation to create the final feature maps. \cite{16sarker2018slsdeep} and \cite{56jahanifar2018segmentation} utilized \textsf{PP} in the decoder to benefit from coarse-to-fine features extracted by different receptive fields from skin lesion images.

Dilated convolutions and skip connections are two other types of multi-scale information extraction techniques, which are explained in Sections~\ref{conv_mod} and \ref{shortcut}, respectively.
   
\paragraph{Attention Modules}
\label{attention_modules}

An explicit way to exploit contextual dependencies in the pixel-wise labeling task is the self-attention mechanism~\citep{hu2018squeeze,fu2019dual}. Two types of attention modules capture global dependencies in spatial and channel dimensions by integrating features among all positions and channels, respectively. \citet{59wang2019automated} and \citet{60sarker2019mobilegan} leveraged both spatial and channel attention modules to recalibrate the feature maps by examining the feature similarity between pairs of positions or channels and updating each feature value by a weighted sum of all other features.  \citet{42singh2019fca} utilized a channel attention block in the proposed factorized channel attention (\textsf{FCA}) blocks, which was used to investigate the correlation of different channel maps for extraction of relevant patterns. Inspired by attention U-Net~\citep{oktay2018attention}, \typo{multiple works} \citep{33abraham2019novel,41song2019dense,65wei2019attention} integrated a spatial attention gate in an encoder-decoder architecture to combine coarse semantic feature maps and fine localization feature maps. \citet{46kaul2019focusnet} proposed FocusNet which utilizes squeeze-and-excitation blocks into a hybrid encoder-decoder architecture. Squeeze-and-excitation blocks model the channel-wise interdependencies to re-weight feature maps and improve their representation power. Experimental results demonstrate that attention modules help the network focus on the lesions and suppress irrelevant feature responses in the background.

\paragraph{Recurrent Convolutional Neural Networks}
\label{rcnn_modules}

Recurrent convolutional neural networks (\textsf{RCNN}) integrate recurrent connections into convolutional layers by evolving the recurrent input over time~\citep{pinheiro2014recurrent}. Stacking recurrent convolutional layers (\textsf{RCL}) on top of the convolutional layer feature extractors ensures capturing spatial and contextual dependencies in images while limiting the network capacity by sharing the same set of parameters in \textsf{RCL} blocks. In the application of skin lesion segmentation, \citet{111attia2017skin} utilized recurrent layers in the decoder to capture spatial dependencies between deep-encoded features and recover segmentation maps at the original resolution. $\nabla ^ N$-Net~\citep{73alom2019skin}, RU-Net, and R2U-Net~\citep{8alom2019recurrent} incorporated \textsf{RCL} blocks into the \textsf{FCN} architecture to accumulate features across time in a computationally efficient way and boosted the skin lesion boundary detection. \citet{7azad2019bi} deployed a non-linear combination of the encoder feature and decoder feature maps by adding a bi-convolutional \textsf{LSTM} (\textsf{BConvLSTM}) in skip connections. \textsf{BConvLSTM} consists of two independent \typo{convolutional \textsf{LSTM} modules (}\textsf{ConvLSTMs}\typo{)} which process the feature maps into two directions of backward and forward paths and concatenate their outputs to obtain the final output. Modifications to the traditional pooling layers were also proposed, \typo{using} a dense pooling strategy \citep{55nasr2019dense}.

\subsubsection{Multiple Network Models}
\label{multiple_network_models}

Motivations for models comprising more than one \textsf{DL} sub-model are diverse, ranging from alleviating training noise and exploiting a diversity of features learned by different models to exploring synergies between multi-task learners. After examining the literature (Fig.~\ref{fig:arch_taxonomy}), we further classified the works in this section into standard ensembles and multi-task models. We also discuss generative adversarial models, which are intrinsically multi-network \typo{models}, in a separate category.

\paragraph{Standard Ensembles}
\label{standard_ensembles}
Ensemble models are widely used in machine learning, motivated by the hope that the complementarity of different models may lead to more stable combined predictions~\citep{sagi2018ensemble}. Ensemble performance is contingent on the quality and diversity of the component models, which can be combined at the feature level (early fusion) or the prediction level (late fusion). The former combines the features extracted by the components and learns a meta-model on them, while the latter pools or combines the models' predictions with or without a meta-model. 

All methods discussed in this section employ late fusion, except for an approach loosely related to early fusion~\citep{tang2019}, which explores various learning-rate decay schemes, and builds a single model by averaging the weights learned at different epochs to bypass poor local minima during training. Since the weights correspond to features learned by the convolution filters, this approach can be interpreted as feature fusion.

Most works employ a single \textsf{DL} architecture with multiple training routines, varying configurations more or less during training~\citep{71canalini2019skin}. The changes between component models may involve network hyperparameters\typo{: }number of filters per block and their size~\citep{Codella17}\typo{;} optimization\typo{ and }regularization hyperparameters\typo{: }learning rate, weight decay~\citep{44tan2019evolving}\typo{;} the training set\typo{: }multiple splits of a training set~\citep{93yuan2017automatic, 9yuan2019improving}, separate models per class~\citep{19bi2019step}\typo{;} preprocessing\typo{: }different color spaces~\citep{74pollastri2019augmenting}\typo{;} different pretraining strategies to initialize feature extractors~\citep{71canalini2019skin}\typo{;} or different ways to initialize the network parameters~\citep{38cui2019ensemble}. Test-time augmentation may also be seen as a form of inference-time ensembling~\citep{53chen2018multi,75liu2019skin,56jahanifar2018segmentation} that combine\typo{s} the outputs of multiple augmented images to generate a more reliable prediction.

\citet{19bi2019step} trained a separate \textsf{DL} model for each class, as well as a separate classification model. For inference, the classification model output is used to weight  the outputs of the category-specific segmentation networks. In contrast, \citet{51soudani2019image} trained a meta “recommender” model to dynamically choose, for each input, a segmentation technique from the top five scorers in the \textsf{ISIC} 2017 challenge, although their proposition was validated on a very small test set ($10\%$ of \textsf{ISIC} 2017 test set).

Several works \typo{also} ensemble different \typo{model} architectures \typo{for skin lesion segmentation}. \citet{2goyal2019automatic} investigate multiple fusion approaches to avoid severe errors from individual models, comparing the average-, maximum- and minimum-pooling of their outputs. A common assumption is that the component models of the ensemble are trained independently, but \citet{84bi2017dermoscopic} cascaded the component models, i.e., used the output of one model as the input of the next (in association with the actual image input). Thus, each model attempts to refine the segmentation obtained by the previous one. They consider not only the final model output, but all the outputs in the cascade, making the technique a legitimate ensemble.

\paragraph{Multi-task Models} 
\label{multi_task_models}
Multi-task models jointly address more than one goal, in the hope that synergies among the tasks will improve overall performance~\citep{zhang2017survey}. This can be particularly helpful in medical image analysis, wherein aggregating tasks may alleviate the issue of insufficient data or annotations. For skin lesions, a few multi-task models exploiting segmentation and classification have been proposed~\citep{53chen2018multi,li2018multi,yang2018skin,81xie2020mutual,145jin2021cascade}.

The synergy between tasks may appear when their models share common relevant features. \citet{li2018multi} assume that all features are shareable between the tasks\typo{,} and train a single fully convolutional residual network to assign class probabilities at the pixel level. They aggregate the class probability maps to estimate both lesion region and class by weighted averaging of probabilities for different classes inside the lesion area. \citet{yang2018skin} learn an end-to-end model formed by a shared convolutional feature extractor followed by three task-specific branches (one to segment skin lesions, one to classify them as melanoma \typo{versus} non-melanoma, and one to classify them as seborrheic keratosis \typo{versus} non-seborrheic keratosis.)
Similarly, \citet{53chen2018multi} add a common feature extractor and separate task heads, and introduce a learnable gate function that controls the flow of information between the tasks to model the latent relationship between two tasks.

Instead of using a single architecture for classification and segmentation, \citet{81xie2020mutual} and \citet{145jin2021cascade} use three \textsf{CNNs} in sequence to perform a coarse segmentation, followed by classification and, finally, a fine segmentation. Instead of shared features, these works exploit sequential guidance, in which the output of each task improves the learning of the next. 
While \citet{81xie2020mutual} feed the output of each network to the next, assuming that the classification network is a diagnostic category and a class activation map~\citep{zhou2016}, \citet{145jin2021cascade} introduce feature entanglement modules, which aggregate features learned by different networks.

All multi-task models discussed so far have results suggesting complementarity between classification and segmentation, but there is no clear advantage among these models.
The segmentation of dermoscopic features (e.g., networks, globules, regression areas) combined with the other tasks is a promising avenue of research, which could bridge classification and segmentation, by fostering the extraction of features that “see” the lesion as human specialists do.

We do not consider in the hybrid group, two-stage models in which segmentation is used as ancillary preprocessing to classification~\citep{yu2017,Codella17,106gonzalez2018dermaknet,121al2020multiple}, since without mutual influence (sharing of losses or features) or feedback between the two tasks, there is no opportunity for synergy. 

\citet{1vesal2018multi} stressed the importance of object localization as an ancillary task for lesion delineation, in particular deploying Faster-\textsf{RCNN}~\citep{ren2015faster} to regress a bounding box to crop the lesions before training a SkinNet segmentation model. While this two-stage approach considerably improves the results, it is computationally expensive (a fast non-\textsf{DL}-based bounding box detection algorithm was proposed earlier by \citet{Celebi09c}). \citet{14goyal2019skin} employed \textsf{ROI} detection with a deep extreme cut to extract the extreme points of lesions (leftmost, rightmost, topmost, bottommost pixels) and feed them\typo{,} in a new auxiliary channel\typo{,}  to a segmentation model.

\paragraph{Generative Adversarial Models}
\label{GAN_models}
We discussed \textsf{GAN}s for synthesizing new samples, their main use in skin lesion analysis, in Section~\ref{subsec:aug}. In this section, we are interested in \textsf{GAN}s not for generating additional training samples, but for directly providing enhanced segmentation models. Adversarial training encourages high-order consistency in predicted segmentation by implicitly looking into the joint distribution of class labels and ground-truth segmentation masks.

\citet{peng2019}, \citet{61tu2019dense}, \citet{lei2020skin}, and \citet{izadi2018generative} use a U-Net-like generator that takes a dermoscopic image as input, and outputs the corresponding segmentation, while the discriminator is a traditional \textsf{CNN} which attempts to discriminate pairs of image and generated segmentation from pairs of image and ground-truth. The generator has to learn to correctly segment the lesion in order to fool the discriminator. \citet{29jiang2019decision} use the same scheme, with a dual discriminator. \citet{lei2020skin} also employ a second discriminator that takes as input only segmentations (unpaired from input images).

\typo{Since t}he discriminator may trivially learn to recognize the generated masks due to the presence of continuous probabilities, instead of the sharp discrete boundaries of the ground-truths\typo{,} \citet{65wei2019attention} \typo{and} \citet{61tu2019dense} address this by pre-multiplying both the generated and real segmentations with the (normalized) input images before feeding them to the discriminator. 

We discuss adversarial loss functions further in Section~\ref{loss_adv}.

\subsubsection{Hybrid Feature Models}
\label{hybrid_feature_models}
Although the major strength of \textsf{CNN}s is their ability to learn meaningful image features without human intervention, a few works tried to combine the best of both worlds, with strategies ranging from employing pre\typo{-} or postprocessing to enforce prior knowledge to adding  hand-crafted features.
Providing the model with prior knowledge about the expected shape of skin lesions---which is missing from \textsf{CNN}s---may improve the performance. \citet{mirikharaji2018star} encode shape information into an additional regularization loss, which penalizes segmentation maps that deviate from a star-shaped prior (Section~\ref{subsec:starloss}).

Conditional random fields (\textsf{CRF}s) use pixel-level color information models to refine the segmentation masks output by the \textsf{CNN}. While both \citet{21tschandl2019domain} \typo{and} \citet{adegun2020} consider a single \textsf{CNN}, \citet{118qiu2020inferring} combine the outputs of multiple \textsf{CNN}s into a single mask, before feeding it together with the input image to the \textsf{CRF}s. \cite{66unver2019skin} use GrabCut~\citep{rother2004grabcut} \typo{to} obtain the segmentation mask given the dermoscopy image and a region proposal obtained by the YOLO~\citep{redmon2016} network. 
These methods regularize the \textsf{CNN} segmentation, which is mainly based on textural patterns, with expected priors based on the color of the pixels.

Works that combine hand-crafted features with \textsf{CNN}s follow two distinct approaches. The first consists of pre-filtering the input images to increase the contrast between the lesion and the surrounding skin. Techniques explored include local binary patterns (\textsf{LBP}s)~\citep{howe2018,141jayapriya2020hybrid}, wavelets~\citep{howe2018}, Laplacian pyramids~\citep{Pour20}, and Laplacian filtering~\citep{Saba19}. The second approach consists of predicting an additional segmentation mask to combine with the one generated by the \textsf{CNN}. \citet{26zhang2019automatic}, for example, use \textsf{LBP}s to consider the textural patterns of skin lesions and guide the networks towards more refined segmentations. \citet{20bozorgtabar2017skin} also employ \textsf{LBP}s combined with pixel-level color information to divide the dermoscopic image into superpixels, which are then scored as part of the lesion or the background. The score mask is then combined with the \textsf{CNN} output mask to compute the final segmentation mask. Despite the limited number of works devoted to integrating deep features with hand-crafted ones, the results so far indicate that this may be a promising \typo{research} direction. 

\revision{\subsubsection{Transformer Models}\label{transfomer_models}}

\revision{Initially proposed for natural language processing~\citep{vaswani2017attention}, Transformers have proliferated in the last couple of years in other areas, including computer vision applications, especially with improvements made over the years for optimizing the computational cost of self-attention~\citep{parmar2018image,hu2019local,ramachandran2019stand,cordonnier2019relationship,zhao2020exploring,dosovitskiy2020image}, and have consequently also been adapted for semantic segmentation tasks~\citep{ranftl2021vision,strudel2021segmenter,zheng2021rethinking}. For medical image segmentation, TransUNet~\citep{chen2021transunet} was one of the first works to use Transformers along with CNNs in the encoder of a U-Net-like encoder-decoder architecture, and \citet{gulzar2022skin} showed that TransUNet outperforms several CNN-only models for skin lesion segmentation. To reduce the computational complexity involved with high-resolution medical images, \cite{cao2021swin} proposed the Swin-Unet architecture that uses self-attention within shifted windows~\citep{liu2021swin}.
For a comprehensive review of the literature of Transformers in general medical image analysis, we refer the interested readers to the surveys by \cite{he2022transformers} and \cite{shamshad2022transformers}.}

\revision{\cite{zhang2021transfuse} propose TransFuse which parallelly computes features from CNN and Transformer modules, with the former capturing low-level spatial information and the latter responsible for modeling global context, and these features are then combined using a self-attention-based fusion module. Evaluation on the ISIC 2017 dataset shows superior segmentation performance and faster convergence. The multi-compound Transformer~\citep{ji2021multi} leverages Transformer-based self-attention and cross-attention modules between the encoder and the decoder components of U-Net to learn rich features from multi-scale CNN features. \cite{wang2021boundary} incorporate boundary-wise prior knowledge in segmentation models using a boundary-aware Transformer (BAT) to deal with the ambiguous boundaries in skin lesion images. More recently, \cite{wu2022fat} introduce a feature-adaptive Transformer network (FAT-Net) that comprised of a dual CNN-Transformer encoder, a light-weight trainable feature-adaptation module, and a memory-efficient decoder using a squeeze-and-excitation module. The resulting segmentation model is more accurate at segmenting skin lesions while also being faster (fewer parameters and computation) than several CNN-only models.}

\subsection{Loss Functions}
\label{sebsec:loss}

A segmentation model \typo{$f$} may be formalized as a function $\hat{y} = f_\theta(x)$, which maps an input image $x$ to an estimated segmentation map $\hat{y}$ parameterized by a (large) set of parameters $\theta$. For skin lesions, $\hat{y}$ is a binary mask separating \typo{the} lesion from the surrounding skin. Given a training set of images $x_i$ and their \typo{corresponding} ground-truth masks $y_i$ $\{(x_i, y_i); i = 1, ... , N\}$, training a segmentation model consists of finding the model parameters $\theta$ that maximize the likelihood of observing those data:
\begin{equation}
\theta^* = \argmax_\theta \sum_{i=1}^N \log \mathrm{P}(y_i|x_i;\theta),
\end{equation}
which is performed indirectly, via the minimization of a loss function between the estimated and \typo{the} true segmentation masks:
\begin{equation}
\theta^* = \argmin_\theta \sum_{i=1}^N \mathcal{L}(\hat{y}_i|y_i) = \argmin_\theta \sum_{i=1}^N \mathcal{L}(f_\theta(x_i)|y_i).
\end{equation}

\begin{figure*}
\centering
\includegraphics[width=6in]{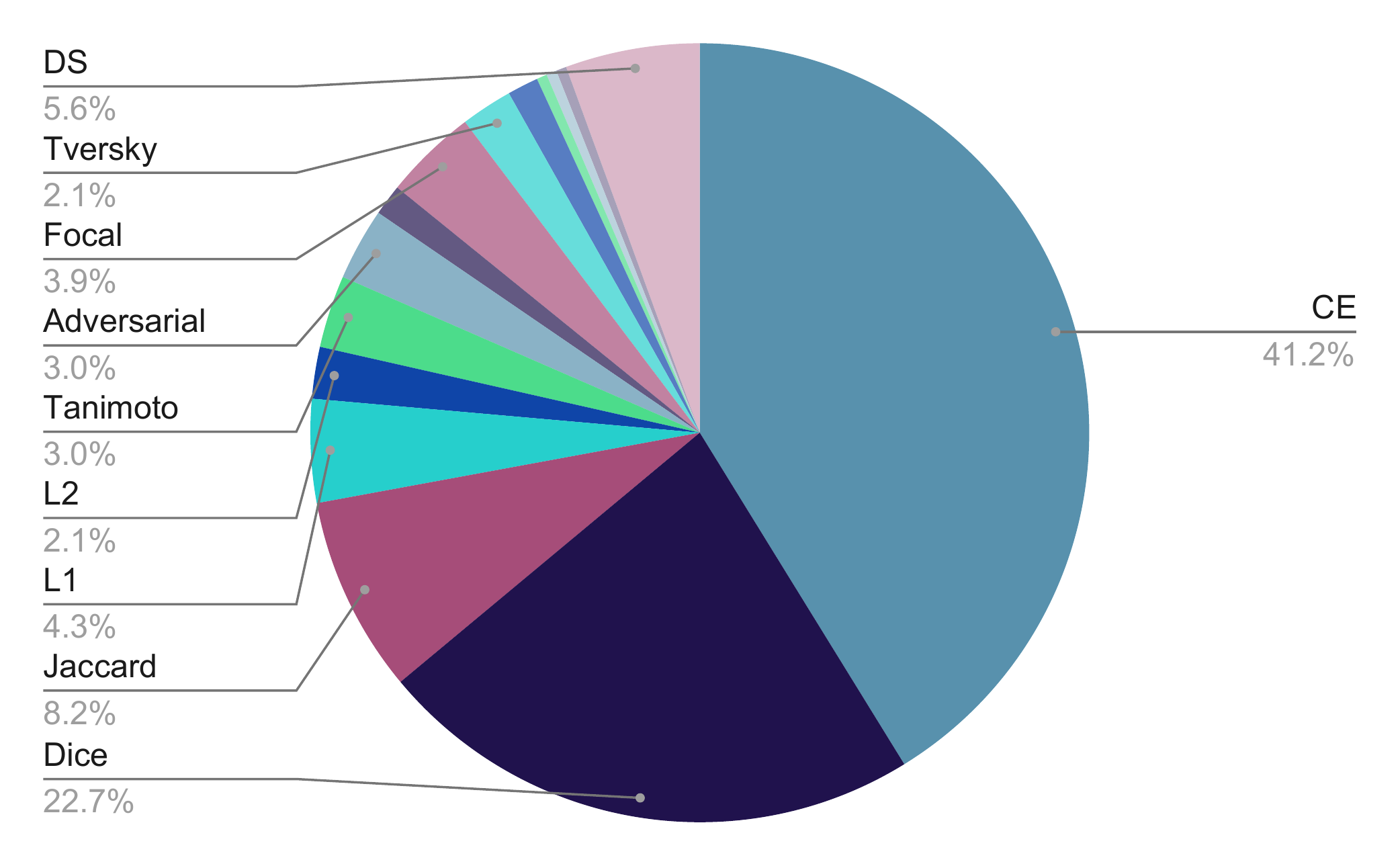}
\caption{\revision{The distribution of loss functions used by the surveyed works in \textsf{DL}-based skin lesion segmentation. Cross-entropy loss is the most popular loss function ($96$ papers), followed by Dice ($53$ papers) and Jaccard ($19$ papers) losses. Of the $177$ surveyed papers, $65$ use a combination of losses, with CE + Dice ($27$ papers) and CE + Jaccard ($11$ papers) being the most popular combinations.}}
\label{loss_distribution}
\end{figure*}

The choice of the loss function is thus critical, as it encodes not only the main optimization objective, but also much of the prior information needed to guide the learning and constrain the search space. \typo{As can been in Table}~\ref{tab:main}\typo{, many} skin lesion segmentation models employ \typo{a combination of} losses to enhance generalization (see Fig.~\ref{loss_distribution}). 

\subsubsection{Losses based on $p$-norms}

Losses based on $p$-norms are the simplest ones, and comprise the \typo{m}ean \typo{s}quared \typo{e}rror (\textsf{MSE}) (for $p=2$) and the \typo{m}ean \typo{a}bsolute \typo{e}rror (\textsf{MAE}) (for $p=1$).

\begin{equation}
    \label{eq:MSE}
    \mathsf{MSE}(X,Y;\theta) = -\sum_{i=1}^{N} \|y_i-\hat{y}_i\|_{2},
\end{equation}

\begin{equation}
    \label{eq:l1}
    \mathsf{MAE}(X,Y;\theta) = -\sum_{i=1}^{N} \|y_i-\hat{y}_i\|_{1}.
\end{equation}

In \textsf{GAN}s, to regularize the segmentations produced by the generator, it is common to utilize hybrid losses containing \textsf{MSE} ($\ell_2$ loss)~\citep{peng2019} or \textsf{MAE} ($\ell_1$ loss)~\citep{peng2019,61tu2019dense,lei2020skin}. The \textsf{MSE} has also been used as a regularizer to match attention and ground-truth maps~\citep{123xie2020skin}.

\subsubsection{Cross-entropy Loss}

Semantic segmentation may be view\typo{ed} as classification at the pixel level, i.e., as assigning a class label to each pixel. From this perspective, minimizing the negative log-likelihoods of pixel-wise predictions (i.e., maximizing their likelihood) may be achieved by minimizing a cross-entropy loss $\mathcal{L}_{ce}$: 
\begin{equation}
\mathcal{L}_{ce}(X,Y;\theta)=-\sum_{i=1}^{N}\sum_
{p\in \Omega_i} y_{ip} \log \hat{y}_{ip}+ (1-y_{ip}) \log (1-\hat{y}_{ip}),~~
\hat{y}_{ip}=P(y_{ip}=1|X(i);\theta),
\label{cross_entropy_loss}
\end{equation}
where $\Omega_i$ is the set of all image $i$ pixels, $P$ is the probability, $x_{ip}$ is $p^{th}$ image pixel in $i^{th}$ image and, $y_{ip} \in \{0,1\}$ and $\hat{y}_{ip} \in [0,1]$ are respectively the true and \typo{the} predicted labels of $x_{ip}$. The cross-en\typo{t}ropy loss appears in the majority of deep skin lesion segmentation works\typo{, }e.g.,~\cite{41song2019dense}\typo{, }\cite{42singh2019fca}\typo{, and }\cite{48zhang2019dsm}. 

Since the gradient of the cross-entropy loss function is inversely proportional to the predicted probabilities, hard-to-predict samples are weighted more in the parameter update equations, leading to faster convergence. 
A variant, \typo{the }weighted cross-entropy \typo{loss}, penalizes pixels and class labels differently. \citet{55nasr2019dense} used pixel weights inversely proportional to their distance to lesion boundaries to enforce sharper boundaries. Class weighting may also mitigate the class imbalance, which, left uncorrected, tends to bias models towards the background class, since lesions tend to occupy a relatively small portion of images. \citet{53chen2018multi}, \citet{14goyal2019skin}, and \citet{72wang2019dermoscopic} apply such a correction, using class weights inversely proportional to \typo{the} class pixel frequency.
\citet{mirikharaji2019learning} weighted the pixels according to annotation noise estimated using a set of cleanly annotated data. All the aforementioned losses treat pixels independently without enforcing spatial coherence, which motivates their combination with other consistency-seeking losses.

\subsubsection{Dice and Jaccard Loss}
\label{Dice_Jaccard_loss}

The Dice score and the Jaccard index are two popular metrics for segmentation evaluation (Section~\ref{subsec:metrics}), measuring the overlap between predicted segmentation and ground-truth. Models may employ differentiable approximations of these metrics\typo{,} known as soft Dice~\citep{3he2017skin,46kaul2019focusnet,82he2018dense, 59wang2019automated} and soft Jaccard~\citep{5venkatesh2018deep,28hasan2020dsnet,60sarker2019mobilegan} to optimize an objective directly related to the evaluation metric. 

For two classes, these losses are defined as follows:
\begin{equation}
\mathcal{L}_{dice}(X,Y;\theta)= 1-\frac{1}{N}\sum_{i=1}^{N}\frac{2\sum_{p \in \Omega}y_{ip}\hat{y}_{ip}}{\sum_{p \in \Omega}y_{ip}+\hat{y}_{ip}},
\label{dice_loss}
\end{equation}

\begin{equation}
\mathcal{L}_{jacc}(X,Y;\theta)= 1-\frac{1}{N}\sum_{i=1}^{N}\frac{\sum_{p \in \Omega}y_{ip}\hat{y}_{ip}}{\sum_{p \in \Omega}y_{ip}+\hat{y}_{ip}-y_{ip}\hat{y}_{ip}}.
\label{jaccard_loss}
\end{equation}

Different variations of overlap-based loss functions address the class imbalance problem in medical image segmentation tasks. The Tanimoto distance loss, $\mathcal{L}_{td}$ is a modified Jaccard loss optimized in some models~\citep{71canalini2019skin,27baghersalimi2019dermonet,93yuan2017automatic}:
\begin{equation}
\mathcal{L}_{td}(X,Y;\theta)= 1-\frac{1}{N}\sum_{i=1}^{N}\frac{\sum_{p \in \Omega}y_{ip}\hat{y}_{ip}}{\sum_{p \in \Omega}y_{ip}^2+\hat{y}_{ip}^2-y_{ip}\hat{y}_{ip}},
\label{Tanimoto_distance}
\end{equation}
which is equivalent to the Jaccard loss when both ${y}_{ip}$ and $\hat{y}_{ip}$ are binary.

The Tversky loss~\citep{33abraham2019novel}, inspired by the Tversky index, is another Jaccard variant that penalizes false positives and false negatives differently to address the class imbalance problem:
\begin{equation}
\mathcal{L}_{tv}(X,Y;\theta)= 1-\frac{1}{N}\sum_{i=1}^{N}\frac{\sum_{p \in \Omega}y_{ip}\hat{y}_{ip}}{\sum_{p \in \Omega}y_{ip}\hat{y}_{ip}+\alpha y_{ip}(1-\hat{y}_{ip})+\beta (1-y_{ip})\hat{y}_{ip}},
\label{tverskey_loss}
\end{equation}
where $\alpha$ and $\beta$ tune the contributions of false negatives and false positives with $\alpha+\beta=1$.

\citet{33abraham2019novel} combined the Tvserky and focal losses~\citep{lin2017}, the latter encouraging the algorithm to focus on the hard-to-predict pixels:
\begin{equation}
    \mathcal{L}_{ftv} = \mathcal{L}_{tv}^{\frac{1}{\gamma}},
\end{equation}
where $\gamma$ controls the relative importance of the hard-to-predict samples.

\subsubsection{Matthews Correlation Coefficient Loss}
\label{Matthews_loss}

Matthews correlation coefficient (\textsf{MCC}) loss is a metric-based loss function based on the correlation between predicted and ground-truth labels~\citep{abhishek2021matthews}. In contrast to the overlap-based losses discussed in Section~\ref{Dice_Jaccard_loss}, \textsf{MCC} considers misclassifying the background pixels by penalizing false negative labels, making it more effective in the presence of skewed class distributions. \textsf{MCC} loss is defined as:

\begin{equation}
\mathcal{L}_{MCC}(X,Y;\theta)=  1 - \frac{1}{N} \sum_{i=1}^{N} 
\frac{\sum_{p \in \Omega}\hat{y}_{ip} y_{ip}  \frac{\sum_{p \in \Omega} \hat{y}_{ip} \sum_{p \in \Omega}y_{ip}}{M_i}} {f(\hat{y}_{i} y_{i})},
\label{mcc_loss}
\end{equation}

\begin{equation}
f(\hat{y}_{i}, y_{i}) = \sqrt{\sum_{p \in \Omega}\hat{y}_{ip}\sum_{p \in \Omega}{y}_{ip} - \frac{\sum_{p \in \Omega}\hat{y}_{ip}(\sum_{p \in \Omega}{y}_{ip})^2}{M_i} -\frac{(\sum_{p \in \Omega}\hat{y}_{ip})^2\sum_{p \in \Omega}{y}_{ip}}{M_i}+(\frac{\sum_{p \in \Omega}\hat{y}_{ip}\sum_{p \in \Omega}{y}_{ip}}{M_i})^2}~,
\label{f_mcc}    
\end{equation}

\noindent
where $M_i$ is the total number of pixels in \typo{the} image $i$.

\subsubsection{Deep Supervision Loss}
\label{Deep_supervision}
In \textsf{DL} models, the loss may apply not only to the final decision layer, but also to the intermediate hidden layers. The supervision of hidden layers, known as deep supervision, guides the learning of intermediate features. Deep supervision also addresses the vanishing gradient problem, leading to faster convergence and improves segmentation performance by constraining the feature space. Deep supervision loss appears in several skin lesion segmentation works~\citep{3he2017skin,22zeng2018multi,32li2018dense,18li2018deeply,82he2018dense,48zhang2019dsm,30tang2019multi}, where it is computed in multiple layers, at different scales. The loss has the general form of a weighted summation of multi-scale segmentation losses:

\begin{equation}
\mathcal{L}_{ds}(X,Y;\theta)= \sum_{l=1}^{m} \gamma_{l} \mathcal{L}_{l}(X,Y;\theta),
\label{Deep_supervision_eq}
\end{equation}

\noindent where $m$ is the number of scales, $\mathcal{L}_{l}$ is the loss at the $l^{th}$ scale, and $\gamma_{l}$ adjusts the contribution of different losses.

\subsubsection{Star-Shape Loss}
\label{subsec:starloss}

In contrast to pixel-wise losses which act on pixels independently and cannot enforce spatial constraints, the star-shape loss \citep{mirikharaji2018star} aims to capture class label dependencies and preserve the target object structure in the predicted segmentation masks. Based upon prior knowledge about the shape of skin lesions, the star-shape loss, $\mathcal{L}_{ssh}$ penalizes discontinuous decisions in the estimated output as follows:
\begin{equation}
\mathcal{L}_{ssh}(X,Y;\theta)=\sum_{i=1}^{N}\sum_{p \in \Omega}\sum_{q \in \ell_{pc}} \mathbbm{1}_{y_{ip}=y_{iq}}\times|y_{ip}-\hat{y}_{ip}|\times |\hat{y}_{ip}-\hat{y}_{iq}|,
\label{star_shape_loss}
\end{equation}
where $c$ is the lesion center, $\ell_{pc}$ is the line segment connecting pixels $p$ and $c$ and, $q$ is any pixel lying on $\ell_{pc}$. This loss encourages all pixels lying between $p$ and $q$ on $\ell_{pc}$ to be assigned the same estimator whenever $p$ and $q$ have the same ground-truth label. The result is a radial spatial coherence from the lesion center.

\subsubsection{End-Point Error Loss}
Many authors consider the lesion boundary the most challenging region to segment. The end-point error loss~\citep{16sarker2018slsdeep, 42singh2019fca} underscores borders by using the first derivative of the segmentation masks instead of their raw values:
\begin{equation}
\mathcal{L}_{epe}(X,Y;\theta) = \sum_{i=1}^{N}\sum_{p \in \Omega}\sqrt{(\hat{y}^0_{ip} - y^0_{ip})^2 + (\hat{y}^1_{ip} - y^1_{ip})^2},
\label{EPE_loss}
\end{equation}
where $\hat{y}^0_{ip}$ and $\hat{y}^1_{ip}$ are the directional first derivatives of the estimated segmentation map in the $x$ and $y$ spatial directions, respectively and, similarly,  $y^0_{ip}$ and $y^1_{ip}$ for the ground-truth derivatives. Thus, this loss function encourages the magnitude and orientation of edges of estimation and ground-truth to match, thereby mitigat\typo{ing} vague boundaries in skin lesion segmentation.

\subsubsection{Adversarial Loss}
\label{loss_adv}

Another way to add high-order class-label consistency is adversarial training. Adversarial training may be employed along with traditional supervised training to distinguish estimated segmentation from ground-truths using a discriminator. The \typo{optimization} objective will weight a pixel-wise loss $\mathcal{L}_s$ matching prediction to ground-truth, and an adversarial loss, as follows:
\begin{equation}
\mathcal{L}_{adv}(X,Y;\theta, \theta_{a}) = \mathcal{L}_s(X,Y;\theta) -\lambda [\mathcal{L}_{ce}(Y,1;\theta_{a})+\mathcal{L}_{ce}(\hat{Y},0;\theta, \theta_{a})],
\label{adv_loss}
\end{equation}
where $\theta_a$ are the adversarial model parameters. The adversarial loss employs a binary cross-entropy loss to encourage the segmentation model to produce indistinguishable prediction maps from ground-truth maps. \typo{The adversarial o}bjective \typo{(Eqn.~}\eqref{adv_loss}\typo{)} is optimized in a mini-max game by simultaneously minimizing it with respect to $\theta$ and maximizing it with respect to $\theta_a$.

Pixel-wise losses, such as cross-entropy~\citep{izadi2018generative, 42singh2019fca,29jiang2019decision}, soft Jaccard~\citep{60sarker2019mobilegan,61tu2019dense,65wei2019attention}, end-point error~\citep{61tu2019dense,42singh2019fca}, \textsf{MSE}~\citep{peng2019} and \textsf{MAE} ~\citep{60sarker2019mobilegan,42singh2019fca,29jiang2019decision} losses have all been incorporated in adversarial learning of skin lesion segmentation. In addition, \citet{87xue2018adversarial} and \citet{61tu2019dense} presented a multi-scale adversarial term to match a hierarchy of local and global contextual features in the predicted maps and ground-truths. In particular, they minimize the \textsf{MAE} of multi-scale features extracted from different layers of the adversarial model.

\subsubsection{Rank Loss}
\label{rank}

Assuming that hard-to-predict pixels lead to larger prediction errors while training the model, rank loss~\citep{81xie2020mutual} is proposed to encourage learning more discriminative information for harder pixels. The image pixels are ranked based on their prediction errors, and the top $K$ pixels with the largest prediction errors from the lesion or background areas are selected. Let $\hat{y}_{ij}^0$ and $\hat{y}_{il}^1$ are respectively the selected $j^{th}$ hard-to-predict pixel of background and $l^{th}$ hard-to-predict pixel of lesion in \typo{the} image $i$, we have:

\begin{equation}
\mathcal{L}_{rank}(X,Y;\theta) = \sum_{i=1}^{N}\sum_{j=1}^{K}\sum_{l=1}^{K} \max\{0, \hat{y}_{ij}^0 - \hat{y}_{il}^1 + margin\},
\label{rank_loss}
\end{equation}

\noindent which encourages $\hat{y}_{il}^1$ to be greater than $\hat{y}_{ij}^0$ plus margin.

Similar to rank loss, narrowband suppression loss~\citep{deng2020weakly} also adds a constraint between hard-to-predict pixels of background and lesion. Different from rank loss, narrowband suppression loss collects pixels in a narrowband along the ground-truth lesion boundary with radius $r$ instead of all image pixels and then selects the top $K$ pixels with the largest prediction errors.

\section{Evaluation}
\label{sec:evaluation}

Evaluation is one of the main challenges for any image segmentation task, skin lesions included~\citep{Celebi15a}. Segmentation evaluation may be subjective or objective~\citep{Zhang08}, the former involving the visual assessment of the results by a panel of human experts, and the latter involving the comparison of the results with ground-truth segmentations using quantitative evaluation metrics.

Subjective evaluation may provide a nuanced assessment of results, but because experts must grade each batch of results, it is usually too laborious to be applied, except in limited settings. In objective assessment, experts are consulted once, to provide the ground-truth segmentations, and that knowledge can then be reused indefinitely. However, due to intra- and inter-annotator variations, it raises the question of whether any individual ground-truth segmentation reflects the ideal “true” segmentation, an issue we address in Section~\ref{subsec:agreement}. It also raises the issue of choosing one or more evaluation metrics (Section~\ref{subsec:metrics}).

\subsection{Segmentation Annotation}
\label{subsec:annotation}

Obtaining ground-truth segmentations is paramount for the objective evaluation of segmentation algorithms. For synthetically generated images (Section~\ref{subsec:aug}), ground-truth segmentations may be known by construction, either by applying parallel transformations to the original ground-truth masks in the case of traditional data augmentation, or by training generative models to synthesize images paired with their segmentation masks. 

For images obtained from real patients, however, human experts have to provide the ground-truth segmentations. Various workflows have been proposed to reconcile the conflicting goals of ease of learning, speed, accuracy, and flexibility of annotation. On one end of the spectrum, the expert traces the lesion by hand, on images of the skin lesion printed on photographic paper, which are then scanned~\citep{Bogo15}. The technique is easy to learn and fast, but the printing and scanning procedure limits the accuracy, and the physical nature of the annotations makes corrections burdensome. On the other end of the spectrum, the annotation is performed on the computer, by a semi-automated procedure~\citep{Codella19}, with an initial border generated by a segmentation algorithm, which is then refined by the expert using an annotation software, by adjusting the parameters of the segmentation algorithm manually. This method is fast and easy to correct, but there might be a learning curve, and its accuracy depends on which algorithm is employed and how much the experts understand it.

By far, the commonest annotation method in the literature is somewhere in the middle, with fully manual annotations performed on a computer. The skin lesion image file may be opened either in a raster graphics editor (e.g., \typo{\textsf{GNU} Image Manipulation Program (}\textsf{GIMP}\typo{)} or Adobe Photoshop), or in a dedicated annotation software~\citep{Ferreira12}, where the expert traces the borders of the lesion using a mouse or stylus, with continuous freehand drawing, or with discrete control points connecting line segments (resulting in a polygon~\citep{Codella19}) or smooth curve segments (e.g., cubic B-splines~\citep{Celebi07b}). This method provides a good compromise, being easy to implement, fast\typo{,} and accurate to perform, after an acceptable learning period for the annotator. 

\subsection{Inter-Annotator Agreement}
\label{subsec:agreement}

Formally, dataset ground-truths can be viewed as samples of an estimator of the true label, which can never be directly observed~\citep{Smyth95}. This problem is often immaterial for classification, when annotation noise is small. However, in medical image segmentation, ground-truths suffer from both biases (systematic deviations from the “ideal”) and significant noise~\citep{Zijdenbos94,Chalana97,Guillod02,Grau04,Bogo15,Lampert16}, the latter appearing as inter-annotator (different experts) and intra-annotator (same expert at different times) variability. 

In the largest study of its kind to date, \citet{Fortina12} measured the inter-annotator variability among $12$ dermatologists with varying levels of experience on a set of $77$ dermoscopic images, showing that the average pairwise \textsf{XOR} dissimilarity (Section~\ref{subsec:metrics}) between annotators was $\approx15\%$, and that in $10\%$ of cases, this value was $>28\%$. They found more agreement among more experienced dermatologists than less experienced ones. Also, more experienced dermatologists tend to outline tighter borders than less experienced ones. They suggest that the level of agreement among experienced dermatologists could serve as an upper bound for the accuracy achievable by a segmentation algorithm, i.e., if even highly experienced dermatologists disagree on how to classify $10\%$ of an image, it might be unreasonable to expect a segmentation algorithm to agree with more than $90\%$ of any given ground-truth on the same image \citep{Fortina12}.

Due to the aforementioned variability issues, whenever possible, skin lesion segmentation should be evaluated against multiple expert ground-truths, a good algorithm being one that agrees with the ground-truths at least as well as the expert agree among themselves~\citep{Chalana97}. Due to the cost of annotation, however, algorithms are often evaluated against a single ground-truth. 

When multiple ground-truths are available, the critical issue is how to employ them. Several approaches have been proposed:
\begin{itemize}
   \item Preferring one of the annotations (e.g., the one by the most experienced expert) and ignoring the others~\citep{Celebi07b}.
   \item Measuring and reporting the results for each annotator separately~\citep{Celebi08}, which might require non-trivial multivariate analyses if the aim is to rank the algorithms.
   \item Measuring each automated segmentation against all corresponding ground-truths and reporting the average result~\citep{Schaefer11}.
   \item Measuring each automated segmentation against an \emph{ensemble ground-truth} formed by combining the corresponding ground-truths pixel-wise using a bitwise \textsf{OR}~\citep{Garnavi11b,Garnavi11c}, bitwise \textsf{AND}~\citep{Garnavi11a}, or a majority voting~\citep{Iyatomi06,Iyatomi08,Norton12}.
\end{itemize}

The ground-truth ensembling process can be generalized using a \emph{thresholded probability map}~\citep{Biancardi10}. First, all ground-truths for a sample are averaged pixel-wise into a \emph{probability map}. Then the map is binarized, with the lesion corresponding to pixels greater than or equal to a chosen threshold. The operations of \textsf{OR}, \textsf{AND}, and majority voting, correspond, respectively to thresholds of $1/n$, $1$, and $(n-\varepsilon)/2n$, with $n$ being the number of ground-truths, and  $\varepsilon$ being a small positive constant. \textsf{AND} and \textsf{OR} correspond, respectively, to the tightest and loosest possible contours, with other thresholds leading to intermediate results. While the optimal threshold value is data-dependent, large thresholds focus the evaluation on unambiguous regions, leading to overly optimistic evaluations of segmentation quality~\citep{Smyth95,Lampert16}.

The abovementioned approaches fail to consider the differences of experience or performance of the annotators~\citep{Warfield04}. More elaborate ground-truth fusion alternatives include shape averaging~\citep{Rohlfing06}, border averaging~\citep{Chen89,Chalana97}, binary label fusion algorithms such as \textsf{STAPLE} \citep{Warfield04}, \textsf{TESD} \citep{Biancardi10}, and \textsf{SIMPLE} \citep{Langerak10}, as well as other more recent algorithms \citep{Peng13,Peng16,Peng17}.

\textsf{STAPLE} (Simultaneous Truth And Performance Level Estimation) has been very influential in medical image segmentation evaluation, inspiring many variants. For each image and its ground-truth segmentations, \textsf{STAPLE} estimates a probabilistic true segmentation through the optimal combination of individual ground-truths, weighting each one by the estimated sensitivity and specificity of its annotator. \textsf{STAPLE} may fail when there are only a few annotators or when their performances vary too much~\citep{Langerak10,Lampert16}, a situation addressed by \textsf{SIMPLE} (Selective and Iterative Method for Performance Level Estimation)~\citep{Langerak10} by iteratively discarding poor quality ground-truths.

\begin{figure}[!ht]
\centering
\includegraphics[totalheight=0.36\textwidth,draft=false]{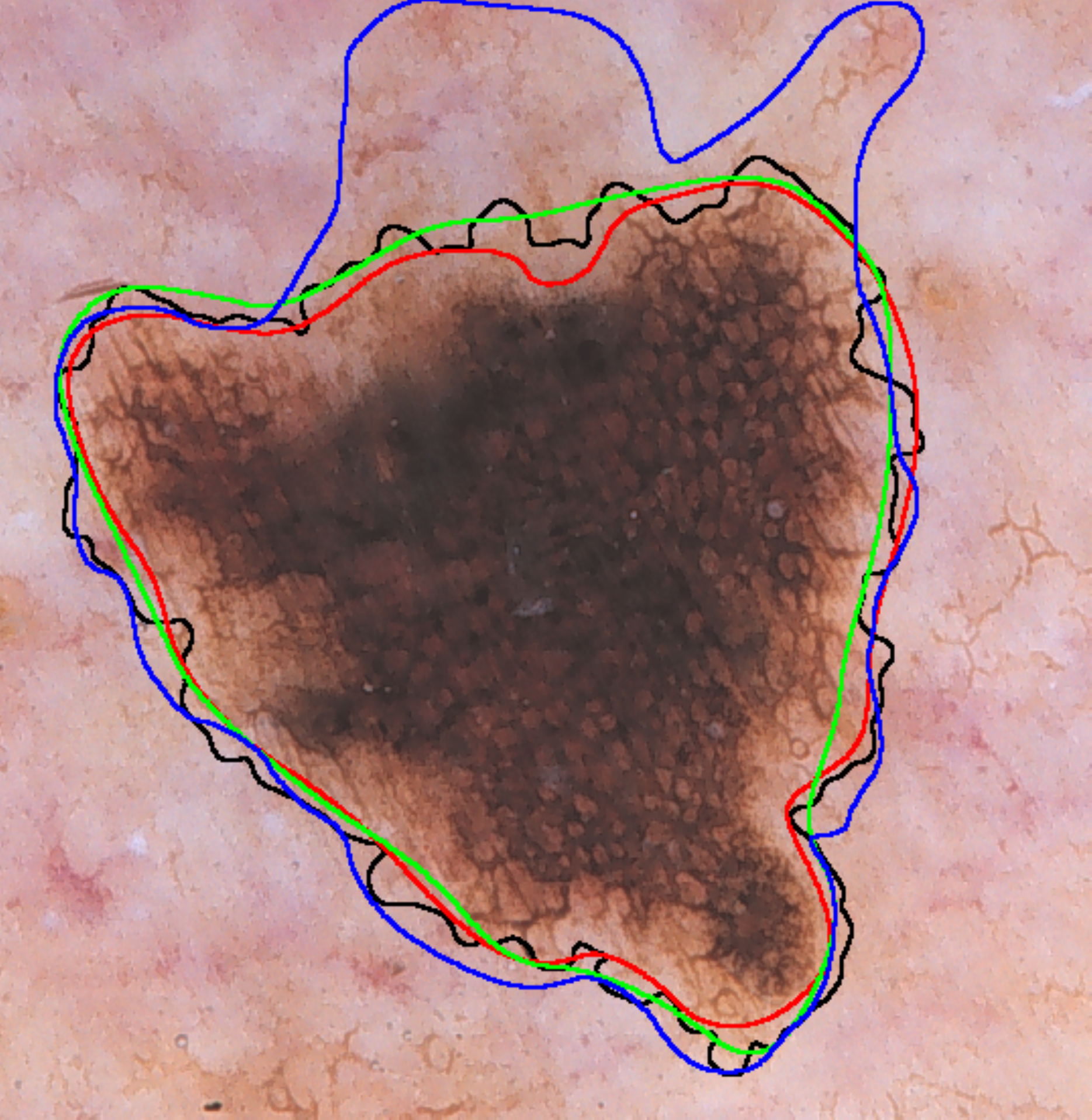}
\caption{ \label{fig:npri} Sample segmentation results demonstrating inter-annotator disagreements. \revision{Note how annotator preferences can affect the manual segmentations, e.g., smooth lesion borders (green), jagged lesion borders (black), oversegmented lesion (blue), etc. Figure taken from \citet{Celebi09a} with permission.}}
\end{figure}

Instead of attempting to fuse multiple ground-truths into a single one before employing conventional evaluation metrics, the metrics themselves may be modified to take into account annotation variability. \citet{Celebi09a} proposed the \emph{normalized probabilistic rand index} (\textsf{NPRI}) \citep{Unnikrishnan07}, a generalization of the \emph{rand index} \citep{Rand71}. 
It penalizes segmentation results more (less) in regions where the ground-truths agree (disagree). Fig.~\ref{fig:npri} illustrates the idea: ground-truths outlined by three experienced dermatologists appear in red, green, and blue, while the automated result appears in black. \textsf{NPRI} does \emph{not} penalize the automated segmentation in the upper part of the image, where the blue border seriously disagrees with the other two~\citep{Celebi09a}. Despite its many desirable qualities, \textsf{NPRI} has a subtle flaw: it is non-monotonic with the fraction of misclassified pixels \citep{Peserico10}. Consequently, this index might be unsuitable for comparing poor segmentation algorithms. 

\subsection{Evaluation Metrics}
\label{subsec:metrics}

We can frame the skin lesion segmentation problem as a binary pixel-wise classification task, where the positive and negative classes correspond to the lesion and the background skin, respectively. Suppose that we have an input image and its corresponding segmentations: an \emph{automated segmentation} (\textsf{AS}) produced by a segmentation algorithm and a \emph{manual segmentation} (\textsf{MS}) outlined by a human expert. We can formulate a number of quantitative segmentation evaluation measures based on the concepts of \emph{true positive}, \emph{false negative}, \emph{false positive}, and \emph{true negative}, whose definitions are given in Table~\ref{tab:pairs}. In this table, actual and detected pixels refer to any given pixel in the \textsf{MS} and the corresponding pixel in the \textsf{AS}, respectively.

\begin{table}[!ht]
\centering
\caption{\label{tab:pairs} Definitions of true positive, false negative, false positive, and true negative \revision{pixels in the context of skin lesion segmentation}.}

\resizebox{0.45\textwidth}{!}{
\setlength{\tabcolsep}{0.5em}
\def\arraystretch{1.85}
\begin{tabular}{ll l l}
 \multicolumn{2}{c}{} & \multicolumn{2}{c}{Detected Pixel} \\
\cline{3-4}
 &  & \multicolumn{1}{|l|}{Lesion $(+)$} & \multicolumn{1}{l|}{Background $(-)$} \\
\cline{2-4}
\multicolumn{1}{l|}{Actual} & \multicolumn{1}{l|}{Lesion $(+)$} & \multicolumn{1}{l|}{True Positive} & \multicolumn{1}{l|}{False Negative} \\
\cline{2-4}
\multicolumn{1}{l|}{Pixel} & \multicolumn{1}{l|}{Background $(-)$} & \multicolumn{1}{l|}{False Positive} & \multicolumn{1}{l|}{True Negative} \\
\cline{2-4}
\end{tabular}}
\end{table}

For a given pair of automated and manual segmentations, we can construct a $2 \times 2$ confusion matrix \revision{(aka a contingency table \citep{pearson1904theory,miller1955analysis})} $\mathsf{C} = \begin{psmallmatrix} \textsf{TP} & \textsf{FN}\\\textsf{FP} & \textsf{TN}\end{psmallmatrix}$, where \textsf{TP}, \textsf{FN}, \textsf{FP}, and \textsf{TN} denote the numbers of true positives, false negatives, false positives, and true negatives, respectively. Clearly, we have $N = \textsf{TP} + \textsf{FN} + \textsf{FP} + \textsf{TN}$, where $N$ is the number of pixels in either image. Based on these quantities, we can define a variety of scalar similarity measures to quantify the accuracy of segmentation~\citep{Baldi00,Japkowicz11,Taha15}:
\begin{itemize}
   \item \revision{Sensitivity (\textsf{SE}) and Specificity (\textsf{SP})~\citep{kahn1942serology,yerushalmy1947statistical,binney2021origin}: \textsf{SE} = $\dfrac{\textsf{TP}}{\textsf{TP} + \textsf{FN}}$ \& \textsf{SP} = $\dfrac{\textsf{TN}}{\textsf{TN} + \textsf{FP}}$}
   \item \revision{Precision (\textsf{PR}) and Recall (\textsf{RE})~\citep{kent1955machine}: \textsf{PR} = $\dfrac{\textsf{TP}}{\textsf{TP} + \textsf{FP}}$ \& \textsf{RE} = $\dfrac{\textsf{TP}}{\textsf{TP} + \textsf{FN}}$}
 \item $\text{Accuracy (\textsf{AC})} = \dfrac{\textsf{TP} + \textsf{TN}}{\textsf{TP} + \textsf{FN} + \textsf{FP} + \textsf{TN}}$ 
   \item $\text{F-measure (\textsf{F})}$~\citep{vanRijsbergen79} = $\dfrac{2 |\textsf{AS} \cap \textsf{MS}|}{|\textsf{AS}| + |\textsf{MS}|} = \dfrac{2 \cdot \textsf{PR} \cdot \textsf{RE}}{\textsf{PR} + \textsf{RE}} = \dfrac{2 \textsf{TP}}{2 \textsf{TP} + \textsf{FP} + \textsf{FN}}$
   \item $\text{G-mean (\textsf{GM})}$~\citep{Kubat98} = $\sqrt{\textsf{SE} \cdot \textsf{SP}}$ 
   \item $\text{Balanced Accuracy (\textsf{BA})}$ ~\citep{Chou78} = $\dfrac{\textsf{SE} + \textsf{SP}}{2}$  
   \item $\text{Jaccard index (\textsf{J})}$~\citep{Jaccard01} =
         $\dfrac{|\textsf{AS} \cap \textsf{MS}|}{|\textsf{AS} \cup \textsf{MS}|} =
         \dfrac{\textsf{TP}}{\textsf{TP} + \textsf{FN} + \textsf{FP}}$  
   \item $\text{Matthews Correlation Coefficient (\textsf{MCC})}$~\citep{Matthews75} = $\dfrac{\textsf{TP} \cdot \textsf{TN} - \textsf{FP} \cdot \textsf{FN}}{\sqrt{(\textsf{TP} + \textsf{FP}) (\textsf{TP} + \textsf{FN})(\textsf{TN} + \textsf{FP})(\textsf{TN} + \textsf{FN})}}$ 
\end{itemize}

For each similarity measure, the higher the value, the better the segmentation. Except for \textsf{MCC}, all of these measures have a unit range, that is, $[0,1]$. The $[-1, 1]$ range of \textsf{MCC} can be mapped to $[0,1]$ by adding one to it and then dividing by two. Each of these unit-range similarity measures can then be converted to a unit-range dissimilarity measure by subtracting it from one. Note that there are also dissimilarity measures with no corresponding similarity formulation. A prime example is the well-known \textsf{XOR} measure \citep{Hance96} defined as follows:
\begin{equation}
   \text{\textsf{XOR}} =
   \dfrac{|\textsf{AS} \oplus \textsf{MS}|}{|\textsf{MS}|} =
   \dfrac{|\left( \textsf{AS} \cup \textsf{MS} \right) - \left( \textsf{AS} \cap \textsf{MS} \right)|}{|\textsf{MS}|} =
   \dfrac{\textsf{FP} + \textsf{FN}}{\textsf{TP} + \textsf{FN}}. 
\end{equation}

It is essential to notice that different evaluation measures capture different aspects of a segmentation algorithm's performance on a given dataset, and thus there is no universally applicable evaluation measure \citep{Japkowicz11}. This is why most studies employ multiple evaluation measures in an effort to perform a comprehensive performance evaluation. Such a strategy, however, complicates algorithm comparisons, unless one algorithm completely dominates the others with respect to all adopted evaluation measures.

Based on their observation that experts tend to avoid missing parts of the lesion in their manual borders, \citet{Garnavi11b} argue that true positives have the highest importance in the segmentation of skin lesion images. The authors also assert that false positives (background pixels incorrectly identified as part of the lesion) are less important than false negatives (lesion pixels incorrectly identified as part of the background). Accordingly, they assign a weight of $1.5$ to \textsf{TP} to signify its overall importance. Furthermore, in measures that involve both \textsf{FN} and \textsf{FP} (e.g., \textsf{AC}, \textsf{F}, and \textsf{XOR}), they assign a weight of 0.5 to \textsf{FP} to emphasize its importance over \textsf{FN}. Using these weights, they construct a \emph{weighted performance index}, which is an arithmetic average of six commonly used measures, namely \textsf{SE}, \textsf{SP}, \textsf{PR}, \textsf{AC}, \textsf{F}, and (unit complement of) \textsf{XOR}. This scalar evaluation measure facilitates comparisons among algorithms.

In a \typo{follow-up} study, \citet{Garnavi11c} parameterize the weights of \textsf{TP}, \textsf{FN}, \textsf{FP}, and \textsf{TN} in their  weighted performance index and then use a constrained non-linear program to determine the optimal weights. They conduct experiments with five segmentation algorithms on $55$ dermoscopic images. They conclude that the optimized weights not only lead to automated algorithms that are more accurate against manual segmentations, but also diminish the differences among those algorithms.

We make the following key observations about the popular evaluation metrics and how they have been used in the skin lesion segmentation literature:

\begin{itemize}
	\item Historically, \textsf{AC} has been the most popular evaluation measure owing to its simple and intuitive formulation. However, this measure tends to favor the majority class, leading to overly optimistic performance estimates in class-imbalanced domains. This drawback prompted the development of more elaborate performance evaluation measures, including \textsf{GM}, \textsf{BA}, and \textsf{MCC}.
	
	\item \textsf{SE} and \textsf{SP} are especially popular in medical domains\revision{, tracing their usage in serologic test reports in the early 1900s~\citep{binney2021origin}}. \textsf{SE} (aka \emph{True Positive Rate}) quantifies the accuracy on the positive class, whereas \textsf{SP} (aka \emph{True Negative Rate}) quantifies the accuracy on the negative class. These measures are generally used together because it is otherwise trivial to maximize one at the expense of the other (an automated border enclosing the corresponding manual border will attain a perfect \textsf{SE}, whereas in the opposite case, we will have a perfect \textsf{SP}). Unlike \textsf{AC}, they are suitable for class-imbalanced domains. \textsf{BA} and \textsf{GM} combine these measures into a single evaluation measure through arithmetic and geometric averaging, respectively. Unlike \textsf{AC}, these composite measures are suitable for class-imbalanced domains \citep{Luque20}.

	\item \textsf{PR} is the proportion of examples assigned to the positive class that actually belongs to the positive class. \textsf{RE} is equivalent to \textsf{SE}. \textsf{PR} and \textsf{RE} are typically used in information retrieval applications, where the focus is solely on relevant documents (positive class). \textsf{F} combines these measures into a single evaluation measure through harmonic averaging. This composite measure, however, is unsuitable for class-imbalanced domains \citep{Zou04,Chicco20,Luque20}.  
   
   \item \textsf{MCC} is equivalent to the \emph{phi coefficient}, which is simply the \emph{Pearson correlation coefficient} applied to binary data \citep{Chicco20}. \textsf{MCC} values fall within the range of $[-1,1]$ with $-1$ and $1$ indicating perfect misclassification and perfect classification, respectively, while $0$ indicating a classification no better than random \citep{Matthews75}. Although it is biased to a certain extent \citep{Luque20,Zhu20}, this measure appears to be suitable for class-imbalanced domains \citep{Boughorbel17,Chicco20,Luque20}. 

   \item \textsf{J} (aka \emph{Intersection over Union} \revision{\citep{jaccard1912distribution}}) and \textsf{F} (aka \emph{Dice coefficient} \revision{aka \emph{S{\o}rensen-Dice coefficient}} \citep{Dice45,sorensen1948method}) are highly popular in medical image segmentation \citep{Crum06}. These measures are monotonically related as follows: $J = F/(2 - F)$ and $F = 2J/(1 + J)$. Thus, it makes little sense to use them together. There are two major differences between these measures:
      \begin{inparaenum}[(i)]
         \item $(1 - J)$  is a proper distance metric, whereas $(1 - F)$ is \emph{not} (it violates the triangle inequality).
	 \item It can be shown \citep{Zijdenbos94} that if \textsf{TN} is sufficiently large compared to \textsf{TP}, \textsf{FN}, and \textsf{FP}, which is common in skin lesion segmentation, $F$ becomes equivalent to \emph{Cohen's kappa} \citep{Cohen60}, which is a chance-corrected measure of inter-observer agreement.
      \end{inparaenum}
  
  \item Among the seven composite evaluation measures given above, \textsf{AC}, \textsf{GM}, \textsf{BA}, and \textsf{MCC} are symmetric, that is, \typo{they are }invariant to class swapping, while \textsf{F}, \textsf{J}, and \textsf{XOR} are asymmetric.

  \item \textsf{XOR} is similar to \emph{False Negative Rate}, that is, the unit complement of \textsf{SE}, with the exception that \textsf{XOR} has an extra additive \textsf{TN} term in its numerator. While \textsf{XOR} values are guaranteed to be nonnegative, they do \emph{not} have a fixed upper bound, which makes aggregations of this measure difficult. \textsf{XOR} is also biased against small lesions \citep{Celebi09a}. Nevertheless, owing to its intuitive formulation, \textsf{XOR} was popular in skin lesion segmentation until about 2015 \citep{Celebi15a}.

  \item The 2016 \revision{and 2017} \textsf{ISIC} Challenge\revision{s} \citep{Gutman16,Codella18} adopted five measures: \textsf{AC}, \textsf{SE}, \textsf{SP}, \textsf{F}, and \textsf{J}, with the participants ranked based on the last measure. The 2018 \textsf{ISIC} Challenge \revision{\citep{Codella19}} featured a \emph{thresholded Jaccard index}, which returns the same value as the original \textsf{J} if the value is greater than or equal to a predefined threshold and zero otherwise. Essentially, this modified index considers automated segmentations yielding \textsf{J} values below the threshold as complete failures. The \typo{challenge} organizers set the threshold equal to $0.65$ based on an earlier study \citep{Codella17} that determined the average pairwise \textsf{J} similarities among the manual segmentations outlined by three expert dermatologists. \revision{Since the majority of papers in this survey ($168$ out of $177$ papers) use the \textsf{ISIC} datasets (Fig.~\ref{data_eval_freq}), we list the \textsf{J} for all the papers in Table~\ref{tab:main} wherever it has been reported in the corresponding papers. For papers that did not report \textsf{J} and instead reported \textsf{F}, we list the computed \textsf{J} based on \textsf{F} and denote it with an asterisk.}

  \item Some of the aforementioned measures (\typo{i.e.}, \textsf{GM} and \textsf{BA}) have \emph{not} been used in a skin lesion segmentation study yet.
  
  \item The evaluation measures discussed above are all region-based and thus fairly insensitive to border irregularities~\citep{Lee03}, \typo{i.e.}, indentations\typo{,} and protrusions along the border. Boundary-based evaluation measures~\citep{Taha15} have \emph{not} been used in the skin lesion segmentation literature much except \typo{for} the symmetric Hausdorff metric~\citep{Silveira09}, which is known to be sensitive to noise~\citep{Huttenlocher93} and biased in favor of small lesions~\citep{Bogo15}.
\end{itemize}

\section{\revision{Discussion and Future Research}}
\label{sec:conc_future}
In this paper, we presented an overview of \textsf{\revision{DL}}-based skin lesion segmentation algorithms. A lot of work has been done in this field since the first application of \textsf{CNN}s on these images in 2015~\citep{Codella15}. In fact, the number of skin lesion segmentation papers published over the past \revision{$8$} years (2015--\revision{2022}) is \revision{more than thrice} those published over the previous $17$ years (1998--2014)~\citep{Celebi15a}.

\revision{However, despite the large body of work, skin lesion segmentation remains an open problem, as evidenced by the \textsf{ISIC} 2018 Skin Lesion Segmentation Live Leaderboard~\citep{ISICLiveLeaderboard}. The live leaderboard has been open and accepting submissions since 2018, and even after the permitted usage of external data, the best thresholded Jaccard index (the metric used to rank submissions) is $83.6\%$. Additionally, the release of the \textsf{HAM10000} lesion segmentations~\citep{tschandl2020human,ham10ksegmentations} in 2020 shows that progressively larger skin lesion segmentation datasets continue to be released.} We believe that the following aspects of skin lesion segmentation \revision{via deep learning are worthy of future work}:

\begin{itemize}
   \item Mobile dermoscopic image analysis: \revision{With the availability of} various inexpensive dermoscopes designed for smartphones\typo{,} mobile dermoscopic image analysis is of great interest worldwide, especially in regions where access to dermatologists is limited. Typical \textsf{\revision{DL}}-based image segmentation algorithms have millions of weights. In addition, classical \textsf{CNN} architectures are known to \typo{exhibit} difficulty dealing with certain image distortions such as noise and blur~\citep{Dodge16}\revision{, and DL-based skin lesion diagnosis models have been demonstrated to be susceptible to similar artifacts: various kinds of noise and blur, brightness and contrast changes, dark corners~\citep{maron2021benchmark}, bubbles, rulers, ink markings, etc.~\citep{katsch2022comparison}}. Therefore, the current dermoscopic image segmentation algorithms may not be ideal for execution on \revision{typically} resource-constrained mobile \revision{and edge} devices\revision{, needed for patient privacy so that uploading skin images to remote servers is avoided}. Leaner \textsf{\typo{DL}} architectures\typo{, }e.g., MobileNet~\citep{Howard19}, ShuffleNet~\citep{Zhang18}, EfficientNet~\citep{Tan19a}, MnasNet~\citep{Tan19b}, and UNeXt~\citep{valanarasu2022unext}\typo{,} should be investigated in addition to the robustness of such architectures with respect     to image noise and blur.
   
   \item Datasets: To train more accurate and robust deep neural segmentation architectures, we need larger, more diverse, and more representative skin lesion datasets with multiple manual segmentations per image. Additionally, as mentioned in Section~\ref{subsec:datasets}\typo{,} several skin lesion image classification datasets do not have the corresponding lesion mask annotations, and given their popularity in skin image analysis tasks, they may be good targets for manual delineations. For example, the PAD-UFES-20 dataset~\citep{pacheco2020pad} consists of clinical images of skin lesions captured using smartphones, and obtaining ground-truth segmentations on this dataset would help advance skin image analysis on mobile devices. \revision{Additionally, a recent study conducted by \citet{daneshjou2021lack} found that as little as 10\% of the AI-based studies for dermatological diagnosis included skin tone information for at least one dataset used, and that several studies included little to no images of darker skin tones, underlining the need to curate datasets with diverse skin tones.}

   \item \revision{Collecting segmentation annotations}: 
   At the time of this writing, the \textsf{ISIC} Archive contains over \revision{$71,000$} publicly available images. Considering that the largest public dermoscopic image set contained a little over $1,000$ images about \revision{six} years ago, we have come a long way. The more pressing problem now is the lack of manual segmentations for most of these images. Since manual segmentation by medical experts is laborious and costly, crowdsourcing techniques~\citep{Kovashka16} could be explored to collect annotations from non-experts. Experts could then revise these initial annotations\revision{, or methods that tackle the problem of annotation noise~\citep{mirikharaji2019learning,karimi2020deep,li2021superpixel} could be explored}. Note that the utility of crowdsourcing in medical image annotation has been demonstrated in multiple studies~\citep{Foncubierta12,Gurari15,Sharma17,Goel20}. \revision{
   Additionally, keeping in mind the time-consuming nature of manual supervised annotation, an
   alternative is to use weakly-supervised annotation, e.g., bounding-box annotations~\citep{Dai15,Papandreou15}, which are much less time-consuming to collect. For example, for several large skin lesion image datasets that do not have any lesion mask annotations (see Section~\ref{subsec:datasets}), bounding-box lesion annotations can be obtained more easily than dense pixel-level segmentation annotations. In addition, weakly-supervised annotation \citep{bearman2016s,tajbakhsh2020embracing,roth2021going,en2022annotation} is more amenable to crowdsourcing \citep{maier2014can,rajchl2016learning,papadopoulos2017extreme,lin2019block}, especially for non-experts.}

   \item \revision{Handling multiple annotations per image}:
  If the skin lesion image \typo{data}set at hand contains multiple manual segmentations per image, one should consider \revision{either using an algorithm such as \textsf{STAPLE}~\citep{Warfield04} for fusing the manual segmentations (see Section~\ref{sec:evaluation}), or relying on learning-based approaches, either through variants of STAPLE adapted for \textsf{DL}-based segmentation~\citep{kats2019soft,zhang2020learning}, or other methods~\citep{133mirikharaji2020d,melba_2022_031_lemay}}. Such a fusion algorithm can also be used to build an ensemble of multiple automated segmentations.
   
   \item Supervised \typo{s}egmentation \typo{e}valuation \typo{m}easures: Supervised segmentation evaluation measures popular in the skin image analysis literature (see \highlight{Section~\ref{subsec:metrics}}) are often region-based, pair-counting measures. Other region-based measures, such as information-theoretic measures \revision{(e.g., mutual information, variation of information, etc.)} as well as boundary-based measures \revision{e.g., Hausdorff distance} \citep{Taha15} should be explored as well.

   \item Unsupervised \typo{s}egmentation and \typo{u}nsupervised \typo{s}egmentation \typo{e}valuation: Current \textsf{\typo{DL}}-based skin lesion segmentation algorithms are \revision{mostly} based on supervised \revision{learning, as shown in a supervision-level breakdown of the surveyed works (Fig.~\ref{semi_chart})}, meaning that these algorithms require manual segmentations for training \revision{segmentation prediction models}. Nearly all of these segmentation studies employ supervised segmentation evaluation, meaning that they also require manual segmentations for testing. Due to the scarcity of annotated skin lesion images, it may be beneficial to investigate unsupervised \textsf{DL}~\citep{Ji19} as well as unsupervised segmentation evaluation~\citep{Chabrier06,Zhang08}.
   
   \item Systematic \typo{e}valuations: Systematic evaluations that have been performed for skin lesion classification~\citep{Valle2020,Bissoto_2021_CVPR,Perez18} are, so far, nonexistent in the skin lesion segmentation literature. For example\typo{,} statistical significance analysis are conducted on the results of a few prior studies in skin lesion segmentation\typo{, }e.g., \citet{Fortina12}.

   \item Fusion of \typo{h}and-\typo{c}rafted and \typo{d}eep \typo{f}eatures: Can we integrate the deep features extracted by \revision{\textsf{DL} models} and hand-crafted features synergistically? \revision{For example, exploration of shape and appearance priors of skin lesions that may be beneficial to incorporate, via loss terms~\citep{nosrati2016incorporating,el2021high,ma2021loss}, in deep learning models for skin lesion segmentation, similar to star-shape~\citep{mirikharaji2018star} and boundary priors~\citep{wang2021boundary}.}

   \item Loss of \typo{s}patial \typo{r}esolution: The use of repeated subsampling in \textsf{CNN}s leads to coarse segmentations. Various approaches have been proposed to minimize the loss of spatial resolution, including fractionally-strided convolution (or deconvolution)~\citep{long2015fully}, atrous (or dilated) convolution~\citep{Chen17}, and conditional random fields~\citep{Krahenbuhl11}. More research needs to be conducted to determine appropriate strateg\typo{ies} for skin lesion segmentation \typo{that effectively minimize or avoid the loss of spatial resolution}.

   \item Hyperparameter \typo{t}uning: Compared to traditional machine learning classifiers (e.g., nearest neighbors, decision trees, and support vector machines), deep neural networks have a large number of hyperparameters related to their architecture, optimization, and regularization. An average \textsf{CNN} classifier has about a dozen or more hyperparameters~\citep{Bengio12} and tuning these hyperparameters systematically is a laborious undertaking. \emph{Neural architecture search} is an active area of research~\citep{Elsken19}\typo{,} and some of these model selection approaches have already been applied to \revision{semantic} segmentation~\citep{Liu19} \revision{and medical image segmentation~\citep{weng2019unet}}.
   
   \item \revision{Reproducibility of results: \cite{kapoor2022leakage} define research in ML-based science to be reproducible if the associated datasets and the code are publicly available and if there are no problems with the data analysis, where problems include the lack of well-defined training and testing partitions of the dataset, leakage across dataset partitions, features selection using the entire dataset instead of only the training partition, etc. Since several skin lesion segmentation datasets come with standardized partitions (Table~\ref{datasets}), sharing of the code can lead to more reproducible research~\citep{colliot2022reproducibility}, with the added benefit to researchers who release their code to be cited significantly more~\citep{vandewalle2012code}. In our analysis, we found that only $38$ of the $177$ surveyed papers ($21.47\%$) had publicly accessible code (Table~\ref{tab:main}), a proportion similar to a smaller-scale analysis by \citet{renard2020variability} for medical image segmentation. Another potential assessment of a method's generalization performance is its evaluation on a common held-out test set, where the ground truth segmentation masks are private, and users submit their test predictions to receive a performance assessment. For example, the \textsf{ISIC} 2018 dataset's test partition is available through a live leaderboard~\citep{ISICLiveLeaderboard}, but it is rarely used. We found that out of $71$ papers published in 2021 and 2022 included in this survey, $36$ papers reported results on the \textsf{ISIC} 2018 dataset, but only 1 paper \citep{170saini2021bsegnet} used the online submission platform for evaluation.}

   \begin{figure}
    \centering
    \begin{minipage}{0.75\textwidth}
        
        \captionsetup{width=1.0\linewidth}
        \includegraphics[width=1\linewidth]{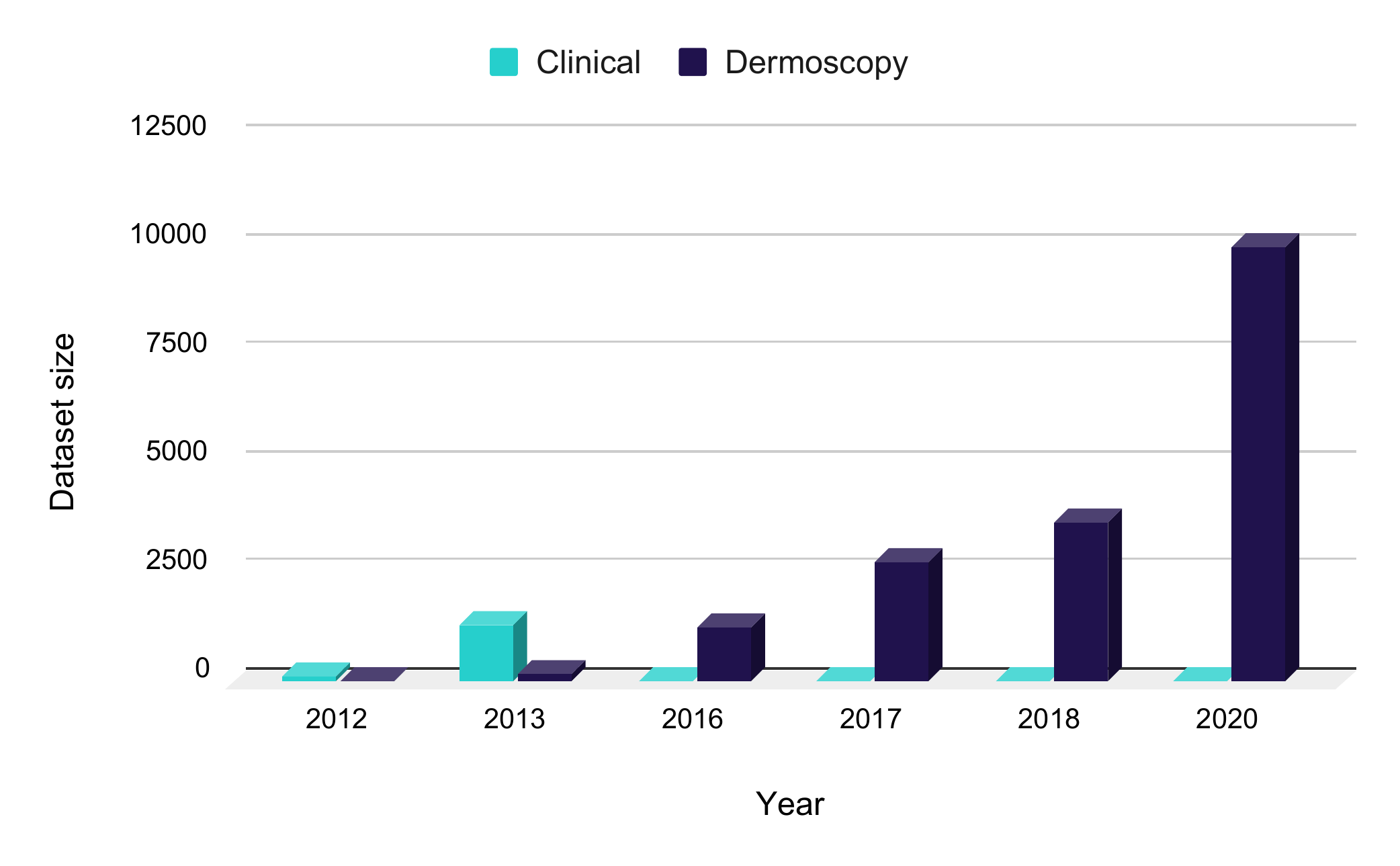}
        \caption{Number of skin lesion images with ground-truth segmentation maps per year categorized based on modality\revision{. It is evident that while the number of dermoscopic skin lesion images has been constantly on the rise, the number of clinical images has remained unchanged for the past several years.}}
        \label{data_mod_chart}
    \end{minipage}
\end{figure}

  \item Research on \typo{c}linical \typo{i}mages: Another limitation is the limited number of benchmark datasets of clinical skin lesion images with expert pixel-level annotations. Fig.~\ref{data_mod_chart} shows that while the number of dermoscopic image datasets with ground-truth segmentation masks has been increasing over the last few years, only a few datasets with clinical images are available. In contrast to dermoscopic images requiring a special tool that is not always utilized even by dermatologists~\citep{engasser2010dermatoscopy}, clinical images captured by digital cameras or smartphones have the advantage of easy accessibility, which can be utilized to evaluate the priority of patients by their lesion severity level\revision{, i.e., triage patients}. \revision{As shown in Fig.~\ref{data_eval_freq} and Table~\ref{tab:main},} most of the deep skin lesion segmentation models are trained \revision{and evaluated} on dermoscopic images\revision{, primarily because of the lack of large-scale clinical skin lesion image segmentation datasets (Table~\ref{datasets})}, leaving the need to develop automated tools for non-specialists unmet.
  
  \item Research on \typo{t}otal \typo{b}ody \typo{i}mages: While there has been some research towards detecting and tracking skin lesions over time in 2D wide-field images~\citep{Mirzaalian2016,li2016skin,Korotkov2019,Soenksen2021,huang2022dicom} and in 3D total body images~\citep{Bogo2014a,zhao2022skin3d,ahmedt2023monitoring}, simultaneous segmentation of skin lesions from total body images~\citep{sinha2023dermsynth3d} would help with early detection of melanoma~\citep{halpern2003total,hornung2021value}, thus improving patient outcomes.
  
  \item \revision{Effect on downstream tasks: End-to-end systems have been proposed for skin images analysis tasks that directly learn the final tasks (e.g., predicting the diagnosis~\citep{Kawahara19} or the clinical management decisions~\citep{abhishek2021predicting} of skin lesions), and these approaches present a number of advantages such as computational efficiency and ease of optimization. On the other hand, skin lesion diagnosis pipelines have been shown to benefit from the incorporation of prior knowledge, specifically lesion segmentation masks~\citep{yan2019melanoma}. Therefore, it is worth investigating how lesion segmentation, often an intermediate step in the skin image analysis pipeline, affects the downstream dermatological tasks.}

  \item \revision{From binary to multi-class segmentation:} 
  \revision{While the existing work in skin lesion segmentation is mainly binary segmentation, future work may explore multi-class settings. For example,}
  automated detection and delineation of clinical dermoscopic features \revision{(e.g., globules, streaks, pigment networks)} within a skin lesion may lead to superior classification performance. \revision{ Further, }dermoscopic feature extraction\revision{, a task in the ISIC 2016~\citep{Gutman16} and 2017~\citep{Codella18} challenges,} can be formulated as a multi-class segmentation problem \citep{kawahara2018fully}. \revision{The multiclass formulation} can \revision{then} be addressed by \textsf{DL} models\revision{, and can be used either as an intermediate step for improving skin lesion diagnosis or used directly in diagnosis models for regularizing attention maps \citep{yan2019melanoma}. Similarly, multi-class segmentation scenarios may also include multiple skin pathologies on one subject, especially in images with large fields of view, or segmentation of the skin, the lesion(s), and the background, especially in in-the-wild images with diverse backgrounds, such as those in the Fitzpatrick17k dataset~\citep{groh2021evaluating}}.

  \item Transferability of \typo{m}odels: As the majority of skin lesion datasets are from fair-skinned patients, the generalizability of deep models to populations with diverse skin complexions is questionable. \revision{With the emergence of dermatological datasets with diverse skin tones~\citep{groh2021evaluating,daneshjou2021disparities} and methods for diagnosing pathologies fairly~\citep{bevan2022detecting,wu2022fairprune,pakzad2022circle,du2022fairdisco}, it is important to assess the transferability of \textsf{DL}-based skin lesion segmentation models to datasets with diverse skin tones.}

\end{itemize}

\section{Acknowledgements}

The authors would like to acknowledge Ben Cardoen and Aditi Jain for help with proofreading the manuscript and with creating the interactive table, respectively.
Z. Mirikharaji, K. Abhishek, and G. Hamarneh are partially funded by the BC Cancer Foundation - BrainCare BC Fund, the Natural Sciences and Engineering Research Council of Canada (NSERC RGPIN-06752), and the Canadian Institutes of Health Research (CIHR OQI-137993). A. Bissoto is partially funded by FAPESP 2019/19619-7. E. Valle is partially funded by CNPq 315168/2020-0. S. Avila is partially funded by CNPq PQ-2 315231/2020-3, and FAPESP 2013/08293-7. A. Bissoto and S. Avila are also partially funded by Google LARA 2020. The RECOD.ai lab is supported by projects from FAPESP, CNPq, and CAPES. C. Barata is funded by FCT project and multi-year funding [CEECIND/00326/2017] and LARSyS - FCT Plurianual funding 2020-2023. M. E. Celebi was supported by the US National Science Foundation under Award No. OIA-1946391. Any opinions, findings, and conclusions or recommendations expressed in this material are those of the authors and do not necessarily reflect the views of the National Science Foundation.

\newpage

\begin{center}
\singlespacing
\begin{tiny}
\begin{longtable}{|c|c|c|c|c|c|c|c|c|c|}

\caption{\textsf{DL} models for skin lesion segmentation. Performance measure reported is the Jaccard index computed on the dataset\typo{,} shown in boldface. The score is asterisked if it is computed based on the reported Dice index. The following abbreviations are used: Ref.: reference, Arch.: architecture, Seg.: segmentation, J: Jaccard index,  \textsf{CDE} : cross-data evaluation. the highlighted dataset and \textsf{PP}: postprocessing, con.: connection and conv.: convolution, \textsf{CE}: cross-entropy, \textsf{WCE}: weighted cross-entropy, \textsf{DS}: deep supervision, \textsf{EPE}: end point error, $\ell_1$: $\ell_1$ norm,  $\ell_2$: $\ell_2$ norm and \textsf{ADV}: adversarial loss. \revision{Please see the corresponding sections for more details: Section~\ref{subsec:architecture} for model architectures, Section~\ref{sebsec:loss} for loss functions, and Section~\ref{sec:evaluation} for model evaluation. 
An interactive version of this table is available online at \url{https://github.com/sfu-mial/skin-lesion-segmentation-survey}.}}
\label{tab:main}\\

\hline
 Ref. &Venue & Data & Arch. modules & Seg. loss & J & \textsf{CDE} & Augmentation & \textsf{PP} & code\\
\hline
\hline
\endfirsthead
\multicolumn{1}{c}%
{{\tablename \thetable{} -- continued from previous page}} \\
\hline
 Ref. &Venue & Data & Arch. modules & Seg. loss & J & \textsf{CDE} & Augmentation & \textsf{PP} & code\\
\hline
\hline 
\endhead

\hline
\multicolumn{1}{l}{{Continued on next page}} \\ 
\endfoot

\endlastfoot

\cite{85jafari2016skin} & \makecell{peer-reviewed\\conference} & DermQuest & image pyramid & - & - &\xmark & - &\cmark & \xmark\\[3em]

\cite{3he2017skin} & \makecell{peer-reviewed\\conference}  & \makecell{ISIC2016\\\textbf{ISIC2017}} & \makecell{residual con.\\ skip con.\\image pyramid} & \makecell{Dice\\ CE\\ DS} & 75.80\%  & \xmark & rotation & \cmark &\xmark\\[3em]

\cite{20bozorgtabar2017skin} & \makecell{peer-reviewed\\journal} & ISIC2016 & - & - & 80.60\% & \xmark & rotation &\xmark&\xmark\\[3em]

\cite{76ramachandram2017skin} & \makecell{peer-reviewed\\journal} & ISIC2017 & - & CE & 79.20\% & \xmark & \makecell{rotation, flipping\\color jittering} &\xmark &\xmark\\[3em]

\cite{yu2017} & \makecell{peer-reviewed\\journal} &\makecell{ISIC2016} & \makecell{skip con.\\residual con.} & - & 82.90\% & \xmark & \makecell{rotation,translation\\random noise\\ cropping} &\xmark &\cmark\\[3em]

\cite{84bi2017dermoscopic} & \makecell{peer-reviewed\\journal} & \makecell{\textbf{ISIC2016}\\PH$^2$} & - & CE & 84.64\% & \cmark & flipping,cropping &\cmark & \xmark\\[3em]

\cite{86jafari2017extraction} & \makecell{peer-reviewed\\journal} & DermQuest & image pyramid & - & - &\xmark & - &\cmark & \xmark\\[3em]

\cite{93yuan2017automatic} & \makecell{peer-reviewed\\journal} & \makecell{\textbf{ISIC2016}\\PH$^2$} & - & Tanimoto & 84.7\% &\cmark & \makecell{flipping, rotation\\scaling,shifting\\contrast norm.} & \cmark &\xmark\\[3em]

\cite{108ramachandram2017lesionseg}& \makecell{non peer-reviewed\\technical report} & \textbf{ISIC2017} & \makecell{dilated conv.} & CE & 64.20\% & \xmark & \makecell{rotation\\flipping} & \cmark & \xmark\\[3em]

\cite{109bozorgtabar2017investigating}& \makecell{peer-reviewed\\conference} & \textbf{ISIC2016} & - & CE & 82.90\% & \xmark & rotations & \cmark & \xmark\\[3em]

\cite{110bi2017semi}& \makecell{peer-reviewed\\conference} & \textbf{ISIC2016} & \makecell{parallel m. s.} & - & 86.36\% & \xmark & \makecell{crops,flipping} & \cmark & \xmark\\[3em]

\cite{111attia2017skin}& \makecell{peer-reviewed\\conference} & \textbf{ISIC2016} & \makecell{recurrent net.} & - & 93.00\% & \xmark & - & \xmark & \xmark\\[3em]

\cite{112deng2017segmentation}& \makecell{peer-reviewed\\conference}  & \textbf{ISIC2016} & \makecell{parallel m. s.} & - & 84.1\% & \xmark & - & \xmark & \xmark\\[3em]

\cite{113mishra2017deep}& \makecell{peer-reviewed\\conference} & \textbf{ISIC2017} & \makecell{skip con.} & Dice & 84.2\% & \xmark & \makecell{rotation\\flipping} & \cmark & \xmark\\[3em]

\cite{114goyal2017multi}& \makecell{peer-reviewed\\conference} & \textbf{ISIC2017} & - & \makecell{CE\\Dice} & - & \xmark & - & \xmark & \xmark\\[3em]

\cite{1vesal2018multi} &\makecell{peer-reviewed\\conference} & \makecell{\textbf{ISIC2017} \\ PH$^2$} & \makecell{dilated conv. \\ dense con. \\skip con.} & \makecell{Dice} & 88.00\% & \cmark & - & \xmark & \xmark\\[3em]

\cite{5venkatesh2018deep} & \makecell{peer-reviewed\\conference} & \makecell{ISIC2017} & \makecell{residual con.\\ skip con.} & Jaccard & 76.40\% & \xmark & \makecell{rotation,flipping\\translation, scaling} &\cmark &\xmark\\[3em]

\cite{yang2018skin} & \makecell{\revision{peer-reviewed}\\\revision{conference}} & \makecell{ISIC2017}  & \makecell{skip con.\\parallel m.s. conv.} & - & 74.10\% & \xmark & rotation,flipping &\xmark &\xmark\\[3em]

\cite{16sarker2018slsdeep} & \makecell{peer-reviewed\\conference} & \makecell{ISIC2016 \\ \textbf{ISIC2017}} & \makecell{skip con.\\residual con.\\dilated conv.\\pyramid pooling} & \makecell{CE\\ EPE} & 78.20\% & \xmark & rotation,scaling &\xmark &\cmark\\[3em]

\cite{17al2018skin} & \makecell{peer-reviewed\\journal} & \makecell{\textbf{ISIC2017}\\ PH$^2$} & - & CE & 77.10\% & \cmark & rotation &\xmark&\xmark\\[3em]

\cite{18li2018deeply} & \makecell{peer-reviewed\\conference} & ISIC2017 & \makecell{skip con.\\residual con.} & DS & 77.23\% & \xmark & flipping, rotation &\xmark&\cmark\\[3em]

\cite{22zeng2018multi} & \makecell{peer-reviewed\\conference} & \makecell{ISIC2017}  & \makecell{dense con.\\skip con.\\image pyramid} & \makecell{CE\\ $\ell_2$\\DS} & 78.50\% & \xmark & flipping, rotation &\cmark &\xmark\\[3em]

\cite{25devries2018leveraging} & \makecell{non peer-reviewed\\technical report} & ISIC2017 & skip con. & CE & 73.00\% & \xmark & flipping, rotation &\xmark&\xmark\\[3em]

\cite{izadi2018generative} & \makecell{peer-reviewed\\conference} & DermoFit & skip con. & \makecell{CE\\ADV} & 81.20\% & \xmark & \makecell{flipping, rotation\\elastic deformation} &\xmark &\cmark\\[3em]

\cite{32li2018dense} & \makecell{peer-reviewed\\journal} & \makecell{ISIC2016 \\ \textbf{ISIC2017}} & \makecell{skip con.\\residual con.\\dense con.} & \makecell{Jaccard\\DS} & 76.50\% & \xmark & - & \xmark & \xmark\\[3em]

\cite{mirikharaji2018star} & \makecell{peer-reviewed\\conference} & \makecell{ISIC2017} & residual con. & \makecell{CE\\Star shape} & 77.30\% & \xmark & - & \xmark & \xmark\\[3em]

\cite{35pollastri2018improving} & \makecell{peer-reviewed\\conference} & ISIC2017 & - & \makecell{Jaccard\\$\ell_1$} & 78.10\% & \xmark & GAN &\cmark &\xmark\\[3em]

\cite{49vesal2018skinnet} & abstract & ISIC2017 & \makecell{dilated conv. \\ dense con. \\skip con.} & Dice & 76.67\% & \xmark & \makecell{rotation, flipping,\\ translation, scaling,\\ color shift} & \xmark &\xmark\\[3em]

\cite{53chen2018multi} & \makecell{peer-reviewed\\conference}& \makecell{ISIC2017} & \makecell{residual con.\\dilated conv.\\parallel m.s. conv.} & WCE & 78.70\% & \xmark &  \makecell{rotation, flipping\\cropping, zooming\\Gaussian noise} &\cmark &\xmark\\[3em]

\cite{56jahanifar2018segmentation} &\makecell{non peer-reviewed\\technical report} & \makecell{ISIC2016 \\ \textbf{ISIC2017} \\ ISIC2018 } & \makecell{skip con.\\pyramid pooling\\parallel m.s. conv.} & Tanimoto & 80.60\% & \cmark & \makecell{flipping, rotation\\zooming,translation\\shearing,color shift\\intensity scaling\\adding noises\\contrast adjust.\\sharpness adjust.\\disturb illumination\\hair occlusion} &\cmark &\xmark\\[3em]

\cite{mirikharaji2018deep} & \makecell{peer-reviewed\\conference} & ISIC2016 & skip con. & CE & 83.30\% & \xmark & flipping,rottaion &\xmark &\xmark\\[3em]

\cite{79bi2018improving} & \makecell{non peer-reviewed\\technical report} & \makecell{ISIC2018} & residual con. & CE & 83.12\% &\xmark & GAN &\xmark &\xmark\\[3em]

\cite{82he2018dense} & \makecell{peer-reviewed\\journal} & \makecell{ISIC2016 \\ \textbf{ISIC2017}} & \makecell{skip con.\\residual con.\\image pyramid} & \makecell{CE\\Dice\\DS} & 76.10\% & \xmark & rotation &\cmark &\xmark\\[3em]

\cite{87xue2018adversarial}& \makecell{peer-reviewed\\conference} & \makecell{ISIC2017} & \makecell{skip con.\\residual con.\\global conv.\\GAN} & \makecell{$\ell_1$\\DS\\ADV} & 78.50\% &\xmark & \makecell{cropping\\color jittering} &\xmark &\xmark\\[3em]

\cite{159ebenezer2018automatic} & \makecell{non peer-reviewed\\technical report}  & \makecell{\textbf{ISIC 2018}} & \makecell{skip con.} & Dice & 75.6\% & \xmark & \makecell{rotation\\flipping\\zooming} & \cmark & \cmark\\[3em]

\cite{2goyal2019automatic} & \makecell{peer-reviewed\\journal}& \makecell{\textbf{ISIC2017} \\ PH$^2$} & \makecell{dilated conv.\\parallel m.s. conv.\\ separable conv.} & - & 79.34\% & \cmark & - &\cmark& \xmark \\[3em]

\cite{7azad2019bi} & \makecell{peer-reviewed\\conference} & \makecell{ISIC2018}&\makecell{skip con.\\dense con.\\recurrent \typo{CNN}} & CE & 74.00\% & \xmark & - &\xmark &\cmark\\[3em]

\cite{8alom2019recurrent} &\makecell{peer-reviewed\\journal} & \makecell{ISIC2017}& \makecell{skip con.\\ residual con.\\recurrent \typo{CNN}} & CE & 75.68\% & \xmark & - & \xmark &\xmark\\[3em]

\cite{9yuan2019improving} & \makecell{peer-reviewed\\journal} & ISIC2017 & - & Tanimoto & 76.50\% & \xmark & \makecell{rotation,flipping\\ shifting, scaling\\random normaliz.} &\cmark &\xmark\\[3em]
\cite{14goyal2019skin} & \makecell{peer-reviewed\\conference} & \makecell{\textbf{ISIC2017}\\ PH$^2$}& \makecell{dilated conv.\\ parallel m.s. conv.} & WCE & 82.20\% & \cmark & - &\xmark &\xmark\\[3em]

\cite{19bi2019step} & \makecell{peer-reviewed\\journal} & \makecell{ISIC2016\\\textbf{ISIC2017}\\PH$^2$} & \makecell{skip con.\\residual con.\\}& CE & 77.73\% & \cmark& flipping, cropping &\cmark&\xmark\\[3em]

\cite{21tschandl2019domain} & \makecell{peer-reviewed\\journal} & ISIC2017 & skip con. & \makecell{CE\\Jaccard} & 76.80\% & \xmark & flipping, rotation &\cmark&\xmark\\[3em]

\cite{24li2019transformation} & \makecell{peer-reviewed\\journal} & \makecell{ISIC2017}  & \makecell{skip con.\\dense con.\\\revision{semi-supervised}\\\revision{ensemble}} & \makecell{CE\\ $\ell_1$} & 79.80\% & \xmark & \makecell{flipping,rotating \\scaling} &\cmark &\xmark\\[3em]

\cite{26zhang2019automatic} & \makecell{peer-reviewed\\journal} & \makecell{ISIC2016\\ \textbf{ISIC2017}} & skip con. & CE & 72.94\% & \xmark & - &\xmark &\xmark\\[3em]

\cite{27baghersalimi2019dermonet} & \makecell{peer-reviewed\\journal} & \makecell{ISIC2016 \\ \textbf{ISIC2017} \\ PH$^2$}  & \makecell{skip con.\\residual con.\\dense con.} & Tanimoto & 78.30\% & \cmark & flipping,cropping & \xmark&\xmark\\[3em]

\cite{29jiang2019decision} & \makecell{peer-reviewed\\conference} & \makecell{ISIC2017} & \makecell{residual con.\\dilated conv.\\GAN} & \makecell{ADV\\$\ell_2$} & 76.90\% & \xmark & rotation,flipping &\xmark &\xmark\\[3em]

\cite{30tang2019multi} & \makecell{peer-reviewed\\conference} & \makecell{ISIC2016} & skip con. & \makecell{Tanimoto\\DS} & 85.34\% & \xmark & rotation,flipping & \xmark&\xmark\\[3em]

\cite{31bi2019improving} & \makecell{peer-reviewed\\conference} & ISIC2017 & residual con. & CE & 77.14\% & \xmark & GAN & \xmark & \xmark\\[3em]

\cite{33abraham2019novel} & \makecell{peer-reviewed\\conference} & \makecell{ISIC2018} & \makecell{skip con.\\image pyramid\\attention} & \makecell{TV\\Focal} & 74.80\% & \xmark & - &\xmark& \cmark\\[3em]

\cite{38cui2019ensemble} & \makecell{peer-reviewed\\conference} & ISIC2018 & \makecell{dilated conv.\\ parallel m.s. conv.\\ separable conv.} & - & 83.00\% & \xmark & - &\xmark &\xmark\\[3em]

\cite{41song2019dense} & \makecell{peer-reviewed\\conference} & \makecell{ISIC2017} & \makecell{skip con.\\residual con.\\dense con.\\attention mod.} & \makecell{CE\\Jaccard} & 76.50\% & \xmark & - &\xmark &\xmark\\[3em]

\cite{42singh2019fca} & \makecell{peer-reviewed\\journal} & \makecell{ISIC2016 \\ \textbf{ISIC2017} \\ISIC2018} & \makecell{skip con.\\residual con.\\factorized conv.\\attention mod.\\GAN} & \makecell{CE\\$\ell_1$\\EPE} & 78.65\% & \xmark & - & \xmark&\cmark\\[3em]

\cite{44tan2019evolving} & \makecell{peer-reviewed\\journal} & \makecell{\textbf{ISIC2017}\\DermoFit\\PH$^2$} & \makecell{dilated conv.} & Dice & 62.29\%$^*$ & \cmark & - &\cmark &\xmark\\[3em]

\cite{46kaul2019focusnet} & \makecell{peer-reviewed\\conference} & \makecell{ISIC2017} & \makecell{skip con.\\residual con.\\attention mod.} & Dice & 75.60\% & \xmark & channel shift &\xmark &\xmark\\[3em]

\cite{47de2019skin} & \makecell{peer-reviewed\\conference} & \makecell{\textbf{ISIC2017}\\Private} & skip con. & \makecell{CE\\Dice} & 76.07\% & \xmark & \makecell{flipping, shifting\\ rotation\\color jittering} &\cmark &\xmark\\[3em]

\cite{48zhang2019dsm} & \makecell{peer-reviewed\\journal} & \makecell{\textbf{ISIC2017} \\ PH$^2$} & \makecell{skip con.\\residual con.\\parallel m.s. conv.} & \makecell{CE\\Dice\\DS} & 78.50\% & \cmark & \makecell{flipping, rotation\\ whitening\\contrast enhance.} & \cmark&\xmark\\[3em]

\cite{51soudani2019image} & \makecell{peer-reviewed\\journal} & \makecell{ISIC2017} & residual con. & CE & 78.60\% & \xmark & rotation, flipping &\xmark &\xmark\\[3em]

\cite{mirikharaji2019learning} & \makecell{peer-reviewed\\conference} & \makecell{ISIC2017} & \makecell{skip con.} & WCE & 68.91\%$^*$ & \xmark & - &\xmark & \xmark\\[3em]

\cite{55nasr2019dense} &\makecell{peer-reviewed\\journal} & \makecell{DermQuest} & \makecell{dense con.\\} & WCE & 85.20\% & \xmark &  \makecell{rotation,flipping\\cropping}&\xmark &\xmark\\[3em]

\cite{59wang2019automated} & \makecell{peer-reviewed\\conference} & \makecell{\textbf{ISIC2017} \\ ISIC2018} & \makecell{skip con.\\residual con.\\parallel m.s. conv.\\attention mod.} & WDice & 77.60\% & \xmark & copping, flipping &\xmark & \xmark\\[3em]

\cite{60sarker2019mobilegan} & \makecell{non peer-reviewed\\technical report} & \makecell{\textbf{ISIC2017} \\ ISIC2018} & \makecell{factrized conv.\\attention mod.\\GAN} & \makecell{CE\\Jaccard\\$\ell_1$,ADV} & 77.98\% & \xmark & \makecell{flipping\\gamma reconst.\\contrast adjust.}&\xmark &\xmark\\[3em]

\cite{61tu2019dense} & \makecell{peer-reviewed\\journal} & \makecell{\textbf{ISIC2017} \\ PH$^2$} & \makecell{skip con.\\residual con.\\dense con.\\GAN} & \makecell{Jaccard\\EPE, $\ell_1$\\DS, ADV} & 76.80\% & \cmark & flipping &\xmark &\xmark\\[3em]

\cite{65wei2019attention} & \makecell{peer-reviewed\\journal} & \makecell{ISIC2016 \\ \textbf{ISIC2017} \\ PH$^2$} & \makecell{skip con.\\ residual con.\\attention mod.\\GAN} & \makecell{Jaccard\\$\ell_1$\\ADV} & 80.45\% & \cmark & \makecell{rotation,flipping\\color jittering} &\xmark & \xmark\\[3em]

\cite{66unver2019skin} & \makecell{peer-reviewed\\journal} & \makecell{\textbf{ISIC2017}\\PH$^2$} & - & $\ell_2$ & 74.81\% & \cmark & - &\cmark & \xmark\\[3em]

\cite{69al2019deep} & \makecell{peer-reviewed\\conference} & ISIC2017 & - & - & 77.11\% & \xmark & rotation,flipping &\xmark & \xmark\\[3em]

\cite{71canalini2019skin} & \makecell{peer-reviewed\\conference} & ISIC2017 & \makecell{dilated conv.\\parallel m.s. conv.\\ separable conv.} & \makecell{CE\\Tanimoto} & 85.00\% &\xmark &  \makecell{rotating, flipping\\shifting, shearing\\scaling\\color jittering} &\cmark &\xmark\\[3em]

\cite{72wang2019dermoscopic} & \makecell{peer-reviewed\\conference} & ISIC2017 & \makecell{residual con.} & WCE & 78.10\% &\xmark & flipping, scaling &\xmark &\xmark\\[3em]

\cite{73alom2019skin} & \makecell{peer-reviewed\\conference} & ISIC2018 & \makecell{skip con.\\residual con.\\recurrent \typo{CNN}} & CE & 88.83\% &\xmark & flipping &\xmark &\xmark\\[3em]

\cite{74pollastri2019augmenting} & \makecell{peer-reviewed\\journal} & ISIC2017 & - & Tanimoto & 78.90\% &\xmark & \makecell{GAN\\flipping,rotation\\shifting, scaling\\color jittering} &\xmark &\xmark\\[3em]

\cite{75liu2019skin} & \makecell{peer-reviewed\\conference} & ISIC2017 & \makecell{skip con.\\dilated conv.} & CE & 75.20\% & \xmark & \makecell{scaling, cropping\\ rotation, flipping\\image deformation}  &\xmark &\xmark\\[3em]

\cite{80abhishek2019mask2lesion} & \makecell{peer-reviewed\\conference} & \makecell{\textbf{ISIC2017}\\PH$^2$} & skip con. & - & 68.69\%$^*$ & \cmark & \makecell{rotation,flipping\\GAN} &\xmark &\cmark\\[3em]

\cite{127shahin2019deep}& \makecell{peer-reviewed\\conference} & \makecell{\textbf{ISIC2018}} & \makecell{skip con.\\image pyramid}  & \makecell{Generalized\\Dice} & 73.8\% & \xmark & \makecell{rotation\\flipping\\zooming} & \xmark & \xmark\\[3em]

\cite{128adegun2019enhanced}& \makecell{peer-reviewed\\conference} & \textbf{ISIC2017} & - & Dice & 83.0\% & \xmark & elastic & \xmark & \xmark\\[3em]

\cite{135taghanaki2019improved}& \makecell{peer-reviewed\\conference} & \makecell{ISIC 2017} & \makecell{skip con.} & \makecell{Dice\\$\ell_1$\\SSIM} & 69.35\%$^*$ & \xmark & \makecell{rotation\\flipping\\gradient-based\\perturbation} & \xmark & \xmark\\[3em]

\cite{138saini2019detector}&  \makecell{peer-reviewed\\conference} & \makecell{\textbf{ISIC 2017}\\ISIC 2018\\PH2} & \makecell{skip con.\\multi-task} & Dice & 84.9\% & \xmark & \makecell{rotation, flipping\\shearing, stretch\\crop, contrast} & \xmark & \xmark\\[3em]

\cite{144wang2019bi} & \makecell{peer-reviewed\\journal} & \makecell{ISIC2016\\\textbf{ISIC2017}} & \makecell{skip con.\\residual con.\\dilated conv.} & WCE & 81.47\% &\xmark &\makecell{flipping,scaling} & \xmark & \xmark\\[3em]

\cite{kamalakannan2019self}& \revision{\makecell{peer-reviewed\\journal}} & \revision{ISIC Archive} & \revision{\makecell{skip con.}} & \revision{CE} & \revision{-} & \revision{\xmark} & \revision{-} & \revision{\xmark} & \revision{\xmark}\\[3em]

\cite{28hasan2020dsnet} & \makecell{peer-reviewed\\journal} & \makecell{\textbf{ISIC2017} \\ PH$^2$} & \makecell{skip con.\\dense con.\\separable conv.} & \makecell{CE\\Jaccard} & 77.50\% & \cmark & \makecell{rotation, zooming\\shifting, flipping} & \xmark & \cmark\\[3em]

\cite{40al2020automatic} & \makecell{peer-reviewed\\conference} & \makecell{\textbf{ISIC2018}\\PH$^2$} & \makecell{skip con.} & Dice & 80.00\% & \cmark & \makecell{rotation, zooming\\ flipping,elastic dist.\\ Gaussian dist.\\histogram equal.\\ color jittering} &\xmark &\cmark\\[3em]

\cite{deng2020weakly} & \makecell{peer-reviewed\\conference} & \makecell{\textbf{ISIC2017}\\PH$^2$} & \makecell{dilated conv.\\parallel m.s. conv.\\ separable conv.\\\revision{semi-supervised}} & \makecell{Dice\\Narrowband\\suppression} & 83.9\% & \cmark & rotation &\cmark &\xmark\\[3em]

\cite{81xie2020mutual} & \makecell{peer-reviewed\\journal}& \makecell{\textbf{ISIC2017}\\PH$^2$} & \makecell{dilated conv.\\parallel m.s. conv.\\ separable conv.} & \makecell{Dice\\Rank} & 80.4\% & \cmark & \makecell{cropping,scaling\\rotation, shearing\\shifting,zooming\\whitening, flipping} &\xmark &\cmark\\[3em]

\cite{94zhang2020kappa}& \makecell{peer-reviewed\\conference} & \makecell{SCD \\ ISIC2016 \\ \textbf{ISIC2017} \\ ISIC2018} & \makecell{skip con.} & Kappa Loss & 84.00\%$^*$ & \xmark & \makecell{rotation,shifting\\shearing,zooming\\flipping} & \xmark & \cmark\\[3em]

\cite{95saha2020leveraging}& \makecell{peer-reviewed\\conference} & \makecell{ISIC2017\\\textbf{ISIC2018}} & \makecell{skip con.\\dense con.} & CE & 81.9\% & \xmark & \makecell{color jittering\\rotation\\flipping\\translation} & \xmark & \xmark\\[3em]

\cite{96henry2020mixmodule}& \makecell{peer-reviewed\\conference} & \makecell{\textbf{ISIC2018}} & \makecell{skip con.\\parallel m. s. conv.\\attention mod.} & - & 78.04\% & \xmark & \makecell{color jittering\\rotation,cropping\\flipping,shift} & \xmark & \cmark\\[3em]

\cite{97jafari2020dru}& \makecell{peer-reviewed\\conference} & \makecell{\textbf{ISIC2018}} & \makecell{skip con.\\residual con.\\dense con.} & CE & 75.5\% & \xmark & - & \xmark & \cmark\\[3em]

\cite{98li2020generic}& \makecell{peer-reviewed\\conference} & \makecell{\textbf{ISIC2018}} & \makecell{skip con.\\residual con.\\ensemble\\semi-supervised} & \makecell{CE\\Dice} & 75.5\% & \xmark & - & \xmark & \xmark \\[3em]

\cite{99guo2020complementary}& \makecell{peer-reviewed\\conference} & \makecell{\textbf{ISIC2018}} & \makecell{skip con.\\dilated conv.\\parallel m. s. conv.} & \makecell{Focal\\Jaccard}& 77.60\% & \xmark & - & \xmark & \cmark\\[3em]

\cite{100li2020multi}& \makecell{peer-reviewed\\conference} & \makecell{\textbf{ISIC2018}} & \makecell{skip con.\\residual con.\\self-supervised} & \makecell{MSE\\KLD} & 87.74\%$^*$ & \xmark & - & \xmark & \xmark\\[3em]

\cite{116jiang2020skin}& \makecell{peer-reviewed\\journal} & \makecell{\textbf{ISIC2017}\\PH$^2$} & \makecell{skip con.\\residual con.\\attention mod.} & CE &  73.35\% & \xmark & flipping & \xmark & \xmark\\[3em]

\cite{118qiu2020inferring}& \makecell{peer-reviewed\\journal} & \makecell{\textbf{ISIC2017}\\PH$^2$} & \makecell{ensemble} & - & 80.02\% & \xmark & \makecell{translation\\rotation\\shearing} & \cmark & \xmark\\[3em]

\cite{123xie2020skin}& \makecell{peer-reviewed\\journal} & \makecell{ISIC2016\\\textbf{ISIC2017}\\PH$^2$} & \makecell{attention mod.} & CE & 78.3\% & \xmark & \makecell{rotation\\flipping} & \xmark & \xmark\\[3em]

\cite{124zafar2020skin}& \makecell{peer-reviewed\\journal} & \makecell{\textbf{ISIC2017}\\PH$^2$} & \makecell{skip con.\\residual con.} & CE & 77.2\% & \xmark & rotation & \xmark & \xmark\\[3em]

\cite{131azad2020attention}&  \makecell{peer-reviewed\\conference} & \makecell{\textbf{ISIC 2017}\\ISIC 2018\\PH2} & \makecell{dilated conv.\\attention mod.} & - & 96.98\% & \xmark & - & \xmark & \cmark\\[3em]

\cite{132nathan2020lesion}& \makecell{non peer-reviewed\\technical report} & \makecell{ISIC 2016\\\textbf{ISIC 2017}\\ISIC 2018\\PH2} & \makecell{skip con.\\residual con.} & \makecell{CE\\Dice} & 78.28\% & \xmark & \makecell{rotation, flipping\\shearing, zoom} & \xmark & \xmark \\[3em]

\cite{133mirikharaji2020d}& \makecell{peer-reviewed\\conference} & \makecell{\textbf{ISIC Archive}\\PH2\\DermoFit} & \makecell{skip con.\\residual con.\\ensemble} & \makecell{CE} & 72.11\% & \xmark & - & \xmark & \xmark \\[3em]

\cite{136ozturk2020skin}& \makecell{peer-reviewed\\journal} & \makecell{\textbf{ISIC 2017}\\PH2} & \makecell{residual con.} & \makecell{-} & 78.34\% & \cmark & \makecell{-} & \xmark & \xmark\\[3em]

\cite{137abhishek2020illumination}&  \makecell{peer-reviewed\\conference} & \makecell{\textbf{ISIC 2017}\\DermoFit\\PH2} & \makecell{skip con.} & Dice & 75.70\% & \cmark & \makecell{rotation\\flipping} & \xmark & \cmark\\[3em]

\cite{139kaymak2020skin}&  \makecell{peer-reviewed\\journal} & \makecell{ISIC 2017} & \makecell{-} & - & 72.5\% & \xmark & \makecell{-} & \xmark & \xmark\\[3em]

\cite{140bagheri2020two} &\revision{\makecell{peer-reviewed\\journal}} & \revision{\makecell{\textbf{ISIC2017}\\DermQuest}} & \revision{\makecell{dilated conv.\\parallel m.s. conv.\\ separable conv.}} & \revision{-} &\revision{79.05\%} & \revision{\cmark} & \revision{\makecell{rotation,flipping\\ brightness change\\ resizing}}&\revision{\xmark} &\revision{\xmark} \\[3em]

\cite{141jayapriya2020hybrid} & \makecell{peer-reviewed\\journal} & ISIC2016& \makecell{skip con.\\parallel m.s. conv.} & - & 92.42\% & \xmark & - & \xmark & \xmark \\[3em]

\cite{143wang2020cascaded} & \makecell{non peer-reviewed\\technical report} & \makecell{ISIC2016\\\textbf{ISIC2017}\\PH$^2$} & \makecell{residual con.\\dilated conv.\\ attention mod.} & \makecell{CE\\\typo{Dice}\\DS} & 80.30\% & \cmark & \makecell{flipping, rotation\\cropping} & \xmark & \xmark\\[3em]

\cite{151wang2020donet}& \makecell{non peer-reviewed\\technical report} & \makecell{\textbf{ISIC2018}\\PH$^2$} & \makecell{attention mod.\\skip con.\\parallel m.s. conv.\\recurrent CNN\\} & \makecell{Dice\\Focal \typo{T}versky} & 80.6\% & \xmark & \makecell{rotation\\flipping\\cropping} & \xmark & \xmark\\[3em]

\cite{153ribeiro2020less}& \makecell{peer-reviewed\\conference} & \makecell{ISIC Archive\\PH$^2$\\DermoFit} & \makecell{skip con.\\residual con.\\dilated conv.}  & \makecell{Soft Jaccard\\CE} & - & \cmark & \makecell{Gaussian noise\\color jittering} & \cmark & \cmark\\[3em]

\cite{zhu2020asnet}& \makecell{peer-reviewed\\conference} & ISIC2018 & \makecell{skip con.\\ residual con.\\dilated conv.\\ attention mod.} &\makecell{CE \\ \typo{Dice}}& 82.15\% & \xmark & flipping & \xmark & \xmark \\[3em]

\cite{gu2020net}& \makecell{peer-reviewed\\journal} & \makecell{ISIC 2018} & \makecell{residual con.\\skip con.\\attention mod.} & \makecell{Dice} & 85.32\%$^*$ & \xmark &\makecell{cropping, flipping\\rotation} & \xmark & \cmark\\[3em]

\cite{lei2020skin}& \revision{\makecell{peer-reviewed\\journal}} & \revision{\makecell{\textbf{ISIC 2017}\\ISIC 2018}} & \revision{\makecell{skip con.\\dense con.\\dilated conv.\\GAN}} & \revision{\makecell{CE\\$\ell_1$\\ADV}} & \revision{77.1\%} & \revision{\cmark} &\revision{\makecell{flipping, rotation}} & \revision{\xmark} & \revision{\xmark} \\[3em]

\cite{andrade2020data}& \revision{\makecell{peer-reviewed\\journal}} & \revision{\makecell{\textbf{DermoFit}\\SMARTSKINS}} & \revision{\makecell{residual con.\\dilated conv.\\GAN}} & \revision{\makecell{Dice}} & \revision{81.03\%} & \revision{\xmark} &\revision{\makecell{flipping, brightness\\saturation, contrast, hue\\Gaussian hue}} & \revision{\xmark} & \revision{\xmark} \\[3em]

\cite{wu2020automated}& \revision{\makecell{peer-reviewed\\journal}} & \revision{\makecell{\textbf{ISIC 2017}\\ISIC 2018}} & \revision{\makecell{residual con.\\attention mod.\\multi-scale}} & \revision{\makecell{CE\\Dice}} & \revision{82.55\%} & \revision{\xmark} &\revision{\makecell{flipping, rotation\\scaling, cropping\\sharpening, color\\distribution adj., noise}} & \revision{\xmark} & \revision{\xmark} \\[3em]

\cite{134arora2021automated}& \makecell{peer-reviewed\\journal} & \makecell{ISIC 2018} & \makecell{skip con.\\attention mod.} & \makecell{Dice\\Tversky\\Focal Tversky} & 83\% & \xmark & flipping & \cmark & \xmark\\[3em]

\cite{145jin2021cascade} & \makecell{peer-reviewed\\journal} & \makecell{\textbf{ISIC2017}\\ ISIC2018} & \makecell{skip con.\\residual con.\\attention mod.}& \makecell{\typo{Dice}\\ Focal} & 80.00\% & \xmark & \makecell{flipping, rotation\\ affine trans.\\ scaling, cropping} & \xmark & \cmark\\[3em]

\cite{146hasan2021dermodoctor} & \makecell{\revision{peer-reviewed}\\\revision{journal}} & \makecell{ISIC 2016\\ \textbf{ISIC 2017}} & \makecell{skip con.\\residual con.\\ separable conv.} & \makecell{\typo{Dice}\\ CE} & 66.66\%$^*$& \xmark & \makecell{flipping, rotation\\shifting, zooming\\intensity adjust.} & \xmark & \xmark\\[3em]

\cite{147kosgiker2021segcaps} & \makecell{peer-reviewed\\journal} & \makecell{\textbf{ISIC 2017}\\ PH$^2$} & - & \makecell{MSE\\CE} & 90.25\% & \xmark & - & \xmark & \xmark \\[3em]

\cite{162bagheriskin}& \makecell{peer-reviewed\\journal} & \makecell{ISIC2016\\ISIC2017\\\textbf{ISIC2018}\\PH$^2$\\DermQuest} &\makecell{parallel m.s. conv.\\dilated conv.}  & \makecell{Dice\\CE} & 85.04\% & \cmark & \makecell{rotation\\flipping\\color jittering} & \xmark & \xmark\\[3em]

\cite{170saini2021bsegnet} & \makecell{peer-reviewed\\conference} & \makecell{\textbf{ISIC2017}\\ISIC2018\\PH$^2$} & \makecell{pyramid pooling\\residual con.\\skip con.\\dilated conv.\\attention mod.} & Dice & 85.00\% & \cmark & \makecell{rotation,shearing\\color jittering} & \xmark & \xmark\\[3em]

\cite{171tong2021ascu} & \makecell{peer-reviewed\\journal} & \makecell{ISIC2016\\ISIC2017\\\textbf{PH$^2$}} & \makecell{skip con.\\attention mod.} & CE & 84.2\% & \cmark & \makecell{flipping} & \xmark & \xmark\\[3em]

\cite{173bagheri2021skin} & \makecell{peer-reviewed\\journal} & \makecell{\textbf{Derm\typo{Q}uest}\\ISIC2017\\PH$^2$} & \makecell{ensemble} & \makecell{CE\\Focal} & 86.53\% & \cmark & \makecell{rotation\\flipping\\color jittering} & \cmark & \xmark\\[3em]

\cite{163ren2021serial} & \makecell{peer-reviewed\\journal} & ISIC2017 & \makecell{dense con.\\dilated conv.\\separable conv.\\attention mod.} & \makecell{\typo{Dice}\\CE} & 76.92\% & \xmark & flipping, rotation & \xmark & \xmark\\[3em]

\cite{liu2021skin}& \makecell{peer-reviewed\\journal} & ISIC2017 & \makecell{residual con.\\dilated conv.\\pyramid pooling} & WCE & 79.46\% & \xmark & \makecell{flipping, cropping\\rotation\\image deformation }& \xmark & \xmark\\[3em]

\cite{khan2021pmed}& \makecell{peer-reviewed\\journal} & ISIC2018 & \makecell{skip con.\\image pyramid} & Dice & 85.10\% & \xmark & - & \xmark &\cmark\\[3em]

\cite{redekop2021uncertainty}& \makecell{peer-reviewed\\conference} & ISIC2017 & - & - & 68.77$\%^*$ & \xmark & - & \xmark& \xmark \\[3em]

\cite{kaul2021focusnet++}& \makecell{peer-reviewed\\conference} & ISIC2018 & \makecell{skip con.\\residual con.\\attention mod.} &\makecell{CE\\Tversky\\adaptive\\logarithmic}& 82.71\% &\xmark & - &\xmark & \cmark\\[3em]

\cite{abhishek2021matthews}& \makecell{peer-reviewed\\conference} & \makecell{\textbf{ISIC2017}\\PH$^2$\\DermoFit} & skip con. &  MCC &75.18\%  & \xmark & flipping, rotation & \xmark & \cmark\\[3em]

\cite{tang2021introducing}& \makecell{peer-reviewed\\journal} & ISIC2018 & skip con. & CE & 78.25\% & \xmark & \makecell{-} &\xmark & \xmark\\[3em]

\cite{xie2021semi}& \makecell{peer-reviewed\\conference} & ISIC2018 & \makecell{dilated conv.\\\revision{semi-supervised}}  & \makecell{CE\\ KL div.} &  82.37\% & \xmark & \makecell{scaling,rotation\\ elastic transformation}& \xmark & \xmark\\[3em]

\cite{poudel2021deep}& \makecell{peer-reviewed\\journal} & \makecell{ISIC2017} & \makecell{skip con.\\ attention mod.} & CE & 87.44\% & \xmark &\makecell{scaling, flipping\\rotation\\Gaussian noise\\median blur} & \xmark & \xmark\\[3em]

\cite{csahin2021robust}& \makecell{peer-reviewed\\journal} & \makecell{ISIC2016\\\textbf{ISIC 2017}} & \makecell{skip con.\\Gaussian process} & - & 74.51\% & \xmark &\makecell{resize\\rotation\\reflection} & \cmark & \xmark\\[3em]

\cite{sarker2021slsnet}& \makecell{peer-reviewed\\journal} & \makecell{\textbf{ISIC 2017}\\ISIC 2018} & \makecell{parallel m.s. conv.\\attention mod.\\GAN} & \makecell{$\ell_1$\\Jaccard} & 81.98\% & \xmark &\makecell{flipping, contrast\\gamma reconstruction} & \xmark & \xmark\\[3em]

\cite{wang2021knowledge}& \makecell{peer-reviewed\\journal} & \makecell{ISIC 2016\\\textbf{ISIC 2017}} & \makecell{residual con.\\skip con.\\lesion-based pooling\\feature fusion} & \makecell{CE} & 82.4\% & \xmark &\makecell{flipping, scaling\\cropping} & \xmark & \xmark\\[3em]

\cite{sachin2021performance}& \makecell{book\\chapter} & \makecell{ISIC 2018} & \makecell{residual con.\\skip con.} & \makecell{-} & 75.96\% & \xmark &\makecell{flipping, scaling\\color jittering} & \xmark & \xmark\\[3em]

\cite{wibowo2021lightweight}& \makecell{peer-reviewed\\journal} & \makecell{\textbf{ISIC 2017}\\ISIC 2018\\PH2} & \makecell{BConvLSTM\\separable conv.\\residual con.\\skip con.} & \makecell{Jaccard} & 80.25\% & \xmark &\makecell{distortion, blur\\color jittering\\contrast\\gamma sharpen} & \cmark & \cmark\\[3em]

\cite{gudhe2021multi}& \makecell{peer-reviewed\\journal} & \makecell{ISIC 2018} & \makecell{dilated conv.\\residual con.\\skip con.} & \makecell{CE} & 91\% & \xmark &\makecell{flipping, scaling\\shearing, color jittering\\Gaussian blur\\Gaussian noise} & \xmark & \cmark\\[3em]

\cite{khouloud2021w}& \makecell{peer-reviewed\\journal} & \makecell{ISIC 2016\\\textbf{ISIC 2017}\\ISIC 2018\\PH2} & \makecell{feature pyramid\\residual con.\\skip con.\\attention mod.} & \makecell{-} & 86.92\%$^*$ & \xmark &\makecell{-} & \xmark & \xmark\\[3em]

\cite{gu2021kcbac}& \makecell{peer-reviewed\\conference} & \makecell{ISIC 2017} & \makecell{asymmetric conv.\\skip con.} & \makecell{DS} & 79.4\% & \xmark &\makecell{cropping, flipping\\rotation} & \xmark & \xmark\\[3em]

\cite{zhao2021segmentation}& \makecell{peer-reviewed\\journal} & \makecell{ISIC 2018} & \makecell{pyramid pooling\\attention mod.\\residual con.\\skip con.} & \makecell{CE\\Dice} & 86.84\% & \xmark &\makecell{cropping} & \xmark & \xmark\\[3em]

\cite{tang2021afln}& \makecell{peer-reviewed\\journal} & \makecell{ISIC 2016\\\textbf{ISIC 2017}\\ISIC 2018} & \makecell{attention mod.\\residual con.\\skip con.\\ensemble\\pyramid pooling} & \makecell{Focal} & 80.7\% & \xmark &\makecell{copying} & \xmark & \xmark\\[3em]

\cite{zunair2021sharp}& \makecell{peer-reviewed\\journal} & \makecell{ISIC 2018} & \makecell{sharpening kernel\\residual con.} & \makecell{CE} & 79.78\% & \xmark &\makecell{-} & \xmark & \cmark\\[3em]

\cite{li2021superpixel}& \revision{\makecell{peer-reviewed\\conference}} & \revision{\makecell{ISIC 2017}} & \revision{\makecell{skip con.}} & \revision{\makecell{CE\\KL div.}} & \revision{71.12\%*} & \revision{\xmark} &\revision{\makecell{-}} & \revision{\xmark} & \revision{\cmark} \\[3em]

\cite{zhang2021self}& \revision{\makecell{peer-reviewed\\conference}} & \revision{\makecell{ISIC 2016}} & \revision{\makecell{skip con.\\residual con.\\feature fusion\\semi-supervised\\self-supervised}} & \revision{\makecell{CE\\Dice}} & \revision{80.49\%} & \revision{\xmark} &\revision{\makecell{flipping, rotation\\zooming, cropping}} & \revision{\xmark} & \revision{\cmark} \\[3em]

\cite{xu2021dc}& \revision{\makecell{peer-reviewed\\conference}} & \revision{\makecell{ISIC 2018}} & \revision{\makecell{Transformer\\multi-scale}} & \revision{\makecell{Dice}} & \revision{89.6\%} & \revision{\xmark} &\revision{\makecell{flipping, rotation}} & \revision{\xmark} & \revision{\xmark} \\[3em]

\cite{ahn2021spatial}& \revision{\makecell{peer-reviewed\\conference}} & \revision{\makecell{PH$^2$}} & \revision{\makecell{self-supervised\\clustering}} & \revision{\makecell{CE\\Spatial loss\\Consistency loss}} & \revision{71.53\%*} & \revision{\xmark} &\revision{\makecell{-}} & \revision{\xmark} & \revision{\cmark} \\[3em]

\cite{zhang2021transfuse}& \revision{\makecell{peer-reviewed\\conference}} & \revision{\makecell{ISIC 2017}} & \revision{\makecell{skip con.\\feature fusion\\Transformer}} & \revision{\makecell{CE\\Jaccard}} & \revision{79.5\%} & \revision{\xmark} &\revision{\makecell{rotation, flipping\\color jittering}} & \revision{\xmark} & \revision{\cmark} \\[3em]

\cite{ji2021multi}& \revision{\makecell{peer-reviewed\\conference}} & \revision{\makecell{ISIC 2018}} & \revision{\makecell{skip con.\\multi-scale\\Transformer}} & \revision{\makecell{CE\\Dice}} & \revision{82.4\%*} & \revision{\xmark} &\revision{\makecell{flipping}} & \revision{\xmark} & \revision{\cmark} \\[3em]

\cite{wang2021boundary}& \revision{\makecell{peer-reviewed\\conference}} & \revision{\makecell{ISIC 2016\\\textbf{ISIC 2018}\\PH$^2$}} & \revision{\makecell{multi-scale\\Transformer}} & \revision{\makecell{CE\\Dice}} & \revision{84.3\%*} & \revision{\cmark} &\revision{\makecell{flipping, scaling}} & \revision{\xmark} & \revision{\cmark} \\[3em]

\cite{yang2021deep}& \revision{\makecell{peer-reviewed\\journal}} & \revision{\makecell{\textbf{ISIC 2018}\\PH$^2$}} & \revision{\makecell{skip con.\\multi-scale\\feature fusion}} & \revision{\makecell{CE\\Dice}} & \revision{94.0\%} & \revision{\xmark} &\revision{\makecell{rotation, flipping\\cropping, HSC\\manipulation, luminance\\and contrast shift}} & \revision{\xmark} & \revision{\xmark} \\[3em]

\cite{tao2021attention}& \revision{\makecell{peer-reviewed\\journal}} & \revision{\makecell{\textbf{ISIC 2017}\\PH$^2$}} & \revision{\makecell{skip con.\\dense con.\\attention mod.\\multi-scale}} & \revision{\makecell{-}} & \revision{78.85\%} & \revision{\xmark} &\revision{\makecell{rotation}} & \revision{\xmark} & \revision{\xmark} \\[3em]

\cite{kim2021simple}& \revision{\makecell{peer-reviewed\\journal}} & \revision{\makecell{\textbf{ISIC 2016}\\PH$^2$}} & \revision{\makecell{residual con.\\skip con.}} & \revision{\makecell{boundary\\aware loss}} & \revision{84.33\%*} & \revision{\xmark} &\revision{\makecell{-}} & \revision{\xmark} & \revision{\xmark} \\[3em]

\cite{dai2022ms}& \makecell{peer-reviewed\\journal} & \makecell{\textbf{ISIC2018}\\PH2} & \makecell{residual con.\\skip con.\\dilated conv.\\image pyramid\\attention mod.} & \makecell{CE\\Dice\\SoftDice} & 83.45\% & \cmark &\makecell{cropping, flipping\\rotation} & \xmark & \xmark\\[3em]

\cite{bi2022hyper}& \makecell{peer-reviewed\\journal} & \makecell{ISIC2016\\\textbf{ISIC2017}\\PH2} & \makecell{residual con.\\skip con.\\attention mod.\\\revision{feature fusion}} & \makecell{CE} & 83.70\% & \cmark &\makecell{cropping, flipping} & \xmark & \xmark \\[3em]

\cite{lin2022contrans}& \revision{\makecell{peer-reviewed\\conference}} & \revision{\makecell{\textbf{ISIC 2017}\\ISIC 2018}} & \revision{\makecell{attention mod.\\Transformer}} & \revision{\makecell{CE\\Jaccard\\DS}} & \revision{77.81\%*} & \revision{\xmark} &\revision{\makecell{flipping, rotation}} & \revision{\xmark} & \revision{\xmark} \\[3em]

\cite{wu2022seatrans}& \revision{\makecell{peer-reviewed\\conference}} & \revision{\makecell{PH$^2$}} & \revision{\makecell{skip con.\\Transformer\\multi-scale}} & \revision{\makecell{CE}} & \revision{70.0\%*} & \revision{\xmark} &\revision{\makecell{-}} & \revision{\xmark} & \revision{\xmark} \\[3em]

\cite{valanarasu2022unext}& \revision{\makecell{peer-reviewed\\conference}} & \revision{\makecell{ISIC 2018}} & \revision{\makecell{skip con.}} & \revision{\makecell{CE\\Dice}} & \revision{81.7\%} & \revision{\xmark} &\revision{\makecell{-}} & \revision{\xmark} & \revision{\cmark} \\[3em]

\cite{basak2022mfsnet}& \revision{\makecell{peer-reviewed\\journal}} & \revision{\makecell{\textbf{ISIC 2017}\\PH$^2$\\HAM10000}} & \revision{\makecell{residual con.\\multi-scale\\attention mod.}} & \revision{\makecell{CE\\Jaccard\\DS}} & \revision{97.4\%} & \revision{\xmark} &\revision{\makecell{-}} & \revision{\xmark} & \revision{\cmark} \\[3em]

\cite{wu2022fat}& \revision{\makecell{peer-reviewed\\journal}} & \revision{\makecell{ISIC 2016\\\textbf{ISIC 2017}\\ISIC 2018\\PH$^2$}} & \revision{\makecell{skip con.\\residual con.\\attention mod.\\Transformer}} & \revision{\makecell{CE\\Dice}} & \revision{76.53\%} & \revision{\xmark} &\revision{\makecell{flipping, rotation\\brightness change\\contrast change\\change in H,S,V}} & \revision{\xmark} & \revision{\cmark} \\[3em]

\cite{liu2022ncrnet}& \revision{\makecell{peer-reviewed\\journal}} & \revision{\makecell{ISIC 2017}} & \revision{\makecell{skip con.\\residual con.\\dilated conv.\\attention mod.}} & \revision{\makecell{CE\\Dice}} & \revision{78.62\%} & \revision{\xmark} &\revision{\makecell{flipping, rotation}} & \revision{\xmark} & \revision{\xmark} \\[3em]

\cite{wang2022net}& \revision{\makecell{peer-reviewed\\journal}} & \revision{\makecell{ISIC 2017}} & \revision{\makecell{skip con.\\residual con.\\Transformer}} & \revision{\makecell{-}} & \revision{84.52\%} & \revision{\xmark} &\revision{\makecell{flipping, rotation}} & \revision{\xmark} & \revision{\cmark} \\[3em]

\cite{zhang2022feature}& \revision{\makecell{peer-reviewed\\conference}} & \revision{\makecell{ISIC 2017}} & \revision{\makecell{skip con.\\feature fusion}} & \revision{\makecell{Dice\\Focal}} & \revision{74.54\%} & \revision{\xmark} &\revision{\makecell{flipping}} & \revision{\xmark} & \revision{\xmark} \\[3em]

\cite{wang2022superpixel}& \revision{\makecell{peer-reviewed\\conference}} & \revision{\makecell{\textbf{ISIC 2017}\\PH$^2$}} & \revision{\makecell{skip con.\\residual con.\\self-supervised}} & \revision{\makecell{Dice}} & \revision{76.5\%} & \revision{\cmark} &\revision{\makecell{rotation, flipping\\color jittering}} & \revision{\xmark} & \revision{\xmark} \\[3em]

\cite{dong2022tc}& \revision{\makecell{peer-reviewed\\journal}} & \revision{\makecell{ISIC 2016\\\textbf{ISIC 2017}\\ISIC 2018}} & \revision{\makecell{residual con.\\skip con.\\Transformer\\feature fusion}} & \revision{\makecell{CE\\Dice}} & \revision{74.55\%} & \revision{\xmark} &\revision{\makecell{-}} & \revision{\xmark} & \revision{\xmark} \\[3em]

\cite{chen2022skin}& \revision{\makecell{peer-reviewed\\journal}} & \revision{\makecell{\textbf{ISIC 2017}\\PH$^2$}} & \revision{\makecell{skip con.\\attention mod.\\recurrent net.}} & \revision{\makecell{CE}} & \revision{80.36\%} & \revision{\cmark} &\revision{\makecell{flipping, rotation\\affine trans.\\masking, mesh distortion}} & \revision{\xmark} & \revision{\xmark} \\[3em]

\cite{kaur2022automatic}& \revision{\makecell{peer-reviewed\\journal}} & \revision{\makecell{ISIC 2016\\\textbf{ISIC 2017}\\ISIC 2018\\PH$^2$}} & \revision{\makecell{dilated conv.}} & \revision{\makecell{CE}} & \revision{81.7\%} & \revision{\cmark} &\revision{\makecell{scaling, rotation\\translation}} & \revision{\xmark} & \revision{\xmark} \\[3em]

\cite{badshah2022resbcu}& \revision{\makecell{peer-reviewed\\journal}} & \revision{\makecell{ISIC 2018}} & \revision{\makecell{residual con.\\BConvLSTM}} & \revision{\makecell{-}} & \revision{94.5\%} & \revision{\xmark} &\revision{\makecell{-}} & \revision{\xmark} & \revision{\xmark} \\[3em]

\cite{alam2022s2c}& \revision{\makecell{peer-reviewed\\journal}} & \revision{\makecell{HAM10000}} & \revision{\makecell{residual con.\\separable conv.}} & \revision{\makecell{Dice}} & \revision{91.1\%} & \revision{\xmark} &\revision{\makecell{-}} & \revision{\xmark} & \revision{\cmark} \\[3em]

\cite{yu2022mca}& \revision{\makecell{peer-reviewed\\journal}} & \revision{\makecell{ISIC 2018}} & \revision{\makecell{skip con.\\attention mod.\\multi-scale}} & \revision{\makecell{-}} & \revision{87.89\%} & \revision{\xmark} &\revision{\makecell{-}} & \revision{\xmark} & \revision{\xmark} \\[3em]

\cite{jiang2022seacu}& \revision{\makecell{peer-reviewed\\journal}} & \revision{\makecell{\textbf{ISIC 2017}\\ISIC 2018}} & \revision{\makecell{skip con.\\attention mod.\\ConvLSTM}} & \revision{\makecell{CE\\Jaccard}} & \revision{80.5\%} & \revision{\xmark} &\revision{\makecell{-}} & \revision{\xmark} & \revision{\xmark} \\[3em]

\cite{ramadan2022color}& \revision{\makecell{peer-reviewed\\journal}} & \revision{\makecell{ISIC 2018}} & \revision{\makecell{skip con.\\attention mod.}} & \revision{\makecell{CE\\Dice\\sens.-spec. loss}} & \revision{91.4\%} & \revision{\xmark} &\revision{\makecell{-}} & \revision{\xmark} & \revision{\xmark} \\[3em]

\cite{zhang2022dynamic}& \revision{\makecell{peer-reviewed\\journal}} & \revision{\makecell{\textbf{ISIC 2017}\\ISIC 2018}} & \revision{\makecell{skip con.\\dense con.\\semi-supervised}} & \revision{\makecell{CE\\contrastive loss}} & \revision{73.89\%} & \revision{\xmark} &\revision{\makecell{scaling, flipping\\color distortion}} & \revision{\xmark} & \revision{\xmark} \\[3em]

\cite{tran2022fully}& \revision{\makecell{peer-reviewed\\journal}} & \revision{\makecell{\textbf{ISIC 2017}\\PH$^2$}} & \revision{\makecell{skip con.\\attention mod.}} & \revision{\makecell{Focal Tversky\\fuzzy loss}} & \revision{79.2\%} & \revision{\xmark} &\revision{\makecell{rotation, zooming\\flipping}} & \revision{\xmark} & \revision{\xmark} \\[3em]

\cite{wang2022skin}& \revision{\makecell{peer-reviewed\\journal}} & \revision{\makecell{ISIC 2017}} & \revision{\makecell{skip con.\\residual con.\\attention mod.}} & \revision{\makecell{CE\\Jaccard}} & \revision{78.28\%} & \revision{\xmark} &\revision{\makecell{rotation, zooming\\resizing, shifting}} & \revision{\xmark} & \revision{\xmark} \\[3em]

\cite{zhao2022self}& \revision{\makecell{peer-reviewed\\conference}} & \revision{\makecell{ISIC 2017}} & \revision{\makecell{skip con.\\self-supervised}} & \revision{\makecell{CE\\Dice}} & \revision{67.08\%*} & \revision{\xmark} &\revision{\makecell{-}} & \revision{\xmark} & \revision{\xmark} \\[3em]

\cite{wang2022cross}& \revision{\makecell{peer-reviewed\\conference}} & \revision{\makecell{PH$^2$}} & \revision{\makecell{few shot\\mask avg. pooling}} & \revision{\makecell{Dice}} & \revision{86.97\%*} & \revision{\xmark} &\revision{\makecell{-}} & \revision{\xmark} & \revision{\xmark} \\[3em]

\cite{wang2022ctcnet}& \revision{\makecell{peer-reviewed\\conference}} & \revision{\makecell{\textbf{ISIC 2017}\\ISIC 2018}} & \revision{\makecell{residual con.\\dilated conv.\\multi-scale\\feature fusion\\Transformer}} & \revision{\makecell{CE\\Jaccard}} & \revision{78.76\%} & \revision{\xmark} &\revision{\makecell{-}} & \revision{\xmark} & \revision{\xmark} \\[3em]

\cite{liu2022skin}& \revision{\makecell{peer-reviewed\\conference}} & \revision{\makecell{\textbf{ISIC 2017}\\ISIC 2018}} & \revision{\makecell{skip con.\\dilated conv.\\multi-scale\\pyramid pooling\\Transformer}} & \revision{\makecell{CE}} & \revision{80.19\%} & \revision{\xmark} &\revision{\makecell{-}} & \revision{\xmark} & \revision{\xmark} \\[3em]

\cite{gu2022net}& \revision{\makecell{peer-reviewed\\journal}} & \revision{\makecell{ISIC 2017}} & \revision{\makecell{skip con.\\global adaptive\\pooling}} & \revision{\makecell{CE\\$\ell_2$}} & \revision{80.53\%} & \revision{\xmark} &\revision{\makecell{scaling, rotation\\flipping}} & \revision{\xmark} & \revision{\xmark} \\[3em]

\cite{khan2022ensemble}& \revision{\makecell{peer-reviewed\\journal}} & \revision{\makecell{\textbf{ISIC 2017}\\PH$^2$}} & \revision{\makecell{residual con.\\attention mod.\\ensemble}} & \revision{\makecell{CE}} & \revision{79.2\%} & \revision{\xmark} &\revision{\makecell{-}} & \revision{\xmark} & \revision{\xmark} \\[3em]

\cite{alahmadi2022semi}& \revision{\makecell{peer-reviewed\\journal}} & \revision{\makecell{\textbf{ISIC 2017}\\ISIC 2018\\PH$^2$}} & \revision{\makecell{skip con.\\feature fusion\\semi-supervised\\Transformer}} & \revision{\makecell{CE\\Dice\\$\ell_2$}} & \revision{82.78\%*} & \revision{\xmark} &\revision{\makecell{-}} & \revision{\xmark} & \revision{\xmark} \\[3em]

\cite{li2022mhau}& \revision{\makecell{peer-reviewed\\journal}} & \revision{\makecell{ISIC 2018}} & \revision{\makecell{skip con.\\residual con.\\dilated conv.\\attention mod.\\pyramid pooling\\multi-scale}} & \revision{\makecell{CE\\Dice}} & \revision{88.92\%} & \revision{\xmark} &\revision{\makecell{flipping, rotation}} & \revision{\xmark} & \revision{\xmark} \\[3em]

\cite{kaur2022skin}& \revision{\makecell{peer-reviewed\\journal}} & \revision{\makecell{ISIC 2016\\\textbf{ISIC 2017}\\ISIC 2018\\PH$^2$}} & \revision{\makecell{-}} & \revision{\makecell{Tversky}} & \revision{77.8\%} & \revision{\cmark} &\revision{\makecell{rotation, scaling}} & \revision{\xmark} & \revision{\xmark} \\[3em]

\hline

\end{longtable}
\end{tiny}
\end{center}

\bibliographystyle{model2-names.bst}\biboptions{authoryear}
\bibliography{references}

\end{document}